\documentclass[journal]{IEEEtran}
\usepackage{graphicx}
\usepackage{url}
\usepackage{enumitem}
\usepackage{xurl}
\usepackage{multirow}
\usepackage{tikz}
\usepackage{amsfonts}
\usepackage{colortbl}
\usepackage{threeparttable, tablefootnote}
\usepackage{dblfloatfix}
\usepackage{cite}
\usepackage[hidelinks]{hyperref} 

\usepackage{enumitem}

\newcommand\like[1]{\begin{picture}(1,1)
\ifnum0=#1\put(.5,.35){\circle{1}}\else
\ifnum10=#1\put(.5,.35){\circle*{1}}\else
\put(.5,.35){\circle{1}}\put(.5,.35){\circle*{.#1}}
\fi\fi\end{picture}}

\ifCLASSINFOpdf
\else
\fi
\begin{document}
%
\title{\huge A Comprehensive Survey on SmartNICs: Architectures, Development Models, Applications, and Research Directions}
%
%
%

\author{Elie Kfoury, Samia Choueiri, Ali Mazloum, Ali AlSabeh, Jose Gomez, and~Jorge Crichigno
\thanks{E. Kfoury, S. Choueiri, A. Mazloum, A. AlSabeh, J. Gomez and J. Crichigno are with the Integrated Information Technology Department at the University of South Carolina, USA (\{ekfoury, choueiri, amazloum, aalsabeh, gomezgaj\}@email.sc.edu, jcrichigno@cec.sc.edu. \\
This work has been submitted to the IEEE for possible publication. Copyright may be transferred without notice, after which this version may no longer be accessible.
}

}

\maketitle

\begin{abstract}
The end of Moore's Law and Dennard Scaling has slowed processor improvements in the past decade. While multi-core processors have improved performance, they are limited by the application's level of parallelism, as prescribed by Amdahl’s Law. This has led to the emergence of domain-specific processors that specialize in a narrow range of functions. Smart Network Interface Cards (SmartNICs) can be seen as an evolutionary technology that combines heterogeneous domain-specific processors and general-purpose cores to offload infrastructure tasks. Despite the impressive advantages of SmartNICs and their importance in modern networks, the literature has been missing a comprehensive survey. To this end, this paper provides a background encompassing an overview of the evolution of NICs from basic to SmartNICs, describing their architectures, development environments, and advantages over legacy NICs. The paper then presents a comprehensive taxonomy of applications offloaded to SmartNICs, covering network, security, storage, and machine learning functions. Challenges associated with SmartNIC development and deployment are discussed, along with current initiatives and open research issues.

\end{abstract}

\begin{IEEEkeywords}
SmartNIC, Data Processing Unit (DPU), Infrastructure Processing Unit (IPU), Moore's law, application offloading, P4, Application Specific Integrated Circuit (ASIC), Field Programmable Gate Array (FPGA).
\end{IEEEkeywords}

%
\IEEEpeerreviewmaketitle

\section{Introduction}
\begin{figure}[t]
    \centering
  \includegraphics[width=0.48\textwidth]{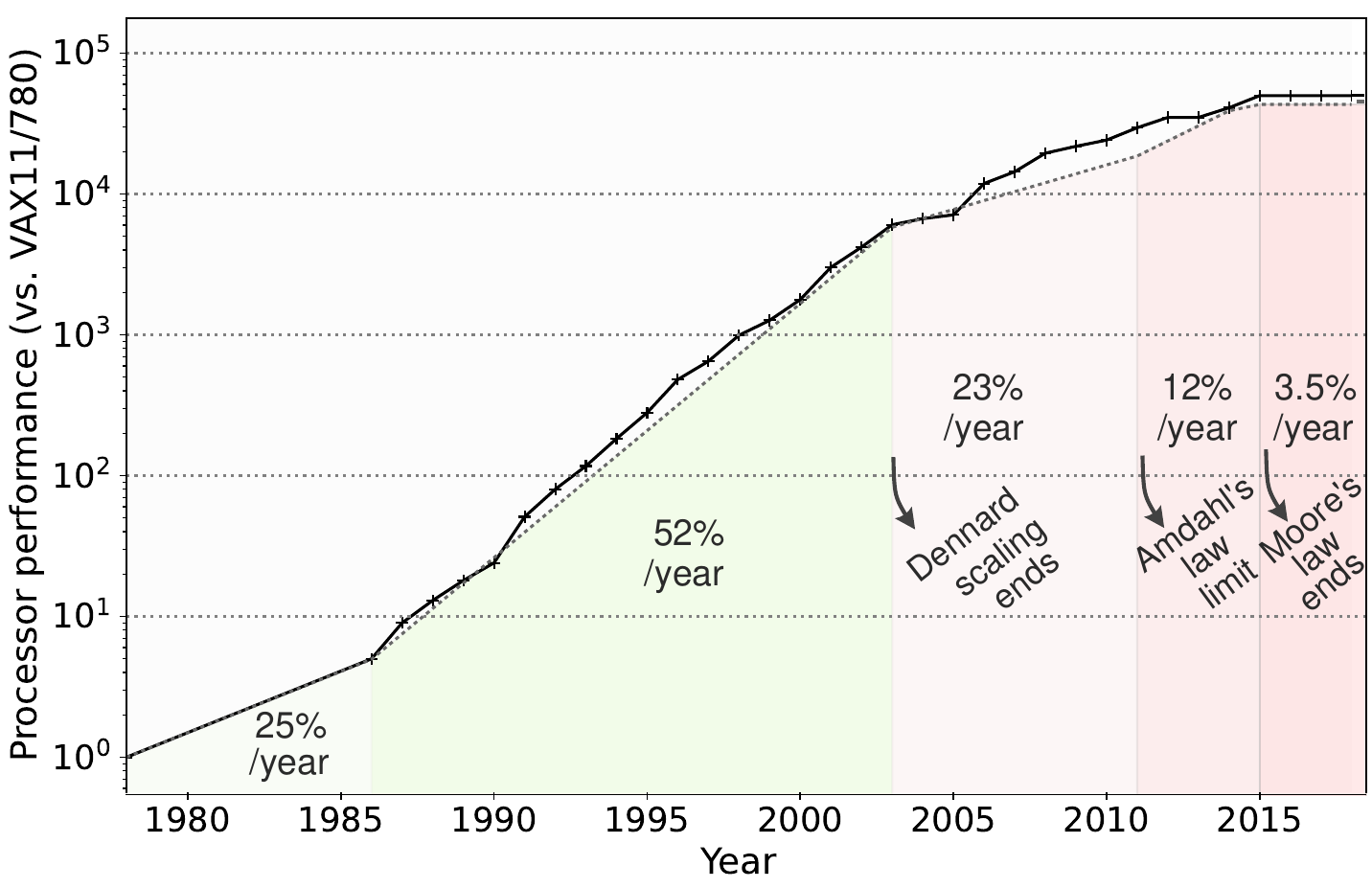}
  \caption{Growth in processor performance over 40 years, relative to the VAX 11/780 as measured by the SPEC integer benchmarks. Reproduced from \cite{hennessy2011computer}.}
  \label{fig:proc_year}  
\end{figure}

%
%
%
%
\IEEEPARstart{I}{n} 1965, Gordon Moore predicted that the number of transistors per chip would double every year \cite{moore1998cramming}, which was updated in 1975 to every two years \cite{moore1975progress}. In 1974, Robert Dennard noted that power density was constant for a given silicon area even as the number of transistors increased because of the smaller dimensions of each transistor. Transistors used less power and the performance of integrated circuits was enhanced by packing more transistors per chip \cite{dennard1974design}. The ability of the microprocessor, or simply processor, to exploit the advances in integrated circuits enabled impressive performance improvements, see Fig. \ref{fig:proc_year} \cite{hennessy2011computer}. Unfortunately, in 2003, the limits of power due to the end of Dennard Scaling slowed processor performance to 23\%. This observation forced the industry to use multiple processors per chip, referred to as cores. While multi-core processors helped improve performance, they have slowed down in the last decade, because of the natural limits prescribed by Amdahl’s Law \cite{amdahl1967validity}: there is a maximum performance benefit from parallelism, as applications also have tasks that must be executed sequentially. Additionally, Moore’s law has recently ended, resulting in the improvements of processors to slow down further.

\begin{figure}[t]
    \centering
  \includegraphics[width=0.489\textwidth]{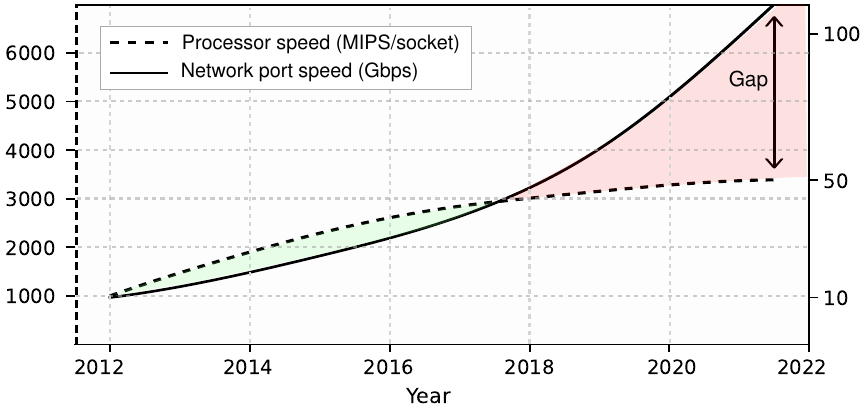}
  \caption{Network port speeds (solid line, right y-axis) versus processor speed (dotted line, left y-axis) over the years. Port speeds are increasing while processor speeds are plateauing. Reproduced from \cite{elinoff2020smartnic}.}
  \label{fig:link_vs_CPU}  
\end{figure}

In today’s world, most data arrive at compute locations as packets from the networks. The traditional communication channel connecting networks and hosts is the Network Interface Card (NIC). In the past NICs were simple hardware-based devices that received the packets from the network and placed them in memory at the host. Packets would then wait for processing time by the general-purpose processor at the host \cite{faircloth2013book}. Although this model was successful for a long time, it has several challenges in current environments:
\begin{itemize}[leftmargin=*]
\itemsep0em
    \item As Moore’s law and Denning scaling ended, simply adding more processing capacity to cope with the increasing amount of traffic is no longer an option.
    \item A large percentage of tasks executed by the processors relate to the infrastructure rather than to the user applications, e.g., TCP/IP tasks, encryption, compression, etc. Such operations use valuable processor cycles that may be used for application tasks instead.
    \item Historical software solutions for packet-related tasks are not efficient in terms of throughput, latency, and energy. While in the past inefficient software solutions were mitigated by the relentless progress of the (hardware) processors, today’s solutions can no longer rely on future improvements in processor performance.
    \item The explosion of network traffic is accompanied by impressive improvement in the physical layer and bandwidth capacity. As network traffic arrives at servers at higher rates, processors are unable to process it on time, and the gap between processor performance and bandwidth is only increasing, see Fig. \ref{fig:link_vs_CPU}.
\end{itemize}

Since the Dennard Scaling ended and the energy budget is no longer increasing, many consider that the only path left to improve energy, performance, and cost is by using domain-specific processors rather than power-hungry general-purpose processors. SmartNICs can be seen as a revolutionary technology developed to address the challenges listed above by combining heterogeneous domain-specific processors that specialize in a narrow range of infrastructure tasks. These include compression/decompression processors,  programmable pipelines, encryption/decryption processors, and others. SmartNICs also include general-purpose processors which are used for managing the system, aiding the domain-specific processors, and enabling users to run control-plane applications. In the context of SmartNICs, the terms accelerators and engines are also used to refer to domain-specific processors \cite{ibanez2019case,avidthink2022report}. Note that the development of domain-specific processors has been successfully used in several domains, including graphics in the 2000s –Graphics Processing Units (GPUs)–, machine learning in mid 2010s –Tensor Processor Units (TPUs)–, networking in the late 2010s –Network Processor Units (NPUs) that adhere to architecture models such as the Protocol Independent Switch Architecture (PISA)–, and genomics in 2018 \cite{ageevexploring,tell2001domain,derek2022hardware}.

The momentum of SmartNICs is reflected in the global Information Technology (IT) ecosystem. Hyperscalers such as Google, Amazon, and Microsoft are designing their own SmartNICs to run infrastructure functions and optimize revenue and performance \cite{google2022,morra2022aws,microsoft2019}. Manufacturers such as Intel, NVIDIA, and AMD are emphasizing the development of SmartNICs for a broad market range, offering Systems on a Chip (SoCs) with programmable domain-specific processors for security, networks, storage, and telemetry \cite{scott2023summit}. Cloud systems such as the Monterey project are redefining cloud architectures by incorporating SmartNICs to run storage, networks, and security services, resulting in substantial improvement in performance while leaving more processor cycles for user applications \cite{ibanez2019case,avidthink2022report}. Research and education networks (RENs) such as the Energy Sciences Network (ESnet) –the high-performance network that carries traffic for the U.S. Department of Energy and research organizations– are upgrading their infrastructures with SmartNICs to enable data-intensive science \cite{esnetsmartnic}. Software vendors are also offloading their solutions to SmartNICs; VMware’s ESXi, vCenter, and NSX -integral components for virtualizing High Performance Computing (HPC) environments- can now be effectively offloaded onto SmartNICs \cite{vmware}. Palo Alto Networks, a leading Next-Generation Firewall (NGFW) vendor, introduced the “Intelligent Traffic Offload” service \cite{paloalto}; this service offloads firewall functions to SmartNICs. Juniper Networks’ virtual router/firewall can also be offloaded to SmartNICs \cite{juniper}. Telecommunication operators are increasingly migrating their core services to run on SmartNICs \cite{telco}. Serverless and edge computing workloads, including Machine Learning (ML) training and inference, can be accelerated using SmartNICs \cite{tootaghaj2022smartnics,zheng2023network}. Testbeds such as FABRIC \cite{baldin2019fabric} and GEANT \cite{geant}, used worldwide for fundamental research, rely on SmartNICs and other programmable devices to allow experimenters to program the data path behavior and process network traffic in novel ways at line rate \cite{geant2023,da2023fabric}.

\begin{figure}[t]
    \centering
  \includegraphics[width=0.489\textwidth]{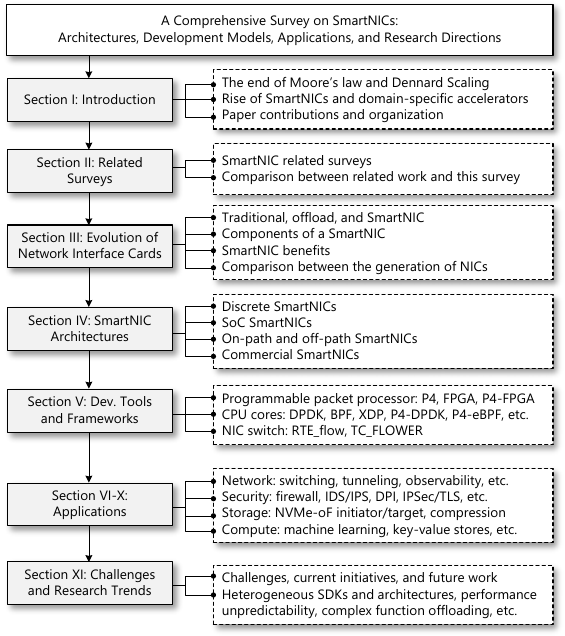}
  \caption{Paper roadmap.}
  \label{fig:paper_roadmap}  
\end{figure}

\begin{table*}[t]
\centering
\caption{Comparison with related surveys.}
\begin{tabular}{|c|c|c|c|c|c|c|c|}
\hline \hline
Paper & \begin{tabular}[c]{@{}c@{}}Evolution and\\ definition\end{tabular} & \begin{tabular}[c]{@{}c@{}}Architectures\\ and models\end{tabular} & \begin{tabular}[c]{@{}c@{}}Development\\ environments\end{tabular} & \begin{tabular}[c]{@{}c@{}}Applications and offloaded \\ workloads taxonomy\end{tabular} & \begin{tabular}[c]{@{}c@{}}Comparisons\\ with regular NICs\end{tabular} & \begin{tabular}[c]{@{}c@{}}Challenges and \\ discussions\end{tabular} & \begin{tabular}[c]{@{}c@{}}Research trends \\ and directions\end{tabular} \\ \hline \hline
\cite{cerovic2018fast} & \like{0} & \like{3} & \like{3} & \like{0} & \like{0} & \like{3} & \like{3} \\ \hline
\cite{freitas2022survey} & \like{0} & \like{3} & \like{3} & \like{0} & \like{0} & \like{3} & \like{3} \\ \hline
\cite{linguaglossa2019survey} & \like{3} & \like{3} & \like{3} & \like{3} & \like{0} & \like{3} & \like{3} \\ \hline
\cite{fei2020paving} & \like{0} & \like{3} & \like{3} & \like{3} & \like{0} & \like{3} & \like{3} \\ \hline
\cite{shantharama2020hardware} & \like{3} & \like{3} & \like{3} & \like{3} & \like{3} & \like{3} & \like{3} \\ \hline
\cite{vieira2020fast} & \like{0} & \like{0} & \like{3} & \like{3} & \like{0} & \like{3} & \like{3} \\ \hline
\cite{rosa2024empowering} & \like{3} & \like{3} & \like{3} & \like{3} & \like{0} & \like{3} & \like{3} \\ \hline
\begin{tabular}[c]{@{}c@{}}This \\ survey\end{tabular} & \like{10} & \like{10} & \like{10} & \like{10} & \like{10} & \like{10} & \like{10} \\ \hline \hline
\end{tabular}
 \begin{tablenotes}
    \centering
        \item[1] \like{10} Covered in this survey \quad \like{0} Not covered in this survey \quad \like{3} Partially covered in this survey
    \end{tablenotes}
\label{table:related_surveys}
\end{table*}

\subsection{Paper Contributions}
Despite the increasing interest in SmartNICs, prior research has only partially covered this technology. As shown in Table \ref{table:related_surveys}, there is currently no updated and comprehensive material on SmartNICs. This paper addresses this gap by providing an overview of the evolution of NICs, starting from traditional basic NICs to SmartNICs. It describes the hardware architectures, technologies, and software development environments used with SmartNICs, as well as the advantages that SmartNICs offer over legacy NICs. The paper then proposes a taxonomy of the functions and applications being offloaded to SmartNICs, illustrating their advantages over the conventional method of executing such applications. Additionally, the paper discusses the challenges associated with SmartNICs and concludes by discussing future perspectives and open research issues.

\subsection{Paper Organization}
The road map of this survey is depicted in Fig. \ref{fig:paper_roadmap}. Section \ref{sec:related_surveys} compares existing surveys on SmartNICs and related technologies and demonstrates the novelty of this work. Section \ref{sec:evolution_NICs} presents an overview of the evolution of NICs, from traditional basic NICs to SmartNICs. It describes the components of SmartNICs and their benefits compared to legacy NICs. Section \ref{sec:architectures} describes the SmartNICs hardware architectures. Section \ref{sec:dev} describes the tools, frameworks, and development environments for SmartNICs, both open-source and vendor-specific.
Section \ref{sec:taxonomy} provides a taxonomy of the applications and infrastructure workloads that are offloaded to SmartNICs. The subsequent sections (Section \ref{sec:security}-\ref{sec:compute}) describe the security, network, storage, and compute functions. Section \ref{sec:challenges} lists challenges associated with SmartNICs. It then discusses current initiatives that overcome the challenges and provides a reflection on open research issues. Section \ref{sec:conclusion} concludes the paper. The abbreviations used in this article are summarized in Table \ref{tab:abbrev}, at the end of the article.

\section{Related Surveys}\label{sec:related_surveys}
Despite the widespread interest from both industry and academia in SmartNICs, there is a noticeable absence of a comprehensive survey that adequately explores their potential and ongoing research endeavors. The existing surveys that are closest to this paper can be divided into 1) packet processing acceleration; and 2) programmable data planes.

\subsection{Surveys on Packet Processing Acceleration}
The existing surveys in this category discuss the advantages of accelerating packet processing, particularly with software technologies. However, while SmartNICs are occasionally mentioned in these surveys, they fail to delve into crucial aspects such as their potential, architectures, applications, etc.


Cerović et al. \cite{cerovic2018fast} discuss various software-based and hardware-based packet accelerators. The survey focuses on server-class networking. It first starts by explaining the problems associated with using the standard Linux kernel for packet processing in high-speed networks and then delves into exploring the different classes of packet accelerators. For the software-based packet accelerators, the survey mainly describes and analyzes Data Plane Development Kit (DPDK) \cite{dpdkHomeDPDK}, PF\_RING \cite{ntopPF_RING}, NetSlices \cite{marian2012netslices}, and Netmap \cite{rizzo2012netmap}. For the hardware-based packet accelerators, it focuses mainly on leveraging GPUs and  Field Programmable Gate Arrays (FPGAs) for optimized and efficient packet processing. The survey does not cover the latest generation of SmartNICs that include CPU cores and domain-specific accelerators. Also, the survey does not cover the applications or the infrastructure workloads that can be offloaded to SmartNICs. 



Freitas et al. \cite{freitas2022survey} describe multiple packet processing acceleration techniques. The survey focuses on packet processing in Linux environments. It categorizes the packet processing acceleration into hardware, software, and virtualization-based. For each category, the survey offers background information and discusses a simple use case. The survey also provides discussions on the host resource usage efficiency, the high packet rate, the system security, and the flexibility/expandability. The survey briefly mentions programmable NICs (another term used for SmartNICs) and their role in accelerating packet processing. It does not cover their development environments, hardware architectures, and the applications/workloads that can be offloaded. 

Linguaglossa et al. \cite{linguaglossa2019survey} focus on software and hardware technologies that accelerate Network Function Virtualization (NFV). It categorizes software acceleration technologies into pure software acceleration and hardware-supported functions in software. It also provides a brief overview of the software acceleration ecosystem which includes DPDK, XDP, Netmap, and PF\_RING. For the hardware technologies, it discusses the offloading functions of traditional NICs (e.g., CRC calculation, checksum computation, and TCP Offload Engine (TOE)) and a subset of the hardware architectures of SmartNICs. Then, it provides a brief overview of the programming abstractions in SmartNICs. The survey has the following limitations: 1) it does not cover all the hardware architectures; 2) it does not cover the development tools and environments; and 3) it does not cover the applications and infrastructure workloads that can be offloaded to SmartNICs.

\begin{figure*}[t]
  \includegraphics[width=\textwidth]{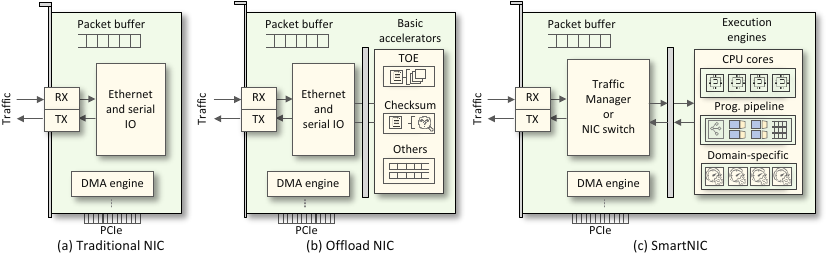}
  \caption{Main functional blocks of (a) traditional NICs, (b) offload NICs, and (c) SmartNICs.}
  \label{fig:nics}  
\end{figure*}

Fei et al. \cite{fei2020paving} also focus on NFV acceleration. The survey classifies NVF acceleration into three high-level categories: computation, communication, and traffic steering. Under the computation category, the survey discusses some hardware offloading architectures which include SmartNICs. The remaining of the survey focuses on software acceleration and how to tune the system to achieve better performance. The survey has the following limitations: 1) it does not cover the hardware architectures used by the latest generation of SmartNICs; 2) it does not cover the development tools and environments; 3) it does not cover the applications and workloads that can be offloaded to the SmartNIC. 

Shantharama et al. \cite{shantharama2020hardware} provide a comprehensive survey on softwarized NFs. The survey classifies the CPU, the memory, and the interconnects as the three main enabling technologies for NFVs. With low-level details, the survey explains how each class operates and how it can be optimized to provide better virtualization support. It also discusses the use of dedicated hardware accelerators (FPGAs, ASICs, etc.) to improve the performance of softwarized NFs. The survey briefly describes some of the applications offloaded to SmartNICS without providing a sufficient overview of the technology, the different available development environments, or the latest enhancement in the field of SmartNICs.


Vieira et al. \cite{vieira2020fast} only focus on the extended Berkeley Packet Filter (eBPF) and the eXpress Data Path (XDP) software acceleration techniques. The survey illustrates the process of enhancing packet processing speed by running eBPF-based applications in the XDP layer of the Linux kernel network stack. It presents a tutorial that includes the compilation and verification processes, the program structure, the required tools, and walk-through example programs. Although the authors mentioned SmartNICs as a target platform for eBPF applications, it does not cover the available architectures of SmartNICs, the development environments, or the applications that can be offloaded to SmartNICs.


Rosa et al. \cite{rosa2024empowering} describe multiple software and hardware techniques to enhance packet processing speed in the cloud. While discussing software-based techniques, the survey focuses on zero-copy data transfers, minimal context switching, and asynchronous processing as the core techniques for network acceleration. After that, it shows how DPDK, XDP, and eBPF are used in the cloud to enable Network Acceleration as a Service (NAaaS). While discussing hardware-based techniques, the survey only focuses on RDMA. The authors only describe SmartNICs as an enabling technology for RDMA and virtualization without describing their different architectures, development environments, of their different capabilities for enhancing network acceleration.

\subsection{Surveys on Programmable Data Planes}
Numerous surveys have covered the general aspects of programmable data planes in the past few years \cite{kfoury2021exhaustive, hauser2023survey, michel2021programmable, kaljic2019survey, da2017data}. Some surveys focused on specific areas such as network security \cite{gao2021review,alsabeh2022survey,chen2023empowering}, ML training and inference \cite{parizotto2023offloading,quan2022ai}, TCP enhancements \cite{gomez2022survey}, virtualization and cloud computing \cite{han2020virtualization}, 5G and telecommunications \cite{brito2023programmable}, rerouting and fast recovery \cite{mazloum2023survey, chiesa2023survey}. All these surveys have discussed some applications developed on SmartNICs. However, their focus is on programmable switches (e.g., Intel's Tofino). Recent advances in SmartNICs are not covered in these surveys.

\subsection{Novelty}
Table \ref{table:related_surveys} summarizes the topics and the features described in the related surveys. It also highlights how this paper differs
from the existing surveys. To the best of the authors’ knowledge, this work is the first to exhaustively explore the whole SmartNIC ecosystem. Unlike previous surveys, this survey provides in-depth discussions on the evolution and definition of SmartNICs, the common architectures used by various SmartNIC models in the market, and the development environments (both open source and proprietary). It then provides a detailed taxonomy covering the applications that are offloaded to SmartNICs, while highlighting the performance gains compared to regular NICs. The survey also presents the challenges associated with programming and deploying SmartNICs, as well as the current and future research trends.

\section{Evolution of Network Interface Cards (NICs)}\label{sec:evolution_NICs}
There are three main generations of NICs: traditional NICs, offload NICs, and SmartNICs. Fig. \ref{fig:nics} shows a simplified diagram of the three NICs.

\subsection{Traditional NICs}
Traditional NICs (Fig. \ref{fig:nics} (a)) are devices that implement basic physical and data-link layer services. These services include serializing/deserializing frames, managing link access, and providing error detection. Typically, these services are executed by a fixed-function component residing on a special-purpose chip within the NIC. On the sending side, the fixed-function component accepts a datagram created by the host, encapsulates it in a link-layer frame, and then transmits the frame into the communication link, following the link-access protocol. On the receiving side, the fixed-function component receives the frame and forwards it to the host via a Peripheral Component Interconnect Express (PCIe) card.

\subsection{Offload NICs}
Offload NICs (Fig. \ref{fig:nics} (b)) incorporate hardware in the form of ASICs and/or FPGAs to execute basic ``infrastructure'' functions\footnote{In this context, infrastructure functions refer to tasks that facilitate data movement to the host and do not involve application data.} that were previously handled by the host. The goal is to free up cycles in the main host’s CPU for application (end-user) tasks rather than infrastructure tasks. Examples of such functions include:
\begin{itemize}[leftmargin=*]
    \item Basic packet processing: parsing and reassembling IP datagrams, computing IP checksum, encapsulating and de-encapsulating TCP segments.
    \item Managing TCP connections on the NIC: connection establishment, checksum and sequence number calculations, TOE, sliding window calculations for segment acknowledgment and congestion control, among others.
    \item Other functions that manipulate TCP/IP header fields to implement basic filtering and traffic classification.
\end{itemize}

Offload NICs allow end users to perform pre-programmed functions on the NIC. However, they do not support the creation and execution of custom applications directly on the NIC. Even with full transport layer offload, application protocols still need to be implemented on the host processor \cite{ageevexploring}.

\subsection{SmartNICs}\label{sec:smartnic_def}
The definition of a SmartNIC is not widely agreed upon. Traditionally, NICs that performed functions beyond basic packet processing were labeled as SmartNICs. Unless otherwise noted, the term SmartNIC in this survey refers to the latest generation of NICs, also known as SoC SmartNICs, Infrastructure Processing Units (IPUs)\footnote{IPU is the terminology used by Intel.}, Data Processing Units (DPUs), and Auxiliary Processing Units (xPUs)\footnote{xPU is used by the Storage Networking Industry Association (SNIA) community.}. 

Fig. \ref{fig:nics} (c) shows a simplified diagram of a SmartNIC. The SmartNIC includes a Traffic Manager (TM) or a NIC switch that performs Quality of Service (QoS) and steers traffic to the NIC execution engines. The NIC execution engines consist of a combination of processors used for custom packet processing and other domain-specific functions. Some SmartNICs (e.g., NIVDIA's BlueField-2 \cite{asic+cpu-bf2}) use a multi-core CPU processor for custom packet processing. Other SmartNICs (e.g., AMD Pensando DSC \cite{asic+cpu-dsc2-200}) use embedded flow engines running a P4 programmable Application Specific Integrated Circuit (ASIC) pipeline. Other SmartNICs (e.g., AMD Xilinx SN1000 \cite{fpga+cpu-sn1000}) use an FPGA for custom packet processing. 
The domain-specific processors are optimized to provide high-performance and energy-efficient processing for a specific set of functions (e.g., cryptography). 
The execution engines have a memory hierarchy that typically consists of an L1 cache, scratchpad, L2 cache, and Dynamic RAM (DRAM). 

The SmartNICs have general-purpose CPU cores for executing control plane functions. The CPU cores also enable SmartNICs to function autonomously and have their own Operating System (OS), such as Ubuntu Linux, which is independent of the host system in which they are running.

\begin{figure}[t]
  \includegraphics[width=0.489\textwidth]{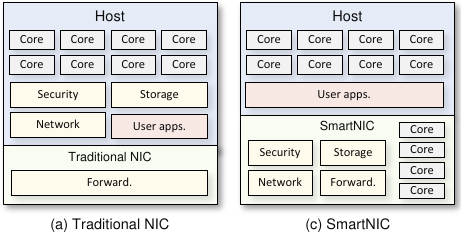}
  \caption{In a deployment with a traditional NIC (a), the host CPU cores execute infrastructure functions and user applications. With SmartNICs (b), the host CPU cores solely execute user applications. The SmartNIC CPU cores assist other accelerators in executing infrastructure functions.}
  \label{fig:deployment}  
\end{figure}

The programmable components of a SmartNIC allows it to execute infrastructure functions, without involving the CPU of the host. Consider Fig. \ref{fig:deployment}. In a deployment with a traditional NIC (a), the host CPU cores execute infrastructure functions (typically classified as network, security, and storage) and user applications; with SmartNICs (b), the host CPU cores solely execute user applications. The SmartNIC CPU cores assist other domain-specific accelerators in executing infrastructure functions. 


\begin{figure}[b]
  \includegraphics[width=0.489\textwidth]{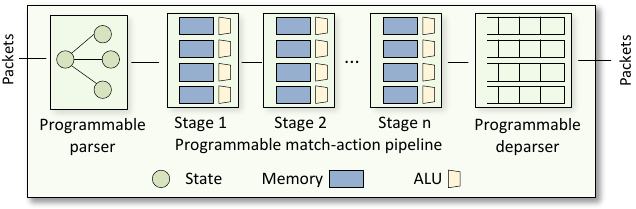}
  \caption{Programmable Pipeline.}
  \label{fig:programmable_packet_processor}  
\end{figure}

\subsubsection{Custom Packet Processing} SmartNICs enable the developers to devise custom packet processing on its execution engines. The packet processing logic can be implemented on CPU cores, FPGAs, or programmable ASIC pipelines. Regardless of the hardware architecture used by the SmartNIC, its packet processing engines include the following components: programmable parser, programmable match-action pipeline, and programmable deparser. These components closely resemble those of the PISA architecture \cite{pisa}, see Fig. \ref{fig:programmable_packet_processor}. The \textit{programmable parser} allows the developer to define headers based on custom or standard protocols and parse them. It is represented as a state machine. The \textit{programmable match-action pipeline} carries out operations on packet headers and intermediate results. Each match-action stage comprises multiple memory blocks (such as tables and registers) and Arithmetic Logic Units (ALUs) that enable concurrent lookups and actions. To address data dependencies and ensure coherent processing, stages are organized sequentially. After processing the packet, the \textit{programmable deparser} reconstructs packet headers and serializes them for transmission.


Although various vendors have their own models for programming the pipeline, there is a common goal across the industry to make them P4-programmable \cite{xing2023unleashing}. P4, originally designed as a domain-specific language for programmable data plane switches, has gained popularity in programming packet data paths due to its simplicity and versatility. 

\subsubsection{Domain-specific Packet Processing} Infrastructure tasks can be broadly categorized into network functions, security functions, and storage functions. These tasks are integral to various networks, including data centers, cloud environments, enterprise networks, and campus networks. Given the specificity of these functions, some are optimized by being implemented directly in hardware to enhance their speed and efficiency. For instance, Transport Layer Security (TLS), a widely used protocol for encrypting application payloads and authenticating users, involves functions like encrypting and decrypting data. Recognizing the repetitive nature of these operations, it is practical to hardcode them into hardware. Hardware-based crypto processors, which have been utilized for some time, are examples of domain-specific processors incorporated into SmartNICs. In addition to improving the speed and efficiency of their respective functions, domain-specific processors free up CPU cores on the host for other computing tasks.

Other examples of domain-specific processors include regular expression (RegEx) used for tasks requiring Deep Packet Inspection (DPI), Non-Volatile Memory Host Controller over Fabrics (NVMe-oF) for remote storage, data compression, data deduplication, Remote Direct Memory Access (RDMA), etc. The application sections of this survey (Section \ref{sec:net_offloads}) will explore more specific use cases that leverage these domain-specific accelerators for various applications.

\subsubsection{Control Plane and Management}
SmartNICs incorporate CPU cores for running control plane functions and for managing the SmartNIC. The CPU cores can also be used for implementing functions that do not fit in ASIC/FPGA execution engines. The CPU cores are typically ARM or MIPS-based. Some advantages of incorporating CPU cores within the SmartNIC are:

\begin{itemize}[leftmargin=*]
    \item Certain infrastructure functions (e.g., key distribution for TLS sessions) require execution in the CPU. The SmartNIC CPU cores can be utilized to perform these functions. This alleviates the burden on the host's CPU cores, allowing them to focus on executing user application functions.
    \item Infrastructure functions will run more efficiently on the CPU cores of the SmartNIC than on the CPU cores of the host since they will be separated from other compute-intensive workloads used by user applications.
    \item The security is improved because the infrastructure functions will be completely isolated from the host. 
    \item The ASIC/FPGA execution engines have limitations on the complexity of operations to be performed on the packets. Such limitations stem from the fact that packets must be processed as quickly as possible to sustain line rate. Having CPU cores on the SmartNIC can help in executing such functions, at the cost of an increase in the latency. 
    
\end{itemize}

\begin{table}[t]
\centering
\caption{Features, Traditional NICs, Offload NICs, and SmartNICs.}
\footnotesize
\begin{tabular}{|l|c|c|c|}
\hline \hline
\multicolumn{1}{|c|}{Feature} & \begin{tabular}[c]{@{}c@{}}Traditional\\ NIC\end{tabular} & \begin{tabular}[c]{@{}c@{}}Offload\\ NIC\end{tabular} & SmartNIC \\ \hline \hline
\begin{tabular}[c]{@{}l@{}}Infrastructure functions \\ separation\end{tabular} & Low & Medium & High \\ \hline
Security Isolation & Low & Low & High \\ \hline
General-purpose CPU & No & No & Yes \\ \hline
Domain-specific processors & Low & Medium & High \\ \hline
Customization of data plane & No & Low & High \\ \hline
\begin{tabular}[c]{@{}l@{}}Flexibility to define new\\ protocols\end{tabular} & No & Low & High \\ \hline
Innovation & Low & Medium & High \\ \hline
Standardized models & Yes & Yes & No \\ \hline
Technology maturity & High & High & Medium \\ \hline \hline
\end{tabular}
\label{table:comparison}
\end{table}

\subsection{SmartNICs Benefits}
SmartNICs offer a wide range of features and benefits that solve modern network challenges.

\begin{itemize}[leftmargin=*]
    \item Infrastructure offloads: Data center infrastructure tasks currently consume up to 30\% of processing capacity \cite{kanev2015profiling}. This phenomenon is commonly known as the \textit{Data Center Tax}. By offloading these tasks to the SmartNIC, the freed-up 30\% of processing capacity becomes available for user applications. This optimization can significantly increase revenue opportunities for cloud providers. This is the main reason why hyperscalers are among the early adopters of this technology.
    \item Application acceleration: By incorporating hardware-based accelerators, SmartNICs demonstrate superior performance per watt compared to host-based applications. This results in reduced latency, enhancing overall efficiency.
    \item Agility and reprogrammability: The process of developing new silicon is time-consuming, expensive, and requires thorough testing. By the time this cycle is completed, rapid technological advancements may have already rendered the hardware obsolete. SmartNICs address this challenge by offering programmable components, allowing for adaptability and timely updates in response to changing technological needs. 
    \item Security isolation: SmartNICs enhance security by isolating the execution of infrastructure functions from the server execution environments.
\end{itemize}

\subsection{Comparison of Traditional, Offload, and SmartNICs}
Table \ref{table:comparison} contrasts the main characteristics of traditional, offload, and SmartNICs. In the latter, the infrastructure functions are separated from the user applications; this isolation improves security by protecting the user applications on the host. The separation is possible due to the presence of CPU cores and domain-specific accelerators on the SmartNIC. Moreover, the data plane (i.e., packet processing) of the SmartNIC is customizable and is defined by the developer's code; this provides flexibility in defining and processing new protocols as well as innovating with new applications. The technology maturity and the standardized architectures for SmartNICs can still be considered low in contrast to traditional and offload NICs.

\newcommand{\cmark}{\ding{51}}%
\newcommand{\xmark}{\ding{55}}%
\usetikzlibrary{trees,positioning,shapes,shadows,arrows}
\usetikzlibrary{calc,shadows.blur}

\tikzset{
  basic/.style  = {draw, minimum height=0.5cm, text width=3.3cm, drop shadow, font=\rmfamily, rectangle},
  root/.style   = {basic, rounded corners=2pt, thin, align=center, fill=white},
  level-2/.style = {basic, rounded corners=2pt, thin,align=center, fill=white, text width=3.5cm,     minimum height=0.5cm},
  level-3/.style = {align=center, font=\rmfamily, text width=2cm, minimum height=0.5cm, dotted, thin, draw},
  level-4/.style = {align=center, font=\rmfamily, minimum height=1.6cm, text width=2.5 cm}
}

\tikzset{no shadows/.style={general shadow/.style=}}

\begin{figure}[t]
\centering
\begin{tikzpicture}[
  level 1/.style={sibling distance=12em, level distance=3.5em, rectangle},
  edge from parent/.style={->,solid,black,thick,sloped,draw}, 
  edge from parent path={(\tikzparentnode.south) -- (\tikzchildnode.north)},
  >=latex, node distance=0.81cm, edge from parent fork down]

\node[root] {\footnotesize{SmartNIC Architectures}}
  child {node[level-2] (c1) {\footnotesize{System on Chip (SoC)}}}
  child {node[level-2] (c2) {\footnotesize{Discrete}}};

\begin{scope}[]
\node [below of = c1, style={level-3}] (c11) {\footnotesize{ASIC + CPU}};
\node [below of = c11, style={level-3}] (c12) {\footnotesize{FPGA + CPU}};

\node [below of = c2, style={level-3}] (c21) {\footnotesize{ASIC}};
\node [below of = c21, style={level-3}] (c22) {\footnotesize{ASIC + FPGA}};
\node [below of = c22, style={level-3}] (c23) {\footnotesize{FPGA}};
\end{scope}

\foreach \value in {1,2}
  \draw[->] (c1.west) |- (c1\value.west);

\foreach \value in {1,...,3}
  \draw[->] (c2.west) |- (c2\value.west);

\end{tikzpicture}
\caption{Taxonomy of SmartNIC architectures based on SoC and discrete categories.}
\label{fig:smartnic_taxonomy}
\end{figure}
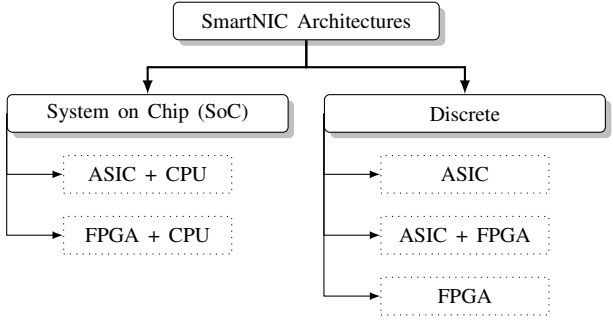

\begin{table}[b]
\centering
\caption{Comparison between various SmartNIC architectures.}
\footnotesize
\begin{tabular}{|c|c|c|c|c|}
\hline \hline
Architecture & Cost & \begin{tabular}[c]{@{}c@{}}Programming \\ Complexity\end{tabular} & Flexibility & Speed \\ \hline \hline
ASIC & Low & Low & Low & High \\ \hline
FPGA & High & High & Medium & High \\ \hline
ASIC + FPGA & High & Medium & Medium & High \\ \hline
ASIC + CPU & Medium & Low & High & Medium \\ \hline
FPGA + CPU & High & High & High & Medium \\ \hline \hline
\end{tabular}
\label{table:arch_compare}
\end{table}

\begin{table*}[t]
\caption{Commercial Discrete SmartNICs from Various Vendors.}
\label{tab:discrete}
\footnotesize
\begin{tabular}{|p{1.5cm}|p{1.5cm}|p{6cm}|p{1.5cm}|p{3.5cm}|p{1.5cm}|}
\hline \hline
\multicolumn{1}{|c|}{Architecture} & \multicolumn{1}{c|}{Vendor} & \multicolumn{1}{c|}{Model} & \multicolumn{1}{c|}{\begin{tabular}[c]{@{}c@{}}PCIe\\ generation\end{tabular}} & \multicolumn{1}{c|}{\begin{tabular}[c]{@{}c@{}}Bandwidth\\ (Gbps)\end{tabular}} & \multicolumn{1}{c|}{\begin{tabular}[c]{@{}c@{}}Technical\\ document\end{tabular}} \\ \hline  \hline
ASIC & Mellanox & ConnectX -5 / 6LX / 6DX / 7 & 3 / 4 / 5 & 50 / 100 / 200 / 400 & \cite{asic-cx5,asic-cx6lx,asic-cx6dx,asic-cx7} \\ \hline  \hline
\multirow{4}{*}{FPGA} & Archonix & VectorPath S7t & 5 & 400 & \cite{fpga-achronix} \\ \cline{2-6} 
& AMD & \begin{tabular}[c]{@{}l@{}}Alveo U50 / U50 LV / U55C / U200 / U250 / U280\end{tabular} & 3 / 4 & 1x100 / 2x100 & \cite{fpga-u50,fpga-u55c,fpga-u200,fpga-u280} \\ \cline{2-6} 
& Napatech & NT200A02 & 3 & 2x100 & \cite{fpga-napatech} \\ \cline{2-6} 
& Silicom & \begin{tabular}[c]{@{}l@{}}N501x / N5110A / FB2CDG1 / N6010/6011   \\ / FB4XXVG TimeSync\end{tabular} & 3 / 4 / 5 & \begin{tabular}[c]{@{}l@{}}4x25 / 2x100 / 4x100 / 2x400\end{tabular} & \cite{fpga-silicomN501,fpga-silicomn5110,fpga-silicomfb2,fpga-silicomn6010,fpga-silicomfb4} \\ \hline  \hline
ASIC+FPGA & NVIDIA & Innova -2 Flex & 4 & 2x100 & \cite{fpga+asic-nvidia} \\ \hline  \hline
\end{tabular}%
\end{table*}

\begin{table*}[t]
\centering
\caption{Commercial SoC SmartNICs from Various Vendors.}
\label{tab:soc}
\footnotesize
\begin{tabular}{|p{1.5cm}|p{2cm}|p{3.5cm}|p{2.25cm}|p{1cm}|p{1cm}|p{1.5cm}|p{1.5cm}|}
\hline \hline
 \multicolumn{1}{|c|}{Architecture} & \multicolumn{1}{c|}{Vendor} & \multicolumn{1}{c|}{Model} & \multicolumn{1}{c|}{Core type} & \multicolumn{1}{c|}{\begin{tabular}[c]{@{}c@{}}CPU\\ cores\end{tabular}} & \multicolumn{1}{c|}{\begin{tabular}[c]{@{}c@{}}PCIe\\ generation\end{tabular}} & \multicolumn{1}{c|}{\begin{tabular}[c]{@{}c@{}}Bandwidth\\ (Gbps)\end{tabular}} & \multicolumn{1}{c|}{\begin{tabular}[c]{@{}c@{}}Technical\\ document\end{tabular}} \\ \hline \hline
\multirow{8}{*}{ASIC+CPU} & AMD Pensando & Giglio / DSC2-(25/100/200) & Arm A72 & 16 & 4 & \begin{tabular}[c]{@{}l@{}}2x25 / 2x100 / \\ 2x200\end{tabular} & \cite{asic+cpu-giglio,asic+cpu-dsc2-200,asic+cpu-dsc2-100-25} \\ \cline{2-8} 
 & Asterfusion & Helium EC2004Y / EC2002P & Arm V8 & 24 & 3 / 4 & 4x24 / 2x100 & \cite{asic+cpu-asterfusion4y,asic+cpu-asterfusion2p} \\ \cline{2-8} 
 & Broadcom & Stingray PS225-H16 & Arm A72 & 8 & 3 & 2x25 & \cite{asic+cpu-broadcom} \\ \cline{2-8} 
 & Intel / Google & E2000 & Arm Neoverse N1 & 16 & 4 & 200 & \cite{asic+cpu-intel} \\ \cline{2-8} 
 & Marvell & LiquidIO III & Arm A72 & 36 & 4 & 5x100 & \cite{asic+cpu-marvell} \\ \cline{2-8} 
 & Netronome & Agilio FX / CX & Arm A72 & 4 & 3 & 2x10 / 2x40 & \cite{asic+cpu-netronomefx,asic+cpu-netronomecx} \\ \cline{2-8} 
 & NVIDIA & BlueField 2 - 2X & Arm A72 & 8 & 4 & 200 & \cite{asic+cpu-bf2} \\ \cline{2-8} 
 & NVIDIA & BlueField 3 -3X & Arm A78 & 16 & 5 & 400 &  \cite{asic+cpu-bf3} \\ \hline \hline
\multirow{3}{*}{FPGA+CPU} & AMD & Alveo U25N / U45N / SN1000 & Arm A53/A42 & 4 / 16 & 3 / 4 & \begin{tabular}[c]{@{}l@{}}2x25 / 1x100 / \\ 2x100\end{tabular} & \cite{fpga+cpu-u25n,fpga+cpu-u45n,fpga+cpu-sn1000} \\ \cline{2-8} 
 & Intel & \begin{tabular}[c]{@{}l@{}}N6000-PL / N6001-PL / \\ C6000X-PL / C5000X-PL\end{tabular} & Arm A53 / Xeon D & 4 / 8 & 3 / 4 & 2x25 / 2x100 & \cite{fpga+cpu-intel} \\ \cline{2-8} 
 & Napatech & NT400D1xSCC / F2070X IPU & Arm A53 / Xeon D & 4 / 8 & 4 & 2x100 & \cite{fpga+cpu-napatech} \\ \hline \hline
\end{tabular}%
\end{table*}

\section{SmartNICs Architectures}\label{sec:architectures}
The definition of a SmartNIC in Section \ref{sec:smartnic_def} targeted SoC SmartNICs. SoC SmartNICs comprise computing units, which include a general-purpose ARM/MIPS multicore processor. It also includes a multi-level onboard memory hierarchy. There is another category of SmartNICs, referred to as discrete SmartNICs. A discrete SmartNIC does not incorporate CPU cores and thus, cannot run autonomously without a host platform. Regardless of whether the SmartNIC is SoC or discrete, its packet processing logic may be ASIC and FPGA. Various SmartNICs available in the market may employ either of these hardware architectures or in some cases, a combination of both. The SmartNICs architecture taxonomy is shown in Fig. \ref{fig:smartnic_taxonomy}. Table \ref{table:arch_compare} summarizes the differences between the various SmartNIC architectures, as described next.

\subsection{Discrete SmartNICs}
Hardware implementations come with tradeoffs in terms of cost, programming simplicity, and adaptability. While an ASIC offers cost-effectiveness and optimal price performance, its flexibility is limited. ASIC-based SmartNICs feature a programmable data path that is relatively straightforward to configure, yet this programmability is constrained by predefined functions within the ASIC, leading to potential limitations in supporting certain workloads. In contrast, an FPGA-based SmartNIC is exceptionally programmable. Given sufficient time, effort, and expertise, it can efficiently accommodate nearly any functionality within the confines of available gates on the FPGA. However, FPGAs are known for being challenging to program and can be costly.

Integrating both ASIC and FPGA within the SmartNIC presents a balanced solution. Common functions are efficiently executed on the ASIC, leveraging its ease of programmability compared to the FPGA. Functions that cannot be programmed on the ASIC will be implemented on the FPGA, providing flexibility, albeit with increased programming complexity. This design provides high packet processing speed but is costly due to the use of FPGA technology.

Table \ref{tab:discrete} shows some popular commercial discrete SmartNICs from various vendors and their specifications.

\subsection{SoC SmartNICs}
Integrating general-purpose CPU cores into the SmartNIC can offer several advantages: 1) it significantly reduces programming complexity, as these cores can be programmed using languages such as C; 2) the flexibility of the system is greatly enhanced, allowing for the implementation of a wide range of programs, including those with complex features like loops and multiplications. This versatility is particularly challenging to achieve on ASIC or FPGA; 3) the management of the SmartNIC will be easier and independent of the host; 4) it will be possible to run an OS and make the SmartNIC autonomous. While the CPU cores allow additional features on the NIC, functions executed on the CPU cores might not achieve line-rate performance and could incur increased latency. Table \ref{tab:soc} shows some popular commercial SoC SmartNICs from various vendors and their specifications.

\begin{figure}[t]
  \includegraphics[width=0.489\textwidth]{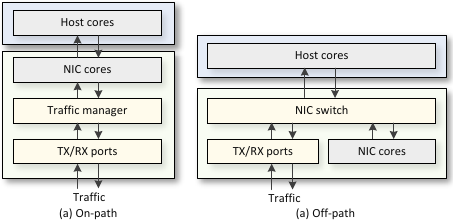}
  \caption{On-path and off-path SmartNICs.}
  \label{fig:onpath_offpath}  
\end{figure} 
\subsection{On-path and Off-path SmartNICs}
Another way to categorize the architectures of SmartNICs is based on how their NIC cores interact with network traffic. There are two categories: \textit{on-path} and \textit{off-path} \cite{liu2019offloading}.

\subsubsection{On-path SmartNICs}
With on-path SmartNICs (Fig. \ref{fig:onpath_offpath} (a)), the NIC cores actively manipulate each incoming and outgoing packet along the communication path. These SmartNICs provide low-level programmable interfaces, allowing for direct manipulation of raw packets. In this design, the offloaded code is closely situated to the network packets, increasing efficiency. However, the drawback is that the offloaded code competes for NIC cores with requests sent to the host. If too much computation is offloaded onto the SmartNIC, it can result in a significant degradation of regular networking requests sent to the host. Additionally, programming on-path NICs can be challenging due to the utilization of low-level APIs.

\begin{table}[b]
\centering
\caption{Characteristics of on-path and off-path SmartNICs.}
\footnotesize
\begin {tabular}{|p{4cm}|p{1.75cm}|p{1.75cm}|}
\hline \hline
\multicolumn{1}{|c|}{Characteristics} & \hfil On-path & \hfil Off-path \\ \hline \hline
NIC switch & \hfil $\times$ & \hfil \checkmark \\ \hline
Operating system & \hfil $\times$ & \hfil \checkmark \\ \hline
Full network stack & \hfil $\times$ & \hfil \checkmark \\ \hline
Programming complexity & \hfil High & \hfil Low \\ \hline
Host performance impact & \hfil High & \hfil Low \\ \hline 
Complex code offloading & \hfil Low & \hfil High \\ \hline \hline
\end{tabular}
\label{table:onpath_offpath}
\end{table}

\subsubsection{Off-path SmartNICs}
Off-path SmartNICs (Fig. \ref{fig:onpath_offpath} (b)) take a different approach by incorporating additional compute cores and memory in a separate SoC located next to the NIC cores. The offloaded code is strategically placed \textit{off} the critical path of the network processing pipeline. The SoC is treated as a second full-fledged host with an exclusive network interface, connected to NIC cores and the host through an embedded switch (sometimes referred to as eSwitch). Based on forwarding rules installed on the embedded switch, the traffic will be delivered to the host or the SmartNIC cores. In contrast to on-path SmartNICs, the offloaded code in off-path SmartNICs does not impact the host's network performance. This clear separation enables the SoC to run a complete kernel (e.g., Linux) with a comprehensive network stack (RDMA), simplifying system development and allowing for the offloading of complex tasks.

Table \ref{table:onpath_offpath} summarizes the differences between the on-path and off-path SmartNICs.

\begin{figure*}[t]
  \includegraphics[width=1\textwidth]{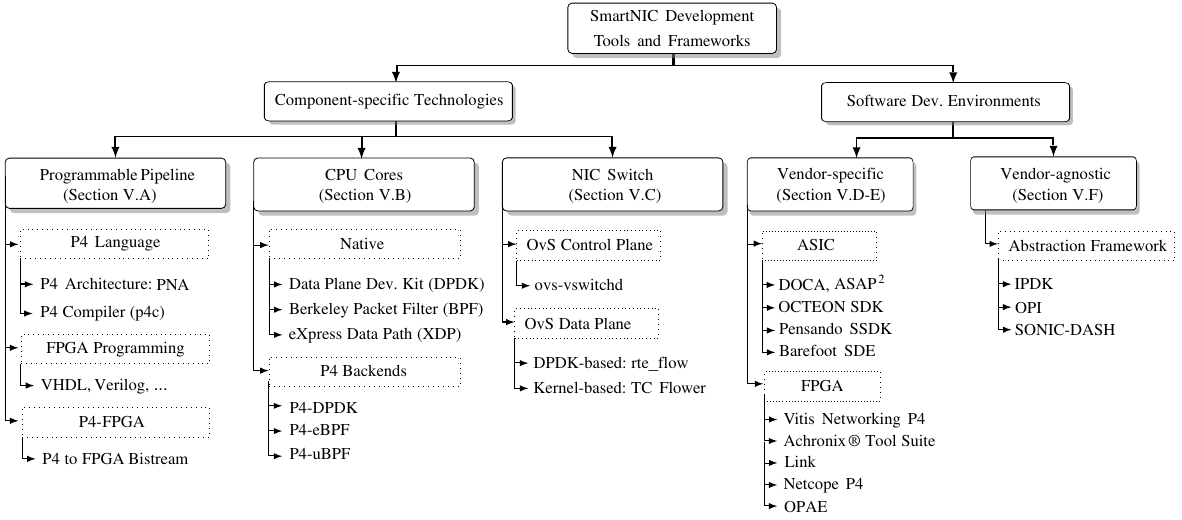}
  \caption{Taxonomy of SmartNIC development tools and frameworks, categorized by component-specific technologies and software development environments.}
  \label{fig:taxonomy_dev}  
\end{figure*}

\section{SmartNICs Development Tools and Frameworks}\label{sec:dev}
This section provides an overview of the development tools and frameworks employed for programming SmartNICs. The taxonomy, illustrated in Fig. \ref{fig:taxonomy_dev}, categorizes them based on the specific component within the SmartNIC being programmed.

\subsection{Programmable Pipeline}
The packet processing logic is commonly built using ASICs or FPGAs. The development of offloaded applications depends on the hardware architecture and the vendor's Software Development Kits (SDKs). 

\subsubsection{P4 Language}
In 2016, the PISA architecture was introduced as a domain-specific processor for networking \cite{pisa}. PISA is programmed using the Programming Protocol-independent Packet Processor (P4) language \cite{bosshart2014p4}. Although P4 was initially intended to program the data plane of PISA-based switches, it has demonstrated its versatility to program data planes for other packet processing devices. Despite the variety of programming models used by various vendors, there is a common goal which is to make their pipeline programmed in P4 \cite{xing2023unleashing}. 

P4 has a reduced instruction set and has the following goals:
\begin{itemize}[leftmargin=*]
    \item Reconfigurability: P4 enables the reconfiguration of the parser and the processing logic in the field.
    \item Protocol independence: P4 ensures that the device remains protocol-agnostic, allowing the programmer to define protocols, parsers, and operations for processing headers.
    \item Target independence: P4 hides the underlying hardware from the programmer, with the compiler considering the device's capabilities when transforming a target-independent P4 program into a target-dependent binary.
\end{itemize}

The initial specification of the P4 language, denoted as P4$_{14}$, was released in 2014 \cite{p4_14}. Subsequently, in 2016, a more refined version known as P4$_{16}$ was drafted \cite{fig_p4}. P4$_{16}$ represents a matured language that extends the capabilities of P4 to a broader range of underlying targets, including ASICs, FPGAs, SmartNICs, and more.

\begin{figure}[b]
  \includegraphics[width=0.489\textwidth]{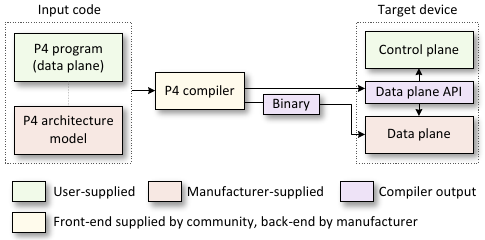}
  \caption{P4 workflow.}
  \label{fig:p4_workflow}  
\end{figure}

Fig. \ref{fig:p4_workflow} shows the workflow of developing a P4 program and deploying it into a target device. The P4 code is written by the user. The code must include a P4 architecture model, which is typically supplied by the device's manufacturer. The code is then compiled by a P4 compiler, which generates two artifacts: 1) the binary that will be deployed in the data plane of the target device; and 2) data plane APIs that will allow the control plane to interact with the data plane (e.g., for generating table entries, manipulate stateful memories, etc.). 

\subsubsection{P4 Architecture}
A P4 architecture is a programming model that defines the capabilities of a target's P4 processing pipeline. P4 programs are specifically designed for a particular P4 architecture, and these programs can be applied to any targets that adhere to the same P4 architecture.

Although the P4 architecture is provided by the manufacturer of the device, it often follows the specifications of open-source architectures. With the emergence of SmartNICs, the community has developed an open-source architecture tailored for programming these NICs. This architecture is the Portable NIC Architecture (PNA) \cite{p42021pna}.

\begin{figure}[t]
  \includegraphics[width=0.489\textwidth]{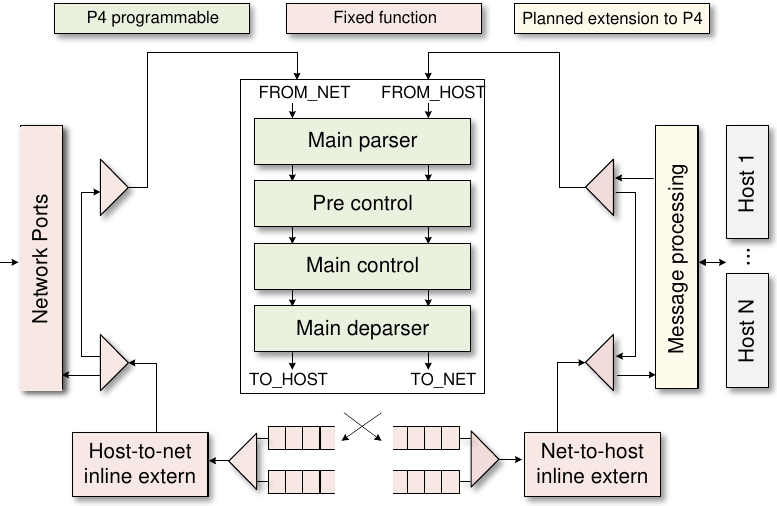}
  \caption{Portable NIC Architecture (PNA).}
  \label{fig:pna}  
\end{figure}

\paragraph{Portable NIC Architecture (PNA)}\label{sec:pna} PNA \cite{p42021pna} is a P4 architecture that defines the structure and common capabilities for SmartNICs. 
PNA has four P4 programmable blocks (main parser, pre-control, main control, and main deparser), and several fixed-function blocks, as shown in Fig. \ref{fig:pna}. The host-to-net and net-to-host externs allow executing functions on the domain-specific accelerators such as encrypting or decrypting IPsec payload. The message processing is responsible for converting between large messages in host memory and network size packets on the network and for dealing with one or more host operating systems, drivers, and/or message descriptor formats in host memory.

\begin{table}[b]
\centering
\caption{Comparison between P4 architectures: PSA and PNA.}
\begin{tabular}{|l|c|c|}
\hline \hline
\multicolumn{1}{|c|}{Feature} & PSA & PNA \\ \hline \hline
Main target devices & Switches & SmartNICs \\ \hline
Table entries modification & $\times$ & \checkmark \\ \hline
Table accessibility & One stage & Multiple stages \\ \hline
\begin{tabular}[c]{@{}l@{}}Non-packet processing (message \\ processing)\end{tabular} & $\times$ & \checkmark \\ \hline
Accelerator invocation & $\times$ & \checkmark \\ \hline
Directionality (host-to-net, net-to-host) & $\times$ & \checkmark \\ \hline
TCP connection tracking & $\times$ & \checkmark \\ \hline
\begin{tabular}[c]{@{}l@{}}Stateful elements (counters,\\ registers, meters, etc.)\end{tabular} & \checkmark & \checkmark \\ \hline \hline
\end{tabular}
\label{table:pna_psa}
\end{table}

PNA has features that are not traditionally supported by other similar P4 architectures including:

\begin{enumerate}[leftmargin=*]
    \item Table entries modification: Other P4 architectures only allow modifying table entries from the control plane. PNA allows modifying the entries in a table directly from the data plane. 
    \item Table accessibility: traditional P4 architectures allow only one operation on a table per stage. With PNA, tables can be accessible by multiple stages, even in different pipelines.
    \item Non-packet processing: PNA facilitates message processing, enabling operations on larger blocks of data to be transferred to and from host memory.
    \item Accelerator invocation: PNA is the only P4 architecture that supports invoking accelerators (e.g., crypto accelerator).
\end{enumerate}

Table \ref{table:pna_psa} compares and contrasts PNA and the Portable Switch Architecture (PSA), an architecture mainly used by switches.

\subsubsection{P4 Compiler}
\begin{figure}[t]
  \includegraphics[width=0.489\textwidth]{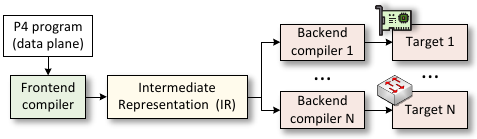}
  \caption{P4 compilation process.}
  \label{fig:compiler}  
\end{figure}

After writing a P4 program, the programmer invokes the compiler to generate a binary that will be deployed on the target device (e.g., the programmable pipeline of the SmartNIC). Consider Fig. \ref{fig:compiler}. The P4 compiler (p4c) has a frontend and a backend. The frontend is universal across all targets and handles the parsing, syntactic analysis, and target-independent semantic analysis of the program. The frontend generates an Intermediate Representation (IR) which is then compiled by the backend compiler for a specific target. The backend is provided by the manufacturer of the device.

\subsubsection{FPGA Programming}
FPGAs consist of an array of configurable logic blocks and programmable interconnects, allowing users to define the functionality of the chip based on their application requirements. FPGA-based SmartNICs follow the same programming workflows as other FPGAs provided by the vendors. This means that the development tools, methodologies, and languages used for programming traditional FPGAs can be applied to SmartNICs as well. FPGA vendors provide software tools that facilitate the programming process. These tools include Integrated Development Environments (IDEs) and compilers that translate Hardware Description Languages (HDLs) such as VHDL and Verilog into configuration files for the FPGA.

\subsubsection{P4-FPGA}\label{sec:p4_fpga}
Programming FPGAs with languages such as VHDL or Verilog can be challenging and time-consuming, especially for newcomers. To address this issue, frameworks have been developed to translate P4 code into FPGA bitstream. P4, being a high-level and user-friendly language ideal for programming datapaths, offers a faster and more efficient alternative for FPGA programming. This approach streamlines the programming process, making it particularly accessible for users without extensive FPGA programming expertise, ultimately enhancing both accessibility and efficiency. However, there are challenges in designing a compiler that translates P4 code to VHDL or Verilog. First, FPGAs are typically programmed using low-level libraries that are not portable across devices. Second, generating an efficient implementation from a source P4 program is difficult since programs vary widely and architectures make different tradeoffs. 

The community has been actively working on developing P4 FPGA compilers. The vendors (e.g., Xilinx \cite{amd2023vivado}, Intel \cite{intel_p4_fpga}) are providing their workflows to generate bitstreams from P4 on their targets. P4-FPGA tools can significantly reduce the engineering effort required to develop packet-processing systems based on devices while maintaining high performance per Lookup Table (LUT) or Random Access Memory (RAM).

\subsection{CPU Cores}
\begin{figure}[t]
  \includegraphics[width=0.489\textwidth]{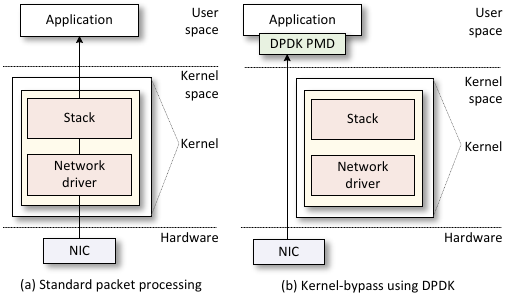}
  \caption{Software packet processing. (a) standard packet processing (interrupt-based), (b) kernel-bypass packet processing (polling mode).}
  \label{fig:std_vs_dpdk}  
\end{figure}

User applications run on CPU cores, whether on the cores on the SmartNIC, or the cores in the host. The steps for an application to process a packet coming from the NIC are shown in Fig. \ref{fig:std_vs_dpdk} (a). When a packet is received, the NIC triggers an interrupt that informs the OS about the packet's location in memory. The OS subsequently transfers the packet to the network stack which then initiates system calls from the OS kernel to deliver the packet to its intended user-level application. These steps induce overheads that dramatically degrade the bandwidth throughput. Today’s NICs have already reached more than 200Gbps [18]. As NICs become faster, the available time for processing individual packets becomes increasingly limited. For instance, with 200Gbps, the time between consecutive 1500-byte packets is as low as 60 nanoseconds (ns). The standard network stack is inadequate to keep up with the high traffic rates.

\subsubsection{Data Plane Development Kit (DPDK)}
DPDK comprises a collection of libraries and drivers designed to enhance packet processing efficiency by bypassing the kernel space and handling packets within user space (see Fig. \ref{fig:std_vs_dpdk} (b)). With DPDK, the ports of the NIC are disassociated from the kernel driver and associated with a DPDK-compatible driver. In contrast to the conventional method of packet processing within the kernel stack using interrupts, the DPDK driver operates as a Poll Mode Driver (PMD). It consistently polls for incoming packets. The utilization of a PMD, combined with the kernel bypass, yields superior packet processing performance. DPDK's APIs can be used in C programs. 

DPDK started as a project by Intel and then became open source. Its community has been growing, and DPDK now supports all major CPU and NICs architectures from various vendors. A list of supported NICs can be found at \cite{dpdk_hardware}. 

\subsubsection{eXpress Data Path (XDP) and extended Berkeley Packet Filter (eBPF)}
When utilizing DPDK, the kernel is bypassed to achieve enhanced performance. However, this comes at the cost of losing access to networking functionalities provided by the kernel. User space applications are then required to re-implement these functionalities. XDP presents a solution to this issue. XDP operates as an eBPF program within the kernel's network code. It introduces an early hook in the RX (receive) path of the kernel, specifically within the NIC driver after interrupt processing. This early hook allows the execution of a user-supplied eBPF program, enabling decisions to be made before the Linux networking stack code is executed. Decisions include dropping packets, passing packets to the normal network stack, and redirecting packets to other ports on the NIC. XDP reduces the kernel overhead and avoids process context switches, network layer processing, interrupts, etc.

\begin{figure}[t]
  \includegraphics[width=0.489\textwidth]{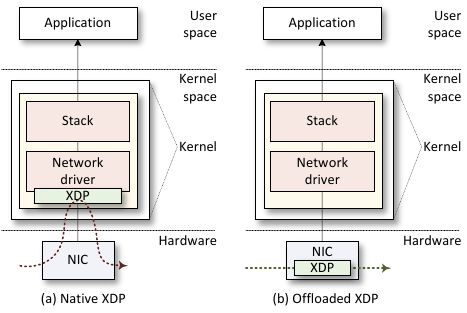}
  \caption{XDP packet processing. (a) Native XDP, slower, (b) Offloaded XDP, faster.}
  \label{fig:xdp}  
\end{figure}

XDP programs have callbacks that will be invoked when a packet is received on the NIC. There are three models for deploying an XDP program:

\begin{itemize}[leftmargin=*]
    \item Generic XDP: XDP programs are incorporated into the kernel within the regular network path. While this method does not deliver optimal performance advantages, it serves as a convenient means to experiment with XDP programs or deploy them on standard hardware that lacks dedicated support for XDP.
    \item Native XDP: The NIC driver loads the XDP program during its initial receive path, see Fig. \ref{fig:xdp} (a). Support from the NIC hardware is required for this mode.
    \item Offloaded XDP: The XDP program is loaded directly by the NIC hardware, bypassing the CPU as a whole, see Fig. \ref{fig:xdp} (b). This requires support from the NIC hardware.
\end{itemize}

\subsubsection{P4 Backends}
Creating P4 programs is generally considered more straightforward compared to writing DPDK or BPF/XDP code. Consequently, there have been efforts to translate P4 into these codes. The P4 compiler (p4c) is equipped with backends specifically designed for generating DPDK, BPF/XDP, and Userspace BPF (uBPF) codes. Table \ref{table:p4_backends} compares the P4 backends.

\begin{table}[b]
\centering
\caption{Comaprison between the P4 backends.}
\footnotesize
\begin{tabular}{|l|c|c|c|}
\hline \hline
\multicolumn{1}{|c|}{Features} & \multicolumn{1}{c|}{P4-DPDK} & \multicolumn{1}{c|}{P4-eBPF/XDP} & \multicolumn{1}{c|}{P4-uBPF} \\ \hline \hline
Userspace & \checkmark & $\times$ & \checkmark \\ \hline
NIC support & \checkmark & \checkmark & \checkmark \\ \hline
P4 Architectures & PNA, PSA & \begin{tabular}[c]{@{}l@{}}[ebpf,xdp]\\ \_model.p4\end{tabular} & ubpf\_model.p4 \\ \hline
Compilation & \begin{tabular}[c]{@{}l@{}}P4\textrightarrow spec\\ \textrightarrow C\textrightarrow so\end{tabular} & \begin{tabular}[c]{@{}l@{}}P4\textrightarrow C\textrightarrow\\ eBPF bytecode\end{tabular} & \begin{tabular}[c]{@{}l@{}}P4\textrightarrow C\textrightarrow \\ uBPF bytecode\end{tabular} \\ \hline
\begin{tabular}[c]{@{}l@{}}Supported \\ Features\end{tabular} & High & Low & Medium \\ \hline \hline 
\end{tabular}
\label{table:p4_backends}
\end{table}

\begin{figure}[t]
  \includegraphics[width=0.489\textwidth]{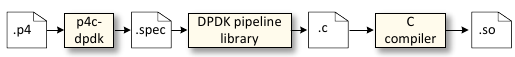}
  \caption{P4 DPDK pipeline.}
  \label{fig:p4dpdk}  
\end{figure}

\paragraph{P4-DPDK} 
The p4c-dpdk backend translates P4$_{16}$ programs into DPDK  Application Programming Interface (API), allowing the configuration of the DPDK software switch (SWX) pipeline \cite{dpdk_swx}. The P4 programs can be written for the PNA or PSA architectures\footnote{Preliminary experiments show that more features are implemented for PNA over PSA for the P4-DPDK target.}. The backend transforms a given P4 program into a representation (.spec) that aligns with the DPDK SWX pipeline (see Fig. \ref{fig:p4dpdk}). The subsequent step involves the generation of a C code from the .spec file. This code includes C functions corresponding to each action and control block. A C compiler then generates a shared object (.so) from the C code. It is important to note that P4-DPDK is not a P4 simulator (e.g., BMv2); it achieves high performance.  

\paragraph{P4-eBPF} 
The expressive powers of P4 and eBPF programming languages differ, yet there is a significant overlap, particularly in network packet processing. The P4 to eBPF compiler translates P4 programs into a restricted subset of C code that is compatible with eBPF. The P4 program only defines the data plane. The control plane is separately implemented; BPF Compiler Collection (BCC) tools simplify this by generating C and Python APIs for the interaction between the data plane and the control plane.
The P4 to eBPF compiler also facilitates the integration of custom C extern functions, enabling developers to extend the P4 program's functionality by incorporating eBPF-compatible C functions. This capability empowers the P4 program with features not natively supported by the P4 language. Upon compilation, the P4 compiler generates a C file and its corresponding header. A subsequent C compiler then generates an eBPF program, loadable into the kernel using the Traffic Control (TC). Once loaded, manipulating tables in the eBPF program can be achieved with the bpftool provided by the kernel. 
\paragraph{P4-uBPF}
uBPF adapts the eBPF processor to run in userspace. The utilization of uBPF is advantageous due to its compatibility with any solution implementing kernel bypass, such as DPDK apps. 

The p4c-ubpf compiler translates P4 programs into uBPF programs. The backend for uBPF predominantly relies on the P4-eBPF compiler, but generates C code compatible with user space BPF implementation. The uBPF backend offers a broader scope compared to the eBPF backend. Beyond simple packet filtering, the P4-uBPF compiler supports P4 registers and programmable actions, encompassing packet modifications and tunneling.
The generated C programs are compiled by the clang compiler, which generates uBPF bytecode. The bytecode is then loaded into the uBPF VM.

\subsection{NIC Switch}
The NIC switch performs QoS traffic control and steers traffic to the NIC execution engines. SmartNICs typically implement the NIC switch following the specifications of the open-source Open vSwitch (OvS). OvS is a software switch originally designed to enable communication among Virtual Machines (VMs). OvS has two major components, the control plane (ovs-switchd) and the data plane, also known as the datapath.

\subsubsection{OvS Control Plane}
\begin{figure}[t]
  \includegraphics[width=0.489\textwidth]{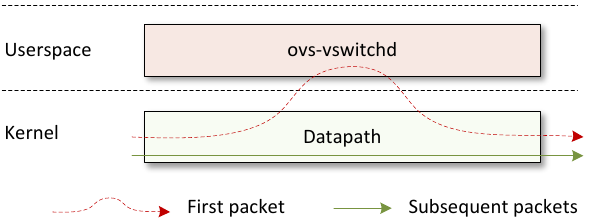}
  \caption{Open vSwitch (OvS) components.}
  \label{fig:ovs_2}  
\end{figure}

Fig. \ref{fig:ovs_2} demonstrates how the OvS components interact to forward packets. After receiving the first packet of a flow, the datapath forwards it to the ovs-switchd. The ovs-switchd then determines how the packet should be handled, and then sends the packet back to the data plane with the desired handling method. It also instructs the data plane to cache the action for handling similar packets. Subsequent packets are then matched and their actions are executed, all in the data plane. Actions may include packet modification, packet sampling, packet dropping, etc. 

The OvS control plane is traditionally executed on the host in the userspace. With SmartNICs, the OvS control plane is executed on the CPU cores of the SmartNICs.

\begin{figure}[b]
  \includegraphics[width=0.489\textwidth]{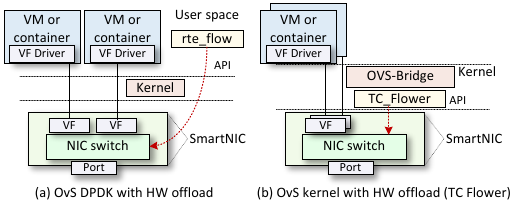}
  \caption{OvS hardware offload. (a) OvS DPDK with hardware offload using rte\_flow; (b) OvS kernel with hardware offload using tc\_flower.}
  \label{fig:ovs_offload}  
\end{figure}

\subsubsection{OvS Data Plane}
The standard OvS switch's datapath is situated in the kernel (see Fig. \ref{fig:ovs_2}), which strains CPU resources and degrades the performance. This performance degradation becomes more pronounced with an increasing number of flows and more complex policy rules, consuming multiple CPU cores for datapath operations and ultimately resulting in the lowest server utilization. To address these issues, many SmartNICs offer support for offloading OvS into their NIC switch. When this feature is utilized, the OvS datapath is moved to the hardware, resulting in superior performance compared to the software-based versions.

\paragraph{OvS-DPDK and rte\_flow} OvS-DPDK enhances OvS by incorporating a DPDK-based datapath in the userspace, surpassing the performance of the standard kernel OvS datapath with reduced latency. OvS-DPDK leverages hardware offload capabilities through rte\_flow \cite{rte_flow}, a DPDK-based API, see Fig. \ref{fig:ovs_offload} (a). This API facilitates the installation of rules into the hardware switch within the SmartNIC (NIC switch). The rte\_flow API enables users to define rules for matching specific traffic, altering the packets, querying related counters, etc. Matching within this context can be based on various criteria such as packet data (including protocol headers and payload) and properties like associated physical port or virtual device function ID. Operations supported by the rte\_flow API include dropping traffic, diverting traffic to specific queues, directing traffic to virtual or physical device functions or ports, performing tunnel offloads, and applying marks, among others.

\paragraph{OvS-Kernel and TC Flower} The OvS-kernel can use the TC Flower \cite{horman2019ovs} to configure rules on the hardware switch integrated into the SmartNIC, see Fig. \ref{fig:ovs_offload} (b). Within the Linux kernel, the TC flower classifier, which is a component of the TC subsystem, offers a means to specify packet matches using a defined flow key. This flow key encompasses fields extracted from packet fields and, if desired, tunnel metadata. TC actions enable the execution of diverse operations on packets, such as drop, modify, output, and various other functionalities.

\begin{figure}[t]
  \centering
  \includegraphics[width=0.48\textwidth]{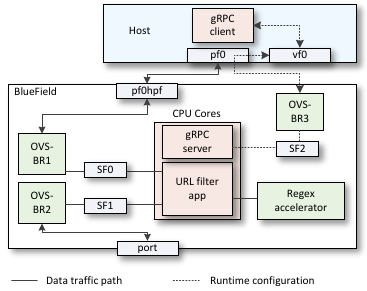}
  \caption{DOCA URL filter reference application.}
  \label{fig:doca_url_filter}  
\end{figure}

\subsection{Vendor-specific SDKs - ASIC}
The following SDKs are proprietary and target ASIC-based SmartNICs. 

\subsubsection{NVIDIA's DOCA}
The Data Center-on-a-Chip Architecture (DOCA), is a software development framework developed by NVIDIA for the BlueField SmartNICs \cite{asic+cpu-bf2}. This framework encompasses various components, including libraries, service agents, and reference applications. Applications developed using DOCA are written in the C programming language and incorporate support for DPDK. This integration ensures that developers have access to all DPDK APIs for efficient packet processing. Additionally, DOCA comes equipped with its own set of libraries designed to streamline interactions with the components on the SmartNIC. For instance, to implement IPsec or perform encryption and decryption, DOCA offers dedicated APIs that developers can easily invoke, simplifying the integration of these functionalities into their applications.

One noteworthy library within DOCA is the DOCA Flow \cite{doca_flow}. This library allows programmers to customize packet processing by defining matching criteria and actions. These match-action units are defined in pipes, which can be chained. Given DOCA's reliance on DPDK, it leverages rte\_flow to transmit rules to the embedded switch (NIC switch). NVIDIA employs its proprietary ASAP$^2$ technology \cite{mellanox2019} for implementing the embedded switch and for efficient traffic offloading to the hardware.

Consider Fig. \ref{fig:doca_url_filter} which shows an example of a DOCA application for Uniform Resource Locator (URL) filtering. The developer must create OvS bridges and connect scalable functions (SF)\footnote{An SF is a lightweight function that has dedicated queues for sending and receiving packets; it is analogous to the virtual function (VF) used in SR-IOV.} to them. Note that the OvS bridge is hardware offloaded. In this specific example, one bridge is used to connect the physical port to the application (OvS-BR2). Another bridge is used to connect the application to the host (OvS-BR1). The incoming packets on the physical port will be forwarded to the application, which runs on the CPU cores. URL filtering involves parsing the application layer because the URL to be visited is located at the HTTP header. The SmartNIC will invoke the regular expression (RegEx) hardware accelerator to scan for the URL, which is significantly faster than scanning using the CPU. A third bridge can be created to enable the user to manage the application (e.g., specifying the URLs to be blocked). BlueField provides gRPC interfaces for the runtime configuration. 

\begin{figure}[t]
  \centering
  \includegraphics[width=0.486\textwidth]{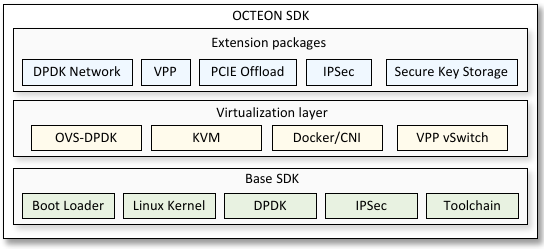}
  \caption{OCTEON SDK layers and modules.}
  \label{fig:octeon_sdk}  
\end{figure}
It is possible to develop DOCA applications without the hardware; however, testing the compiled software must be done on top of a BlueField \cite{doca_wo_hw}. 

\subsubsection{OCTEON SDK}
The OCTEON SDK is a comprehensive suite that integrates a development environment and optimized software modules for building applications on OCTEON family processors. The suite consists of a base SDK, a virtualization layer, and a collection of SDK extension packages designed for specific application functions. The Base SDK relies on a standard Linux environment and user-space DPDK (see Fig. \ref{fig:octeon_sdk}). It facilitates the seamless compilation of DPDK, Linux, or control plane applications on top of it with minimal adjustments. Programmers write C code and invoke libraries for accelerating functions, including compression/decompression, regex matching, encryption/decryption, and more.

In addition to the Base SDK, the suite includes SDK extensions that help users enable complex applications. These extensions consist of pre-optimized, application-specific modules bundled into packages that run on the Base SDK. Notable extensions include OvS-DPDK, Vector Packet Processor (VPP), secure key storage, trusted execution environment, etc. Furthermore, the OCTEON SDK provides a cycle-accurate simulator. This simulator enables the developers to test the behavior of their programs with precision and accuracy in software.

\begin{figure}[t]
  \centering
  \includegraphics[width=0.48\textwidth]{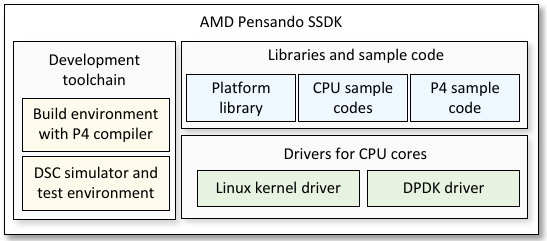}
  \caption{AMD Pensando SSDK.}
  \label{fig:amd_ssdk}  
\end{figure}

\subsubsection{AMD Pensando SSDK}
The AMD Pensando SDK facilitates software development for the AMD Pensando SmartNIC. This comprehensive SDK includes a P4$_{16}$ compiler, debugging tools, a DPDK driver, example codes, and thorough documentation (see Fig. \ref{fig:amd_ssdk}). Specifically, P4$_{16}$ can be used to write code for execution in the programmable pipeline. C and C++ are used to write code for the CPU core complex. Additionally, the SDK allows invoking the SmartNIC's built-in domain-specific accelerators.

Similar to DOCA, developers have the flexibility to compile applications without the SmartNIC hardware. However, unlike DOCA, Pensando SDK provides a simulator, allowing developers to test their ideas before uploading the image to the hardware. This validation capability becomes particularly advantageous when integrating the SDK and simulator into CI/CD-based development and workflows. The simulator boasts machine-register accuracy, ensuring that any code developed for it can be cross-compiled to run seamlessly on the real hardware. The simulator serves as a valuable tool for validation, speeding up development, and simplifying debugging processes within a virtualized environment. 

The reference applications included with the AMD Pensando SDK include a basic skeleton hello world, Software Defined Network (SDN) policy offload with Longest Prefix Matching (LPM), Access Control List (ACL), flow aging, IPsec gateway, and other classic host offload such as TCP Segmentation Offload (TSO), checksum calculation, and Receive Side Scaling (RSS).

\subsubsection{Barefoot SDE / Intel P4 Studio}
The compiler used for programming the pipeline on the Intel IPU has similarities with that used for programming the Tofino switches \cite{asic+cpu-intel}. The compiler was originally developed by Barefoot Networks, which was acquired by Intel in 2021. This compiler was formerly known as Barefoot SDE, and now it has been rebranded as Intel P4 Studio. It is well-established and has undergone extensive revisions and optimizations. Additionally, the compiler is equipped with a Graphical User Interface (GUI) tool (P4 Insight \cite{intel_p4_insight}) that offers comprehensive insights into resource utilization. This includes details such as the location of specific match-action tables, the utilization of hash bits, and the usage of SRAM/TCAM. The public documentation does not provide clear specifics on how the compiler differs between Tofino switches and SmartNICs. 

\subsubsection{SDKs for ASIC SmartNICs Comparison}
Table \ref{table:asic_SDK_comparison} compares the four SDKs. The characteristics compared include the supported SmartNIC models, P4 language support, development feasibility with or without dedicated hardware, availability of simulators or emulators for testing, and the necessity for special licensing. The AMD Pensando and Intel SmartNICs are P4 programmable and thus, their SDKs provide a P4 compiler. The NVIDIA BlueField and the Octeon SDK only support P4 for their CPU cores (e.g., through P4-DPDK). Furthermore, all SDKs except Intel/Barefoot SDE offer development without dedicated hardware, and Pensando SSDK and Octeon SDK provide simulators or emulators for testing purposes. The Pensando SSDK and the Intel SDE require the customer to sign a Non-disclose Agreement (NDA) to get the license for the SDKs. 

\begin{table}[]
\caption{Comparison between the vendor-specific SDKs for ASIC SmartNICs.}
\footnotesize
\begin{tabular}{|l|c|c|c|c|}
\hline \hline
Characteristic & \begin{tabular}[c]{@{}c@{}}NVIDIA\\ DOCA\end{tabular} & \begin{tabular}[c]{@{}c@{}}Octeon \\ SDK\end{tabular} & \begin{tabular}[c]{@{}c@{}}Pensando \\ SSDK\end{tabular} & \begin{tabular}[c]{@{}c@{}}Intel/Barefoot\\ SDE\end{tabular} \\ \hline \hline
\begin{tabular}[c]{@{}l@{}}Supported\\ SmartNICs\end{tabular} & \begin{tabular}[c]{@{}c@{}}BlueField\\ 2/3/X\end{tabular} & \begin{tabular}[c]{@{}c@{}}Marvel \\ LiquidIO\end{tabular} & \begin{tabular}[c]{@{}c@{}}Pensando \\ DSC-200\end{tabular} & \begin{tabular}[c]{@{}c@{}}Intel IPU\\ E2000\end{tabular} \\ \hline
P4 support & $\times$* & $\times$* & \checkmark & \checkmark \\ \hline
\begin{tabular}[c]{@{}l@{}}Development\\ wo/ hardware\end{tabular} & \checkmark & \checkmark & \checkmark & $\times$ \\ \hline
\begin{tabular}[c]{@{}l@{}}Simulator/\\ emulator\end{tabular} & $\times$ & \checkmark & \checkmark & $\times$ \\ \hline
Special licensing & $\times$ & $\times$ & \checkmark & \checkmark \\ \hline \hline
\end{tabular}
\label{table:asic_SDK_comparison}
  \begin{tablenotes}[flushleft]
    \item *While P4 is not the main language used for programming the packet processing engine, it can be used for programming the CPU cores (e.g., with P4 DPDK).
  \end{tablenotes}
\end{table}

\subsection{Vendor-specific SDKs - FPGAs}
The following SDKs are proprietary and target FPGA-based SmartNICs.

\subsubsection{Vitis Networking P4}
Vitis Networking P4 \cite{vitisp4}, developed by AMD Xilinx, is the development environment for their FPGA SmartNICs. This high-level design environment greatly simplifies the creation of packet-processing data planes through P4 programs (see Section \ref{sec:p4_fpga}). The tool's primary function is to translate the P4 design intent into a comprehensive AMD FPGA design solution. The compiler maps the control flow with a custom data plane architecture composed of various engines. This process involves selecting suitable engine types and tailoring each one according to the specified P4 processing requirements. The architecture definition file for Vitis Networking P4 is named xsa.p4. This architecture follows the open-source P4 PNA architecture (see Section \ref{sec:pna}).

\begin{figure}[t]
  \centering
  \includegraphics[width=0.48\textwidth]{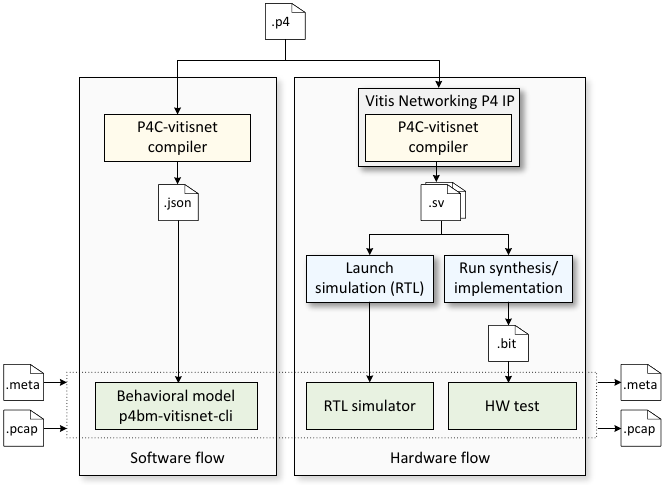}
  \caption{AMD Xilinx's Vitis Networking P4 software and hardware flows.}
  \label{fig:vitis}  
\end{figure}

Fig. \ref{fig:vitis} illustrates the AMD Vivado™ hardware tool flows designed for AMD Vitis™ Networking P4 implementations. There is a flow for the software, which is used for testing the behavior of the P4 program. The other flow is for the hardware. In the software flow, the P4C-vitisnet compiler accepts a P4 file as input and generates an output .json file. This json file is provided to the P4 Behavioral Model. Note that this behavioral model provided by Xilinx closely resembles the behavioral model created by the community, known as BMv2. The software flow also has a Command Line Interface (CLI) which is a control plane application used to interact with the data plane at runtime.  Packets and metadata information can be fed into the behavioral model when testing. 

For the hardware flow, the P4C-vitisnet compiler generates an .sv file. The compiler uses the Vitis Networking P4 IP. The .sv file can be used for launching Register Transfer Level (RTL) simulation on the RTL simulator or for running synthesis/implementation on the hardware. The hardware flow also accepts metadata and packets as inputs.

Xilinx also provides the Xilinx Runtime library (XRT) \cite{fig_xrt}, which is a user-friendly, open-source software stack designed to facilitate communication between the application code and the FPGA device. It offers APIs for Python and C/C++ applications. 

\subsubsection{Intel P4 Suite for FPGA}
\begin{figure}[t]
  \centering
  \includegraphics[width=0.48\textwidth]{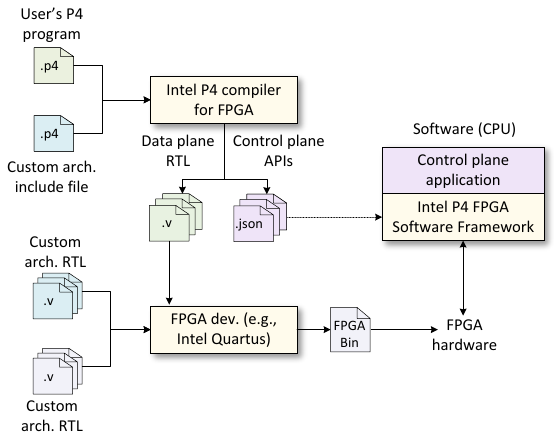}
  \caption{Intel P4 Suite for FPGA workflow.}
  \label{fig:intel_p4_fpga}  
\end{figure}

The Intel P4 Suite for FPGA is a high-level design toolkit that produces packet processing IP for Intel-based FPGAs from P4 codes. This toolkit comprises a compiler responsible for converting P4 into RTL and a software framework equipped with APIs that facilitate the interaction between control plane applications and the data plane. The toolkit's workflow is shown in Fig. \ref{fig:intel_p4_fpga}. Intel's P4 compiler for FPGA accepts as input the P4 program and a custom architecture, and generates RTL for the data plane and APIs for the control plane. The resulting RTL, combined with the architecture's RTL and the shell RTL are then transformed into an FPGA binary using conventional FPGA development environments (e.g., Intel Quartus). Finally, this binary is deployed to the hardware. The control plane APIs are pushed to the Intel P4 FPGA for Software Framework which enables user-defined control plane applications to interact with the hardware.

\begin{figure}[b]
  \centering
  \includegraphics[width=0.48\textwidth]{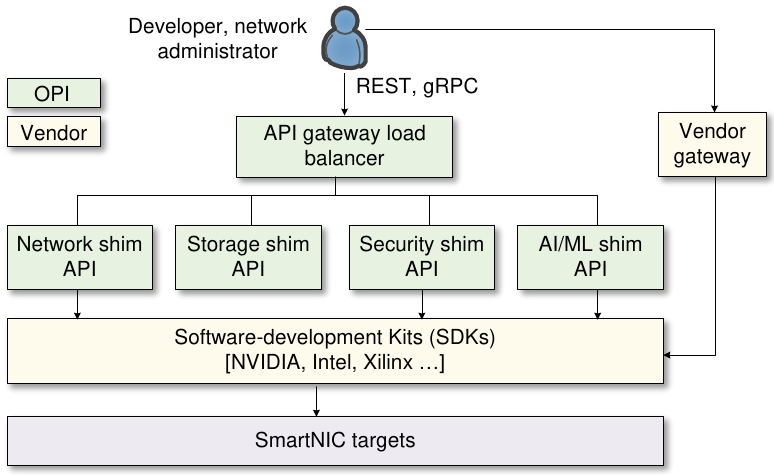}
  \caption{Open Programmable Infrastructure (OPI) architecture. The vendor-specific SDKs and hardware are abstracted via a common API developed by OPI.}
  \label{fig:opi}  
\end{figure}

\begin{figure*}[t]
  \centering
  \includegraphics[width=1\textwidth]{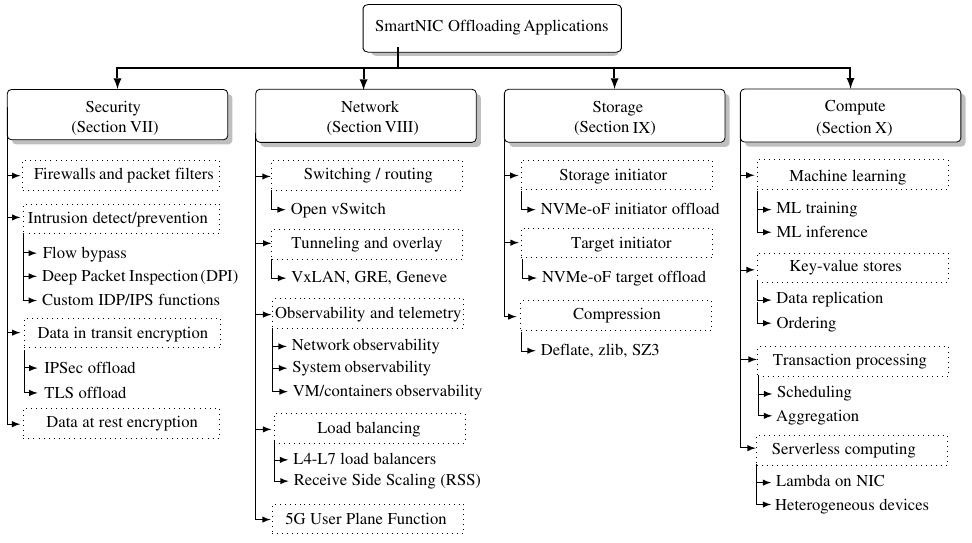}
  \caption{Taxonomy of SmartNIC offloaded applications.}
\label{fig:applications_taxonomy}  
\end{figure*}

\subsubsection{Achronix Tool Suite}
The Achronix Tool Suite (ACE) is used to design Achronix's FPGA SmartNICs. The tool includes place and route functions, timing analysis and bitstream generation and download, synthesis analysis, and in-system debugging using snapshots. The Achronix design flow facilitates ease for FPGA designers by supporting standard RTL (VHDL and Verilog) input and employing industry-standard simulation techniques.

\subsubsection{Napatech Link Toolkit}
The Link toolkit, developed by Napatech, is a collection of software providing plug-and-play features for various SmartNICs from Napatech. The software includes: Link-capture, which facilitates packet capture with nanosecond timestamping and replay with precise inter-frame gap control. Link-Inline accelerates various network and security applications, Link-Virtualization offloads a virtual switch to the SmartNIC, and Link-Storage accelerates virtualized storage.

\subsubsection{OFS and OPAE}
The Open FPGA Stack (OFS) is an open-source solution that offers a hardware and software framework for creating shell design and workload \cite{intelofs}. OFS includes reference shell designs for various Intel FPGA devices, drivers, and software tools. Using OFS, Application-Specific FPGA Interface Managers (FIMs) can be developed, making use of the Open Programmable Acceleration Engine (OPAE) SDK. OPAE, which is a subset of OFS, is a software layer comprising various API libraries. These libraries are used when programming host applications that interact with the FPGA accelerator.

\subsection{Vendor-agnostic }

\subsubsection{Open Programmable Infrastructure (OPI)}
The OPI is a community-driven initiative focused on creating common APIs to configure and manage different SmartNIC targets \cite{opi_project}. Instead of relying on vendor-specific SDKs, developers can use OPI’s standardized APIs to activate services, effectively abstracting the complexities associated with vendor-specific SDKs. Consider Fig. \ref{fig:opi}. The developer uses gRPC and REST APIs to initial calls to the API gateway. The gateway acts as a load balancer between four shim APIs: network, storage, security, and AI/ML. These shim APIs then translate the calls to the hardware accelerators through the vendor-specific SDKs. With such a design, portability can be ensured across various targets. Note that the developers can still execute functions provided by the vendor if they are not available through the OPI APIs.

\subsubsection{Infrastructure Programmer Development Kit (IPDK)}
The IPDK is an open-source, vendor-agnostic framework comprising drivers and APIs tailored for infrastructure offload and management tasks. It is versatile and capable of running on a range of hardware platforms including SmartNICs, CPUs, or switches. Operating within the Linux environment, IPDK leverages established tools like Storage Performance Development Kit (SPDK), DPDK, and P4 to facilitate network and storage virtualization, workload provisioning, root-of-trust establishment, and various offload capabilities inherent to the platform. IPDK is a sub-project of OPI. 

IPDK already supports multiple targets including P4 DPDK, OCTEON SmartNICs, Intel IPU, Intel FPGA, and Tofino-based programmable switch \cite{ipdk_targets}.

IPDK has two main interfaces: 1) Infrastructure Application Interface; and 2) Target Abstraction Interface. The Infrastructure Application Interface serves as the northbound interface of the SmartNIC, encapsulating the diverse range of Remote Procedure Calls (RPCs) supported within IPDK. The Target Abstraction Interface represents an abstraction provided by an infrastructure device (e.g., SmartNIC) that runs infrastructure applications for connected compute instances. These instances could include attached hosts and/or VMs, which may or may not be containerized.

\subsubsection{SONIC-DASH}
SONiC, an open-source operating system for network devices, has experienced significant growth \cite{sonic_dash, fig_dash}. The SONiC community has introduced a new open-source project called DASH (Disaggregated APIs for SONiC Hosts) aiming at being an abstraction framework for SmartNICs and other network devices. It consists of a set of APIs and object models which cover network services for the cloud. The initial objective of DASH is to enhance the performance and connection scale of SDN operations, aiming to achieve a speed increase of 10 to 100 times compared to software-based solutions in today's clouds and enterprise. DASH's ecosystem includes a community of cloud providers, hardware suppliers, and system solution providers.

\section{Offloaded Applications Taxonomy}\label{sec:taxonomy}
This section describes the systematic methodology that was adopted to generate the proposed taxonomy. The results of this literature survey represent derived findings by thoroughly exploring the SmartNIC-related research works published in the last five years. 

Fig. \ref{fig:applications_taxonomy} shows the proposed taxonomy. The taxonomy was meticulously designed to cover the most significant works related to SmartNICs. The aim is to categorize the surveyed works based on various high-level disciplines. The taxonomy provides a clear separation of categories so that a reader interested in a specific discipline can only read the works pertaining to that discipline. 

SmartNICs accelerate various infrastructure applications, categorized primarily into security, networking, and storage functions. It also accelerates various computing workloads including AI/ML inference and training, caching (key-value stores), transaction processing, serverless functions, and others. Each high-level category in the taxonomy is further divided into sub-categories. For instance, various transaction processing works belong to the sub-category ‘‘Transaction processing’’ under the high-level category ‘‘Compute’’. Additionally, the survey offers performance comparisons between applications running on the host and those offloaded to the SmartNICs.

\begin{figure}[t]
  \centering
  \includegraphics[width=0.489\textwidth]{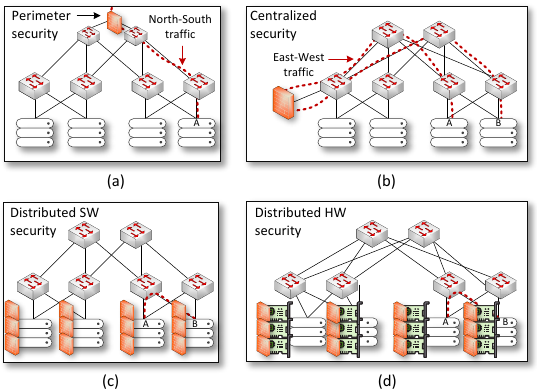}
  \caption{(a) Perimeter-based security. The appliance only inspects NS traffic; (b) Centralized security. The appliance can inspect EW traffic, but the bandwidth overhead is high; (c) Distributed SW firewall. Software-based appliances are attached to the servers and can inspect EW traffic, but the performance is not high. (d) Distributed HW firewall. Appliances are offloaded to SmartNICs on the servers, enabling EW inspection with high performance. }
  \label{fig:security_placement}  
\end{figure}

The subsequent subsections delve into the ongoing developments within each of the aforementioned category, offering insights into the lessons learned from these advancements.

\section{Security}\label{sec:security}
The landscape of data center traffic has undergone a significant transformation with the rise of \textit{cloud-hosted applications} and \textit{microservices} \cite{thones2015microservices}. Traditionally, traffic patterns were characterized by North-South (NS) flows (between internal and external devices) that are protected by dedicated security appliances (e.g., firewall) at the perimeter, see Fig. \ref{fig:security_placement} (a). However, in the past decade, the dynamics have been shifting towards East-West (EW) flows (between internal data center devices), accounting for up to 80\% of total data center traffic \cite{benson2010network, ciscoeastwest}.
Unlike North-South traffic, East-West traffic was relatively unprotected. A workaround for this was to use a centralized security appliance and forward EW traffic to it for inspection\footnote{Sometimes referred to as traffic tromboning.}, see Fig. \ref{fig:security_placement} (b). This results in traffic traversing the intermediary devices (i.e., switches) twice, leading to duplication of both network load and the latency experienced by the two hosts.

This has led to the emergence of Zero Trust and microsegmentation architectures \cite{stafford2020zero}, whose main idea is to decentralize security functionality and move it closer to the resources that require protection. Data centers and cloud providers have shifted to using software-based security functions to protect East-West traffic \cite{basak2010virtualizing}, see Fig. \ref{fig:security_placement} (c). While this shift is advantageous in terms of ease of deployment and cost-effectiveness, it has some drawbacks:

\begin{itemize}[leftmargin=*]
    \item Performance: Packets traverse the regular network stack to be processed by a security function on the general-purpose CPUs. This increases latency and decreases the throughput.
    \item Scalability: The CPU cores often struggle to inspect traffic at high rates, particularly in the absence of software accelerators (e.g., DPDK). This can lead to high packet drop rates.
    \item Isolation: all traffic, including malicious traffic, is sent to the host. This lack of isolation can pose security risks. 
    \item CPU usage: security functions consume a substantial portion of the CPU processing power, particularly during periods of high traffic volume. This can result in performance bottlenecks and service degradation for end-user applications.
\end{itemize}

To mitigate these issues, SmartNICs have been used to offload the security functions from general-purpose CPUs, see Fig. \ref{fig:security_placement} (d). Specifically, SmartNICs have been used to offload firewall functionalities, IDS/IPS, DPI, and data-at-motion and data-at-rest encryption.

\subsection{Firewall}
A firewall monitors incoming and outgoing network traffic and allows or blocks packets based on a set of preconfigured rules. Firewalls typically operate up to layer-4 to perform basic ACL operations. This means that the traffic can be matched against network layer information (e.g., source/destination IP addresses) and transport layer information (e.g., source/destination port numbers). 

Software-based firewalls are widely being used, especially in cloud environments \cite{basak2010virtualizing}. They are typically implemented in conjunction with a virtual switch (e.g., OvS). With software-based firewalls, traffic is inspected using the CPU cores of the host where the firewall is running. This degrades the performance and consumes the compute capacity of the CPU. 

Recall that SmartNICs are equipped with a programmable pipeline or an embedded switch, where match-action rules can be defined. This makes it possible to implement firewalls with basic ACLs directly on the hardware at line rate. The functionality of the firewall can be implemented from scratch by a developer. However, it requires implementing many functions such as connection tracking if stateful inspection\footnote{Stateful inspection is a firewall technology that monitors and evaluates the state of active network connections, making decisions based on the context of the entire communication rather than individual packets.} is needed, flow caching and aging, etc. As an alternative, the hardware offloaded switch on the SmartNICs is being used to implement the firewall functionalities \cite{firewall_offload, firewall_offload_pensando}. The switch rules can be transparently offloaded to the hardware. The developer only needs to specify the rules that allow/block traffic. The connection tracking feature of the switch can be leveraged to enable stateful inspection. As an example, VMware allows offloading the firewall functionalities of its NSX distributed switch to the SmartNIC \cite{vmware}, specifically, the L2-L4 inspection and firewalling. 

\subsection{Intrusion Detection/Prevention System}
Intrusion Detection Systems (IDS) and Intrusion Prevention Systems (IPS) are cybersecurity technologies designed to safeguard networks and hosts from unauthorized access, malicious activities, and security threats. An IDS monitors and analyzes network or system events to identify suspicious patterns or anomalies. It provides real-time alerts or logs for further investigation. On the other hand, an IPS goes a step further by actively preventing or blocking unauthorized activities in real-time. Examples of open-source IDS/IPS include Zeek (formerly known as Bro) \cite{zeekproject}, Suricata \cite{suricataHomeSuricata}, and Snort \cite{snortSnortNetwork}.

IDS and IPS are generally deployed on the general-purpose CPUs of the host. SmartNICs have been offloading IDS/IPS functions to accelerate data processing:

\begin{figure}[t]
  \centering
  \includegraphics[width=0.45\textwidth]{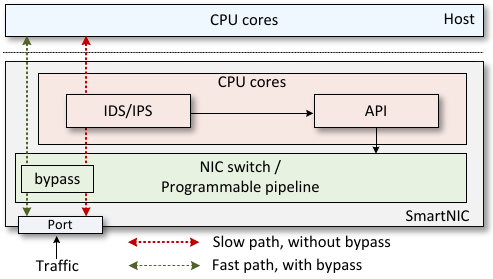}
  \caption{IPS/IDS bypass offload to the SmartNIC.}
  \label{fig:ips}  
\end{figure}

\begin{table*}[t]
\centering
\footnotesize
\caption{Comparison between various works offloading IDS/IPS functions to SmartNICs.}
\begin{tabular}{|l|c|c|c|l|}
\hline \hline
\multicolumn{1}{|c|}{Work} & Hardware used & Targeted attack & SmartNIC used & \multicolumn{1}{c|}{Key points} \\ \hline \hline
Pigasus \cite{zhao2020achieving} & FPGA, CPU & General & Intel Stratix & \begin{tabular}[c]{@{}l@{}}Multi-string matching on FPGA; FPGA is used as a primary \\ engine; CPU is used as a secondary engine; Achieves \\ 100Gbps using five CPU cores.\end{tabular} \\ \hline 
Fidas \cite{chen2022fidas} & FPGA & General & Xilinx & \begin{tabular}[c]{@{}l@{}}Offload rule pattern matching and traffic classification to \\ FPGA; Achieves lower latency and higher throughput than \\ Pigasus.\end{tabular} \\ \hline
Zhao et al. \cite{zhao2021secure} & FPGA, CPU & IoT traffic attacks & N/A & \begin{tabular}[c]{@{}l@{}}FPGA-based traffic analysis; CPU-based threat detection \\ using flow entropy algorithm.\end{tabular} \\ \hline
SmartWatch \cite{panda2021smartwatch} & \begin{tabular}[c]{@{}c@{}}P4 switches (ASIC)\\ SmartNIC (CPU)\end{tabular} & General & \begin{tabular}[c]{@{}c@{}}Netronome\\ Agilio\end{tabular} & \begin{tabular}[c]{@{}l@{}}Coarse-grained traffic analysis on P4 switches; Finer-grained \\ analysis on SmartNIC.\end{tabular} \\ \hline
ONLAD-IDS \cite{wu2022onlad} & CPU cores & General & BlueField & Anomaly detection using ANOVA statistical method. \\ \hline
Tasdemir et al. \cite{tasdemir2023investigation} & CPU cores & SQL attacks & BlueField & \begin{tabular}[c]{@{}l@{}}Natural Language Processing (NLP) for SQL query analysis; \\ ML classifiers for query classification.\end{tabular} \\ \hline
Miano et al. \cite{miano2019introducing} & ASIC, CPU & DDoS & N/A & \begin{tabular}[c]{@{}l@{}}Hardware and software-based packet filtering; Use cases \\ target DDoS mitigation.\end{tabular} \\ \hline \hline
\end{tabular}
\label{table:ids_compare}
\end{table*}
\subsubsection{Offloading IDS/IPS bypass function} 
The IDS/IPS does not need to inspect every packet. Typically, the initial packets within a specific flow contain essential information, making continuous inspection unnecessary. There are situations, such as when dealing with an elephant flow resulting from a large data transfer, where it is imperative not to inspect the packets throughout the flow's lifetime. Additionally, encrypted traffic may also be exempt from inspection.

Current IDS/IPS systems incorporate bypass mechanisms within the software through the kernel datapath \cite{suricatabypass}. While this enhances throughput, the process still depends on software that utilizes CPU cycles to efficiently route packets directly to the user space. Consider Fig. \ref{fig:ips}. With an offloaded IDS/IPS, the bypass function is implemented in the hardware without requiring additional intervention from the IDS/IPS or the host CPU \cite{suricata_offload,paloalto,juniper_srx}. The bypass is carried out on the programmable pipeline or the embedded switch within the SmartNIC. This significantly improves the performance.

\begin{figure}[t]
  \centering
  \includegraphics[width=0.489\textwidth]{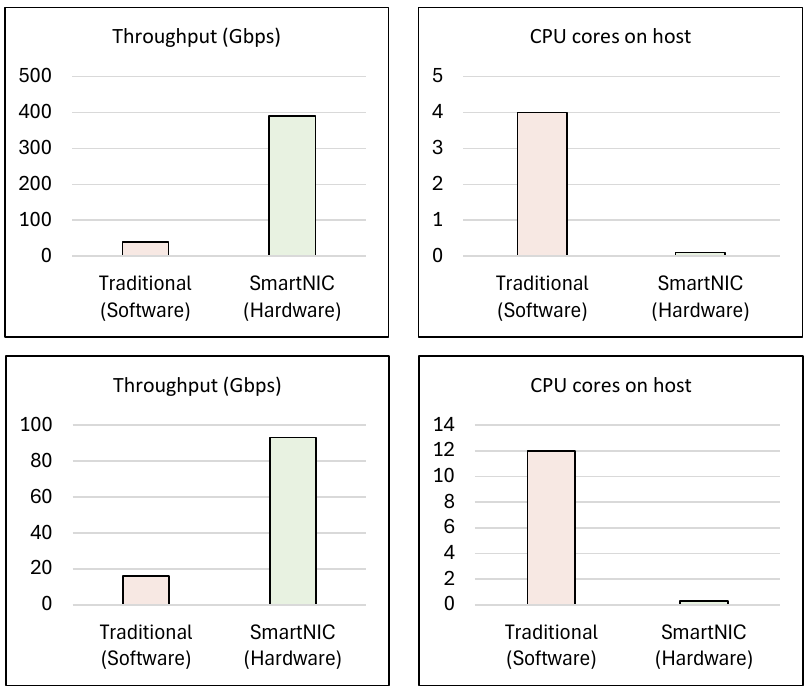}
  \caption{Performance of IPS/IDS with hardware-based bypass (SmartNIC offload) versus software-based bypass. The top row shows the results of offloading the bypass of Suricata IDS while the bottom row shows the result of offloading the bypass of Palo Alto's NGFW. Both were offloaded to NVIDIA's BlueField SmartNIC. Reproduced from \cite{suricata_offload}.}
  \label{fig:ids_res}  
\end{figure}

Fig. \ref{fig:ids_res} shows the throughput (left) and the CPU cores used on the host (right) of software-based IDS/IPS bypass and hardware-based (SmartNIC). The top row shows the results of offloading the bypass of Suricata IDS while the bottom row shows the result of offloading the bypass of Palo Alto's Next Generation Firewall (NGFW). Both were offloaded to NVIDIA's BlueField SmartNIC. The results show that hardware offloading (SmartNIC) can attain near line-rate throughput ($\sim$1200\% better than the software in the case of Suricata and $\sim$430\% in the case of Palo Alto NGFW). Moreover, the CPU cores on the host are almost idle all the time.   

\subsubsection{Offloading Deep Packet Inspection (DPI)}
IDS and IPS often require inspecting the payload within packets to identify potential malicious patterns. One example of this is URL filtering, where the URL in the HTTP header of the packet is matched against a database. This process is used for access control and blocking known harmful websites and phishing pages. The software-driven nature of IDS/IPS in performing URL filtering has notable implications on the performance. This is because, for every packet, the IDS/IPS must parse deep into the packet contents (DPI). In addition to URL filtering, IDS/IPS systems apply DPI for various tasks such as application recognition (sometimes referred to as App-ID), signature matching for malware, etc. 

SmartNICs are now integrating hardware-based RegEx engines. These engines perform pattern matching directly within the hardware, offering improved efficiency compared to traditional software-based approaches. Applications leveraging RegEx matching load a pre-compiled rule set into the engines at runtime. This hardware-driven approach helps alleviate the performance concerns associated with DPI in IDS/IPS, making network security more robust and responsive.

DPI has also been implemented on the hardware from scratch (e.g., using an FPGA). Ceska et al. \cite{cevska2019deep} proposed an FPGA
architecture for regular expression matching that can process network traffic beyond 100Gbps. The system compiles approximate Non-deterministic Finite Automata (NFAs) into a multi-stage architecture. The system uses reduction techniques to optimize the NFAs so that they can fit in the FPGA resources. The system was implemented on Xilinx FPGA. Other works \cite{yang2011high, matouvsek2016high, luchaup2014deep, vcevska2020approximate} have also explored optimizing NFAs for FPGAs.

\subsubsection{Offloading custom IPS/IDS functions}
Zhao et al. \cite{zhao2020achieving} proposed Pigasus, an IDS that uses an FPGA to perform the majority of the IDS functions, and a CPU to perform the secondary functions. Pigasus achieves 100Gbps with 100K+ concurrent connections and 10K+ matching rules, on a single server. It requires on average five CPU cores and a single FPGA-based SmartNIC. The system was tested using Intel Stratix SmartNIC. Another FPGA-based solution proposed by Chen et al. \cite{chen2022fidas} is Fidas, which offloads rule pattern matching and traffic flow rate classification. Fidas achieves lower latency and higher throughput than Pigasus. It was implemented on a Xilinx FPGA. Zhao et al. \cite{zhao2021secure} implemented an FPGA design to analyze Internet of Things (IoT) traffic and summarize it in real time. The CPU then uses a flow entropy algorithm to detect the threats. 

Panda et al. \cite{panda2021smartwatch} proposed SmartWatch, a system that combines P4 switches and SmartNICs to perform IDS/IPS functions. The P4 switches perform coarse-grained traffic analysis while the SmartNIC conducts the finer-grained analysis. The SmartNIC used is Netronome Agilio. Wu et al. \cite{wu2022onlad} implemented an anomaly detection-based IDS on the CPU cores of BlueField SmartNIC. The system uses the Analysis of Variance (ANOVA) statistical method for detecting anomalies. Tasdemir et al. \cite{tasdemir2023investigation} implemented an SQL attack detection system on the BlueField SmartNIC. The system uses NLP and ML classifiers to analyze and classify SQL queries. Miano et al. \cite{miano2019introducing} implemented a DDoS mitigation system by combining hardware-based packet filtering on the SmartNIC and software-based packet filtering using XDP/eBPF.  

Table \ref{table:ids_compare} summarises and compares the aforementioned works that offload custom IDS/IPS functions to the SmartNICs.

\begin{figure}[b]
  \centering
  \includegraphics[width=0.45\textwidth]{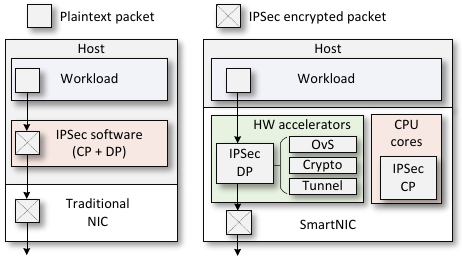}
  \caption{IPSec on the host (a) and IPSec offloaded to the SmartNIC (b). }
  \label{fig:ipsec}  
\end{figure}

\subsection{IPSec offload}
The Internet Protocol Security (IPSec) implements a suite of protocols to establish secure connections between end devices. This is achieved through the encryption and authentication of IP packets. IPsec comprises key modules including 1) \textit{key exchange}, which facilitates the establishment of encryption and decryption keys through a mutual exchange between connected devices; 2) \textit{authentication}, which verifies the trustworthiness of each packet's source; 3) \textit{encryption} and \textit{decryption}, which encrypts/decrypts payload within packets and potentially, based on the transport mode, the packet's IP header. 

IPSec has a data plane (DP) and a control plane (CP). The CP is responsible for the key exchange and session establishment. The DP is used for encapsulating, encrypting, and decrypting packets. Traditionally, the CP and DP of IPSec are executed fully in the host, see Fig. \ref{fig:ipsec} (a). This consumes CPU cores, increases latency, and decreases throughput. Once a packet is encrypted by the IPsec software, it is sent to the network over a traditional NIC. 

The IPsec software can be offloaded to the SmartNIC to enhance security and performance, see Fig. \ref{fig:ipsec} (b). The IPsec crypto operations (encryption/decryption) and encapsulation are executed by the hardware through the domain-specific accelerators. The accelerators include symmetric cryptography algorithms (e.g., Advanced Encryption Standard (AES)), asymmetric cryptography (e.g., RSA, Diffie-Hellman), and a True Random Number Generator (TRNG). This deployment model ensures transparency to the host, securing legacy workloads while benefiting from the offloading capabilities of IPsec.

Diamond et al. \cite{diamond2022securing} measured the performance of IPsec encryption in hardware on the BlueField SmartNIC. The results show that the offloaded IPSec is 10x faster than the fully software-based IPSec. Su et al. \cite{su2023meili} evaluated IPSec using the encryption accelerator on an FPGA SmartNIC. The offloaded IPsec attained $\sim$19x and $\sim$483x throughput improvement at 64B and 1500B packet sizes, respectively.

\begin{figure}[t]
  \centering
  \includegraphics[width=0.485\textwidth]{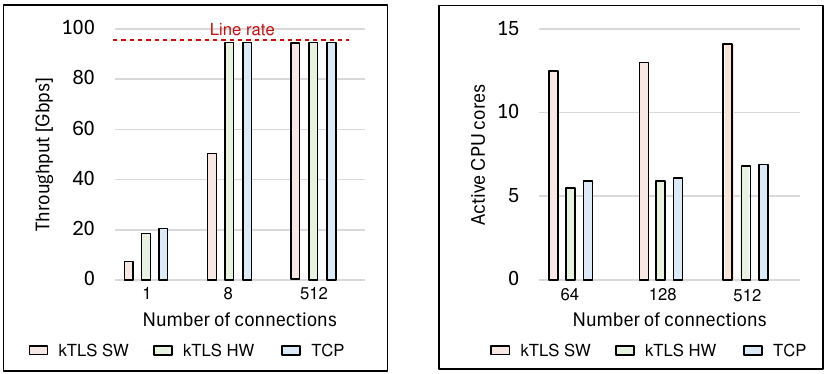}
  \caption{Throughput (a) and CPU core counts (b) of SW kTLS, HW kTLS, and plaintext TCP. }
  \label{fig:tls}  
\end{figure}

\subsection{TLS offload}
The prevalence of HTTPS servers using the TLS protocol exceeds 80\% across all web pages. As the demand for accessing web servers continues to grow steadily, there is a need for an increased rate of bandwidth.

TLS operates on layer 4, on top of TCP. The TLS process, traditionally handled by user-space applications, has evolved with the advent of offloading techniques. The kernel TLS (kTLS) involves the offload of TLS operations into the kernel, while Hardware (HW) kTLS offloads cryptographic functions to the domain-specific accelerators on the SmartNIC. With HW kTLS, the TLS handshake and the error handling (e.g., incorrect sequence number) are performed in software, while packets are encrypted and decrypted in hardware. 

Consider Fig. \ref{fig:tls} which shows the results of benchmarking the performance of HW kTLS with an Nginx server, reproduced from \cite{ktls}. The throughput of kTLS SW is smaller than that of the HW kTLS when the number of connections is small. The HW kTLS achieved line rate with eight connections. The number of CPU cores on the host when using HW kTLS is lower than the SW kTLS and the unencrypted TCP, regardless of the number of connections. 

Kim et al. \cite{kim2020case} explored offloading the TLS handshake to the SmartNIC. Their proof of concept on a BlueField SmartNIC shows that there is a 5.9x throughput improvement over executing the handshake on a single CPU core. Novais et al. \cite{novaisunlocking} performed experimental evaluations to assess the impact of TLS offload on a Chelsio SmartNIC. Their results suggest that hardware offloading improves the throughput, latency, and power consumption. Zhao et al. \cite{zhao2023dis} further detailed experiments on offloading TLS to SmartNICs. Their results suggest that SmartNICs can be beneficial for latency-sensitive tasks, but require caution with computationally heavy loads.

\subsection{Data at Rest Encryption}
SmartNICs can accelerate the encryption of data to be stored. Instead of using the CPU to encrypt the data, the domain-specific processor for encryption is used. Disk encryption protocol like AES-XTS 256/512-bit is used. 

The implementation of encryption offload for storage through SmartNICs offers flexibility across various points in the storage data path. Encryption can occur directly on the storage device (e.g., Just a Bunch of Flash (JBOF)), securing data at rest. Alternatively, it may take place at the backend of the storage controller, ensuring the encryption of data in transit. Another option involves encryption at the initiator, with the initiator retaining control of the keys, and the encrypted data transmitted across the entire storage data path.

\begin{table}[t]
\centering
\caption{Security offloaded service and the used accelerators.}
\begin{tabular}{|l|ccccc|}
\hline \hline
\multicolumn{1}{|c|}{\multirow{2}{*}{\begin{tabular}[c]{@{}c@{}}Security\\ offloaded\\ service\end{tabular}}} & \multicolumn{5}{c|}{Domain-specific accelerator} \\ \cline{2-6} 
\multicolumn{1}{|c|}{} & \multicolumn{1}{c|}{Match-action} & \multicolumn{1}{c|}{RegEx} & \multicolumn{1}{c|}{\begin{tabular}[c]{@{}c@{}}Symmetric\\ Crypto\end{tabular}} & \multicolumn{1}{c|}{\begin{tabular}[c]{@{}c@{}}Asym.\\ Crypto\end{tabular}} & TRNG \\ \hline \hline
\begin{tabular}[c]{@{}l@{}}Stateful\\ firewall\end{tabular} & \multicolumn{1}{c|}{\checkmark} & \multicolumn{1}{c|}{} & \multicolumn{1}{c|}{} & \multicolumn{1}{c|}{} &  \\ \hline
IDS/IPS & \multicolumn{1}{c|}{\checkmark} & \multicolumn{1}{c|}{\checkmark} & \multicolumn{1}{c|}{} & \multicolumn{1}{c|}{} &  \\ \hline
IPSec & \multicolumn{1}{c|}{} & \multicolumn{1}{c|}{} & \multicolumn{1}{c|}{\checkmark} & \multicolumn{1}{c|}{\checkmark} & \checkmark \\ \hline
TLS & \multicolumn{1}{l|}{} & \multicolumn{1}{l|}{} & \multicolumn{1}{c|}{\checkmark} & \multicolumn{1}{c|}{\checkmark} & \checkmark \\ \hline 
\begin{tabular}[l]{@{}l@{}}Storage \\ encryption\end{tabular} & \multicolumn{1}{c|}{} & \multicolumn{1}{l|}{} & \multicolumn{1}{c|}{\checkmark} & \multicolumn{1}{c|}{\checkmark} & \checkmark \\ \hline \hline
\end{tabular}
\label{table:sec_accelerators}
\end{table}

\subsection{Summary and Lessons Learned}
SmartNICs significantly enhance the performance of security inspection. Table \ref{table:sec_accelerators} summarizes the domain-specific accelerators used by the offloaded security functions. The key takeaways are:

\begin{itemize}[leftmargin=*]
    \item Offloading stateful firewall functions to the SmartNIC's embedded switch or programmable pipeline significantly boosts performance. The SmartNIC also allows running the firewall management and control planes on its CPU cores.

    \item The performance of IDS/IPS can be enhanced when their bypass function is offloaded to the hardware. Additionally, IDS/IPS can efficiently apply DPI using the RegEx hardware embedded in the SmartNIC. This is used in various security applications, including URL filtering, signature matching, content filtering, etc.

    \item SmartNICs facilitate the encryption of data-in-flight without compromising performance. Protocol stacks like IPSec or TLS can be easily offloaded to the SmartNIC, without requiring much development. Also, the SmartNIC provides APIs that enable developers to leverage the symmetric, asymmetric, and TRNG crypto processors.

    \item Data-at-rest can be encrypted by the SmartNIC, enabling faster storage encryption compared to the SW-based approach. 

    \item The inherent programmability in SmartNICs opens avenues for developers to  implement custom, novel, and performant security functionalities.
\end{itemize}

\section{Network offloads}\label{sec:net_offloads}
Software-defined networking (SDN) and NFV are transformative technologies that have revolutionized the way networks are designed, deployed, and managed. Virtual switches play crucial roles in enabling the flexibility, scalability, and efficiency that modern networks demand, especially to connect VMs. The networking functions implemented as NFVs on the server strain the CPU, especially in networks with high traffic rates. Recently, SmartNICs have been used to offload the network functions from general-purpose CPUs. For instance, SmartNICs have been used to offload switching/routing, tunneling, measurement and telemetry, and others.

\subsection{Switching}
Virtual switching emerged as a response to the need for hypervisors to seamlessly connect VMs with the external network \cite{pfaff2009extending}. Traditionally, virtual switches were running within the hypervisor, operating in software. However, this approach proved to be CPU-intensive, impacting overall system performance and preventing optimal utilization of available bandwidth \cite{emmerich2014performance, tu2021revisiting}.

Software switches go beyond conventional layer-2 switching and layer-3 routing \cite{pfaff2015design}; they facilitate rule matching on various packet fields and support diverse actions on packets. These actions include forwarding, dropping, marking, and more. 

\subsubsection{Switching offload}SmartNICs, whether they use a NIC switch or a programmable pipeline, have lookups and ALUs implemented in hardware. These components can be used to implement the match-action functions required for switching packets. Instead of re-implementing all the functions required for switching, most SmartNICs allow offloading the datapath of existing software switches, such as OvS \cite{pfaff2009extending}, an open source virtual switch. Note that it is possible to offload the datapath of proprietary switches, such as that of VMware's vSphere Distributed Switch \cite{vmware_switch}. Besides packet switching, virtual switches can handle additional tasks such as Network Address Translation (NAT), tunneling, and QoS functionalities such as rate limiting, policing, and scheduling.

\subsection{Tunneling and Overlay}
Tunneling is a technique that encapsulates and transports one network protocol over another. This is commonly used in virtualized environments to create isolated channels between VMs or between different segments of a virtualized network. Tunneling helps in overcoming the limitations of the underlying physical network and enables the creation of virtual networks that can span across physical boundaries. Various tunneling protocols are used in network virtualization, including Virtual Extensible LAN (VXLAN) \cite{RFC7348}, Geneve \cite{RFC8926}, Generic Routing Encapsulation (GRE) \cite{RFC2784}, etc. This subsection will discuss VXLAN, but the idea generalizes across all other tunneling protocols. 

\begin{figure}[t]
  \centering
  \includegraphics[width=0.489\textwidth]{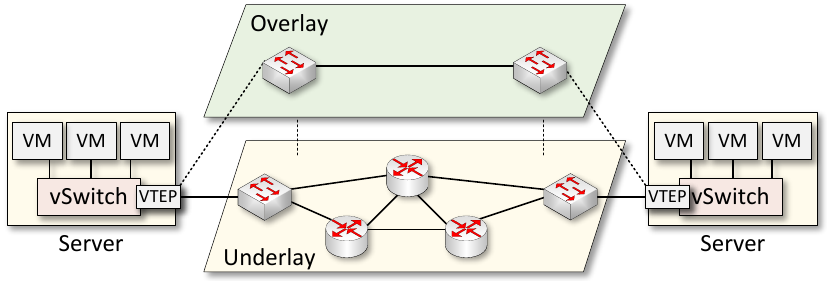}
  \caption{Tunneling. }
  \label{fig:tunnel}  
\end{figure}

\begin{figure}[b]
  \centering
  \includegraphics[width=0.485\textwidth]{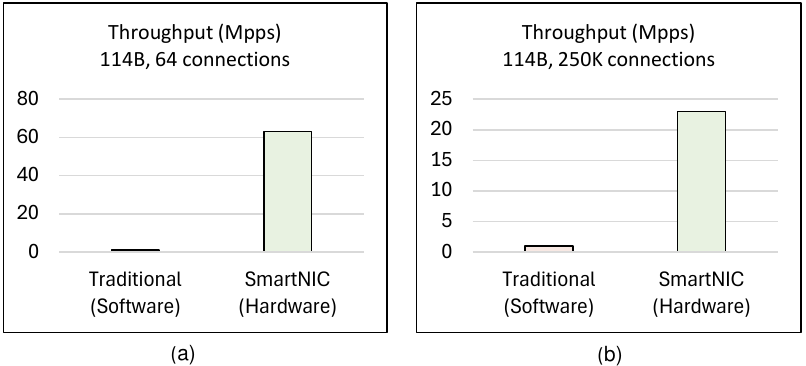}
  \caption{Throughput in million packets per second (Mpps) of software vs SmartNIC tunneling. (a) 114B packets and 64 connections; (b) 114B packets and 250,000 connections. Reproduced from \cite{burstein2021nvidia}. }
  \label{fig:vxlan_perf}  
\end{figure}

VXLAN establishes a virtual network (i.e., overlay network) over an existing layer-3 infrastructure (i.e., underlay network) by creating tunnels between VMs. This overlay scheme enables the scalability of cloud-based services without the necessity to add or reconfigure the existing infrastructure. However, VXLAN introduces an additional layer of packet processing at the hypervisor level. Consider Fig. \ref{fig:tunnel}. Each packet leaving the VM must have an additional header to be transported over the underlay network. A VXLAN Tunnel End Point (VTEP) device encapsulates during packet transmission and decapsulates during packet reception. The VTEP is being implemented on software as part of the hypervisor stack. This process incurs additional CPU overhead \cite{weerasinghe2014cost}. As the number of flows scales up, overloading the CPU with packets for encapsulation/decapsulation can easily lead to network performance bottlenecks in terms of throughput and latency.

SmartNICs can offload tunneling functions from the host CPU \cite{luo2018towards}. They support inline encapsulation/decapsulation of VXLAN and other tunneling protocols. This logic is implemented in the embedded NIC switch \cite{virtual_switch_BlueField} or the programmable pipeline \cite{asic+cpu-dsc2-200}. Tunnels definition, which is part of the control plane, is implemented in software, on the CPU cores of the SmartNIC. This design not only improves throughput and reduces latency for the encapsulation/decapsulation operations, but also frees up CPU cycles on the host for other tasks. Fig. \ref{fig:vxlan_perf} compares the tunneling performance between the software and a BlueField SmartNIC, reproduced from \cite{burstein2021nvidia}. With 114-byte packets and 64 connections, the SmartNIC tunneling is $\sim$60 times higher than the software-based. With 114-byte packets and 250,000 connections, the SmartNIC tunneling is $\sim$20 times higher than the software-based.


\subsection{Observability - Monitoring and Telemetry}
Observability is the ability to collect and extract telemetry information. During a network outage, effective observability facilitates diagnosing and troubleshooting problems. It can also help in detecting malicious events and identifying network performance bottlenecks. 

\begin{figure}[t]
  \centering
  \includegraphics[width=0.485\textwidth]{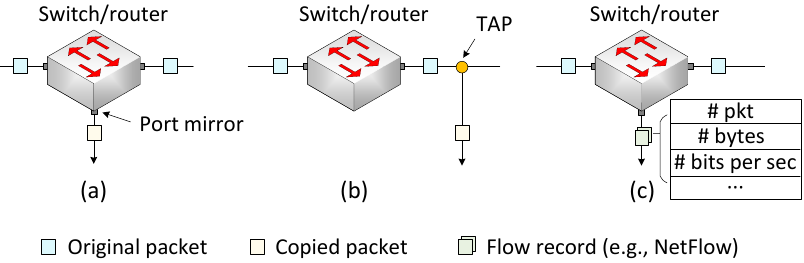}
  \caption{(a) Port mirroring; (b) TAP; (c) NetFlow export.}
  \label{fig:observ}  
\end{figure}

Traditional packet observability solutions are typically implemented in hardware, situated outside the server. Examples include configuring port mirroring (e.g., Switched Port Analyzer (SPAN)) on switches/routers, see Fig. \ref{fig:observ} (a), deploying network TAPs for replicating packets, see Fig. \ref{fig:observ} (b), and exporting flow-based statistics using NetFlow \cite{claise2004cisco} or IPFIX \cite{claise2005ipfix} to a remote collector, see Fig. \ref{fig:observ} (c).

\subsubsection{Offloading Packets Observability to SmartNICs} 
The traditional approaches to packet observability are all supported by SmartNICs. SmartNICs can mirror packets and send them to remote collectors. They can also export telemetry using flow-based telemetry solutions like NetFlow or IPFIX, or using packet-level telemetry streaming such as In-band Network Telemetry (INT) \cite{int} and In-situ OAM \cite{RFC9378}. SmartNICs can also monitor and aggregate telemetry locally, which avoids excessive traffic exports. Furthermore, since they incorporate programmable pipelines, they can be used to implement more complex packet telemetry than the traditional ones. For example, it is possible to implement streaming algorithms such as the Count-min Sketch (CMS) \cite{cormode2009count} to estimate the number of packets per flow in a scalable way, or a Bloom Filter \cite{bloom1970space} to test the occurrence of an element in a set. Such telemetry information can be very useful for a variety of applications (e.g., security \cite{geravand2013bloom}, performance analysis \cite{zeng2023survey}, etc.). 

\subsubsection{Offloading System Observability to SmartNICs} 
The SmartNIC also offers supplementary telemetry data related to the system in which it is located \cite{xpu_accelerator_offload_functions}, such as the host. For example, the SmartNIC can provide telemetry data containing the CPU, memory, and disk usage of the host.

\subsubsection{VM and Containers Observability with SmartNICs} 
\begin{figure}[t]
  \centering
  \includegraphics[width=0.489\textwidth]{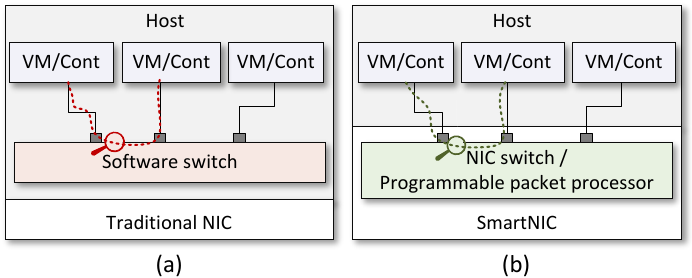}
  \caption{VM and containers observability with (a) software switches and (b) SmartNICs.}
  \label{fig:observ_vm}  
\end{figure}

External approaches to packet observability cannot observe inter-VM/container traffic within the same server. While software-based approaches for monitoring VMs and containers exist, see Fig. \ref{fig:observ_vm} (a), they often burden the CPU, especially with high traffic rates \cite{xpu_accelerator_offload_functions}. SmartNICs provide hardware visibility on traffic between VMs or containers within the same server (see Fig. \ref{fig:observ_vm} (b)), alleviating the CPU burden on the host.

\subsection{Load Balancing}
Load balancers play a crucial role in modern cloud environments by distributing network requests across servers in data centers efficiently. Traditionally, load balancers relied on specialized hardware, but now software-based solutions are prevalent among cloud providers. This shift offers flexibility and allows for on-demand provisioning on standard servers, though it comes with higher provisioning and operational expenses. While software-based load balancers offer greater customization and adaptability compared to hardware-based counterparts, they also entail considerable costs for cloud providers due to server purchase expenses and increased energy consumption.

Load balancers are categorized into two main types: Layer 4 (L4) and Layer 7 (L7). L4 load balancers function at the transport layer of the network stack. They associate a Virtual IP address (VIP) with a list of backend servers, each having its own dynamic IP (DIP) address. Routing decisions made by L4 load balancers are based solely on the packet headers of the transport/IP layers, considering factors such as source and destination IP addresses and ports. Thus, L4 load balancers do not inspect the payload content of the packets. On the other hand, L7 load balancers operate at a higher layer, specifically the application layer. These balancers are more intricate, as they analyze content within the packets, particularly focusing on application-layer protocols like HTTP. The L7 load balancer directs incoming requests to appropriate backend servers based on the specific service being accessed. For instance, differentiation may occur based on URLs.

\subsubsection{Offloading Load Balancing to SmartNICs}
Several works have offloaded load balancing to SmartNICs. Cui et al. \cite{cui2024laconic} proposed Laconic, a system that improves the performance of load balancing due to three key points: 1) Lightweight network stack: unlike traditional L7 load balancers, which heavily rely on the operating system's TCP stack, Laconic opts for a lighter packet forwarding stack on the load balancer itself. This approach minimizes overhead and leverages the end-hosts to achieve the desired end-to-end properties; 2) Lightweight synchronization for shared data structures: Laconic implements a concurrent connection table design based on the cuckoo hash table. This design efficiently manages hash conflicts and reduces the number of entries needing probing during lookups; 3) Acceleration with hardware engines: Laconic optimizes packet processing by transferring common packet rewriting tasks to hardware accelerators. This strategy alleviates the processing burden on the CPU cores of the SmartNIC. Huang et al. \cite{huang2022fglb} offloaded the load balancer to an FPGA-based SmartNIC. The result shows that the system was able to load-balance at 100Gbps.  Chang et al. \cite{chang2023learned} described a scheme that finds an optimal load balancing strategy for a network topology. It uses SmartNICs and programmable switches. Other works \cite{ni2021smartnic, tajbakhsh2022accelerator, durner2019hnlb, zhang2020vms} used variations of the methods above for load balancing
\begin{figure}[t]
  \centering
  \includegraphics[width=0.449\textwidth]{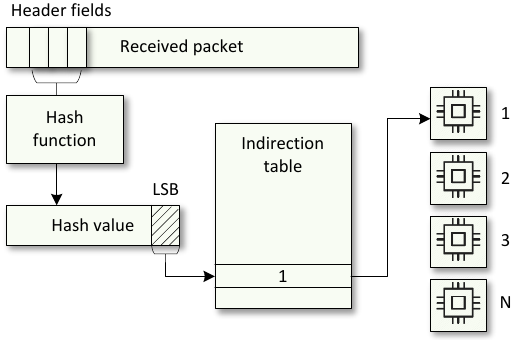}
  \caption{Receive Side Scaling (RSS).}
  \label{fig:rss}  
\end{figure} 

\subsubsection{Receive Side Scaling (RSS)}
SmartNICs commonly include an accelerator for RSS, which is a mechanism to distribute incoming network traffic across multiple CPU cores. To achieve this, the SmartNIC calculates a hash value (Toeplitz hash \cite{krawczyk1995new}) based on header fields (such as the five-tuple) of the received network packet, see Fig \ref{fig:rss}. The hash value's Least Significant Bits (LSBs) are then used as indices for an indirection table, the values of which are used to allocate the incoming data to a specific CPU core.
Some SmartNICs allow steering packets to queues based on programmable filters \cite{rss2}.

\subsection{5G UPF}
\begin{figure}[t]
  \centering
  \includegraphics[width=0.489\textwidth]{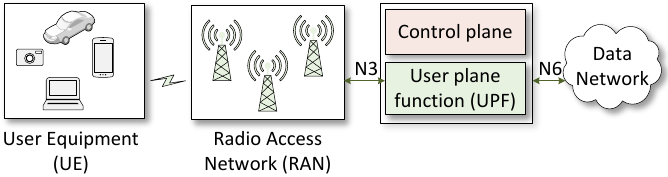}
  \caption{5G network architecture. The packet core is implemented as VNF on general-purpose CPUs. The SmartNIC is being used to offload the UPF functions.}
  \label{fig:5G}  
\end{figure}

The User Plane Function (UPF) in 5G networks represents the data plane within the packet core. It connects the User Equipment (UE) from the Radio Access Network (RAN) to the data network, see Fig. \ref{fig:5G}. The UPF typically performs packet inspection, routing, and forwarding, and QoS enforcement. It processes millions of flows with a high connection rate. 5G networks implement the packet core as VNF running on general-purpose CPUs rather than dedicated appliances. General-purpose CPUs are not capable of guaranteeing high throughput and low latency, which are the requirements and the Key Performance Indicators (KPI) of 5G networks. 

\subsubsection{UPF offload to SmartNIC}
The SmartNIC can be used to offload the UPF functions \cite{napatech_5g}. Specifically, the following functions are offloaded: GTPU tunneling: the encapsulation and decapsulation of packets run at line rate; policing: the SmartNIC will control the bit rates of the devices so that they do not exceed the Maximum Bit Rate (MBR); statistics: the counters and metrics are calculated and used for billing purposes; QoS: the SmartNIC performs Differentiated Services Code Point (DSCP) on flows to enable 5G QoS; Load balancing: the SmartNIC balances the traffic to the corresponding application; Network Address Translation (NAT): the SmartNIC translates IP addresses on traffic; etc.  

Offloading the UPF will not only improve throughput and reduce latency, but it will also boost the number of users per server (7x according to \cite{napatech_5g})  and lower the Capital Expenditure (CapEx) per user.

\subsection{Summary and Lessons Learned}
SmartNICs significantly improve the performance of network functions and reduce their CPU consumption on the hosts. The key takeaways are:
\begin{itemize}[leftmargin=*]
    \item The packet switching functions (i.e., matching header fields and taking actions), can be accelerated with SmartNICs. This is because SmartNICs, whether they use a NIC switch or a programmable pipeline, have lookups and ALUs implemented in hardware.

    \item The performance of tunneling operations (encapsulation/decapsulation) can be significantly improved when offloaded to the SmartNIC. This also frees the CPU cores that were previously used for performing the tunneling operations. 

    \item SmartNICs not only support traditional telemetry solutions but also allow the developer to devise custom fine-grained measurement schemes. They also enable inter-VM/container packet observability and host metrics telemetry.

    \item Offloading the UPF of 5G improves the performance of packet processing, increases the number of users per server, and decreases the per-user CAPEX.  

    \item Instead of re-implementing all the switching functions, SmartNICs allow offloading the datapath of existing software switches. 

    \item Developers can devise custom packet processing algorithms not supported by existing software switches. 
    
\end{itemize}

\section{Storage}\label{sec:storage}
Traditionally, storage devices were directly attached to individual computers or servers. This method provided fast access to data but lacked scalability and centralized management. Network Attached Storage (NAS) emerged as a solution to these limitations. It involves connecting storage devices to a network, allowing multiple users and clients to access the storage resources over the network. NAS provided file-level access to data. Storage Area Network (SAN) provides a high-speed network that connects storage devices to servers, providing block-level access to storage resources. SANs offer higher performance and scalability compared to NAS. 

Traditional remote storage mechanisms establish a connection between a local host initiator and a remote target. This process heavily burdens the host CPU, leading to a significant decrease in overall performance. SmartNICs can be used to offload the processing from the host CPU.

\begin{figure}[t]
  \centering
  \includegraphics[width=0.489\textwidth]{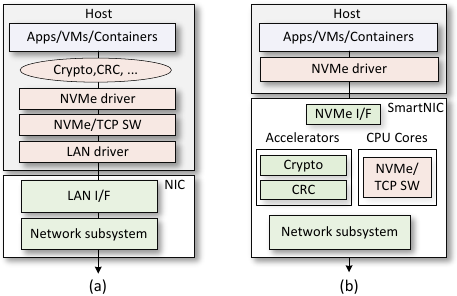}
  \caption{(a) NVMe-oF initiator without offload. (b) NVMe-oF initiator offloaded to the SmartNIC.}
  \label{fig:nvme_initiator}  
\end{figure} 

\subsection{NVMe-oF Initiator}
Non-Volatile Memory Express (NVMe) is an interface specification for accessing a computer's non-volatile storage media usually attached via the PCI Express bus. It is typically used for accessing high-speed storage devices like Solid State Drives (SSDs). NVMe over Fabrics (NVMe-oF) extends NVME to operate over network fabrics such as Ethernet, Fibre Channel, or InfiniBand. The NVMe initiator initiates and manages communication with NVMe targets. It sends commands to NVMe targets to read, write, or perform other operations. The NVMe target refers to the NVMe storage device itself. 

Fig. \ref{fig:nvme_initiator} (a) shows the traditional method of NVMe-oF using the TCP protocol and a regular NIC. The entire NVMe-oF initiator software stack operates on the host. Tasks such as cryptography and CRC computations further strain the host CPU and memory bandwidth.

\subsubsection{NVMe-oF Initiator Offload}
The NVMe-oF initiator functionality can be offloaded to the SmartNIC (Fig. \ref{fig:nvme_initiator} (b)), minimizing the overhead on the host. The SmartNIC exposes a high-performance PCIe interface and NVMe interface to the host. Requests from applications are simply forwarded to a lightweight NVMe driver on the host. The initiator stack on the SmartNIC leverages the hardware accelerators for tasks like inline cryptography and CRC offloading. The TCP stack can either remain on the CPU cores of the SmartNIC or be offloaded to the hardware itself, depending on performance considerations and SmartNIC capabilities. The division of NVMe-oF functions between hardware and software allows for optimization based on performance and SmartNIC capabilities.

Another offload to the SmartNIC is the NVMe-oF RDMA. The NVMe/RDMA data path is implemented in the hardware, with inline cryptography and CRC offloaded. This approach offers a high-performance, low-latency solution.

\begin{figure}[t]
  \centering
  \includegraphics[width=0.489\textwidth]{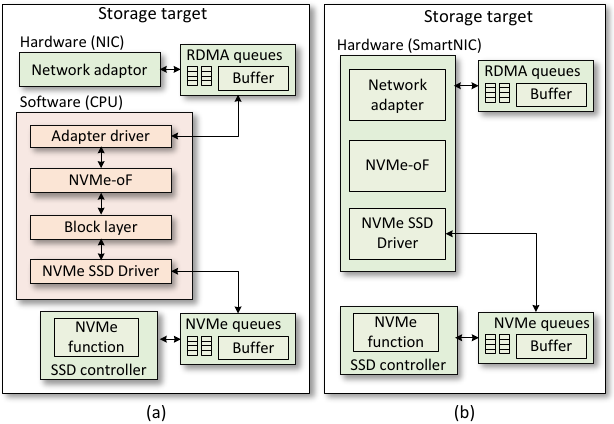}
  \caption{(a) NVMe-oF target without offload. (b) NVMe-oF target with offload. Reproduced from \cite{nvidia_storage}}
  \label{fig:nvme_target}  
\end{figure}

\subsection{NVMe-oF Target}
Another offload opportunity is offloading the storage target functions. On a storage target such as JBOF supporting NVMe-oF, there is a CPU positioned between the network and NVME SSDs, see Fig. \ref{fig:nvme_target} (a). This CPU runs software responsible for converting NVME-over-Fabrics Ethernet or InfiniBand signals to NVME PCIe signals. The software comprises various components, including a network adapter stack, NVME-over-Fabrics stack, operating system block layer, and NVME SSD stack. Both the network adapter and SSD utilize queues and memory buffers to interface with different software stacks.

When a request originates from the network, it arrives at the network adapter as an RDMA SEND with the NVME command encapsulated. The adapter then forwards it to its driver on the target CPU, which further passes it to the NVMe-oF target driver. The NVME command proceeds through the driver for the SSDs and then to the NVME SSD controller. Subsequently, the response follows the reverse path through the software layers.

\subsubsection{NVMe-oF Target Offload}
With the offload, the fast path is shifted to the hardware on the SmartNIC. Instead of burdening CPU cycles with millions of Input/Output Operations per Second (IOPS), the adaptor now handles the load using specialized function hardware. Software stacks remain in place for management traffic. The reduction in latency by removing the software from the data path is by a factor of three \cite{nvidia_storage}. Moreover, the CPU usage with offload is negligible.


\subsection{Compression and Decompression}
The surge in data volumes has caused performance bottlenecks for storage applications. Data compression is a widely adopted technique that mitigates this bottleneck by reducing the data size. It encodes information using fewer bits than the original representation. Notably, machine learning, databases, and network communication rely on compression techniques—both lossless and lossy—to enhance their performance. Data compression is compute-intensive and time-consuming, especially with large sizes of data to be compressed.

\begin{figure}[t]
  \centering
  \includegraphics[width=0.489\textwidth]{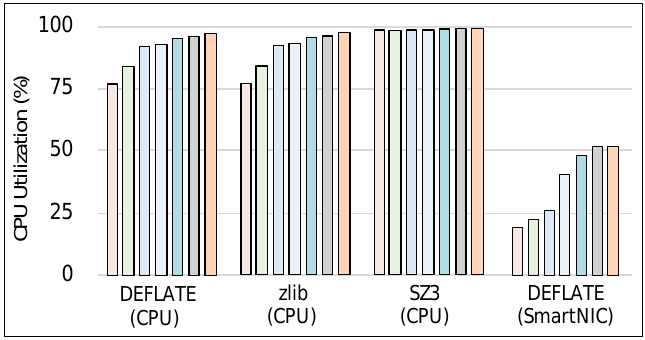}
  \caption{CPU utilization during compression with various algorithms (DEFLATE, zlib, SZ3) on seven datasets. Reproduced from \cite{li2023characterizing}.}
  \label{fig:cpu_compression}  
\end{figure}

\subsubsection{Offloading Compression to SmartNICs}
SmartNICs include onboard hardware accelerators that enable the offloading of compression and decompression tasks from host CPUs. This offloading alleviates the strain on host resources, resulting in savings and improved performance. Fig. \ref{fig:cpu_compression} shows the CPU utilization when compression is executed entirely on the host (denoted as CPU) versus when executed on the compression hardware engine of the SmartNIC (denoted as SmartNIC). The experiment shows results for various compression algorithms (e.g., DEFLATE \cite{RFC1951}, zlib \cite{RFC1950}, SZ3 \cite{liang2022sz3}) over seven datasets. The datasets are sorted in the figure by their sizes in ascending order-- each dataset is a column in the figure. The experiment is reproduced from \cite{li2023characterizing}. When the compression is executed entirely on the host, the CPU usage approaches 100\%, especially with large datasets. With a SmartNIC, there is a significant reduction in the CPU utilization. 

\begin{figure}[t]
  \centering
  \includegraphics[width=0.489\textwidth]{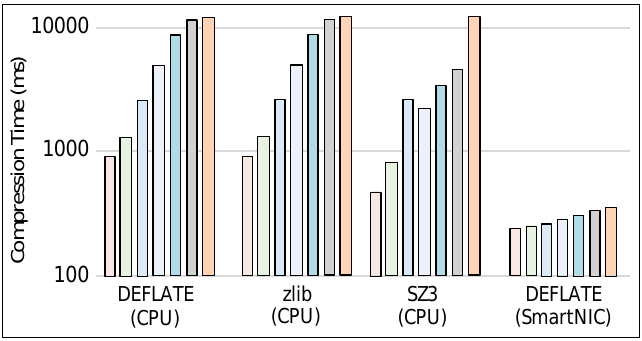}
  \caption{Compression time with various algorithms (DEFLATE, zlib, SZ3) on seven datasets. Reproduced from \cite{li2023characterizing}.}
  \label{fig:compression_time}  
\end{figure}

Fig. \ref{fig:compression_time} shows the compression time needed when executed entirely on the host (denoted as CPU) versus when executed on the compression hardware engine of the SmartNIC. The experiment is reproduced from \cite{li2023characterizing}. With a SmartNIC, there is a significant reduction in the compression time, regardless of the size of the dataset.

\subsection{Summary and Lessons Learned}
Offloading storage functions to the SmartNICs improves the performance.
\begin{itemize}[leftmargin=*]
    \item Due to the hardware accelerators in the SmartNIC (e.g., compression, crypto), storage operations like compression, deduplication, and crypto will run faster than on the host's CPU.
    \item The SmartNIC can be deployed on the initiator or the storage target. In both deployments, the CPU usage on the hosting device is negligible, the latency is minimized, and the number of IOPS is improved.
\end{itemize}


\section{Compute}\label{sec:compute}
This section examines applications offloaded to the SmartNIC that are not specifically tailored to infrastructure functions. Instead, these applications leverage the SmartNIC for accelerated computing tasks.

\subsection{Machine Learning}
State-of-the-art deep ML models have significantly expanded in size, playing a critical role in various domains, including computer vision, Natural Language Processing (NLP), and others \cite{parizotto2023offloading}. The scale of these models has seen a dramatic increase, with the number of parameters growing from 94 million in 2018 \cite{linkedin-ai} to 174 trillion in 2022 \cite{ma2022bagualu}. This exponential growth owes much to advancements in parallel and distributed computing, enabling tasks related to model training to be distributed across multiple computing resources simultaneously. The practice of offloading parts of ML tasks to network resources traces back to the 2000s \cite{moody2003scalable}, a trend that continued with the advent of Software Defined Networking (SDN), where ML primarily operates within the control plane \cite{da2016atlantic}. The recent emergence of programmable data planes (i.e., programmable switches, SmartNICs) has further spurred research and practical applications toward offloading ML phases, such as training and inference, to the hardware. Offloading ML tasks can occur on a single network device or across multiple devices, depending on network requirements and the complexity of the offloaded ML task.

\subsubsection{ML training} 
The training of large ML models can be accelerated by following a distributed approach. This involves computing gradients on each device based on a subset of the data, which are then aggregated to update model parameters. Additionally, optimization of model parameters can be carried out in the data plane to maximize accuracy.

\begin{figure}[t]
  \centering
  \includegraphics[width=0.48\textwidth]{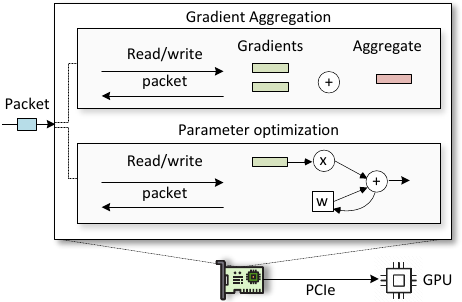}
  \caption{High-level architecture of \cite{itsubo2020accelerating}.}
  \label{fig:itsubo}  
\end{figure}

\begin{figure}[b]
  \centering
  \includegraphics[width=0.48\textwidth]{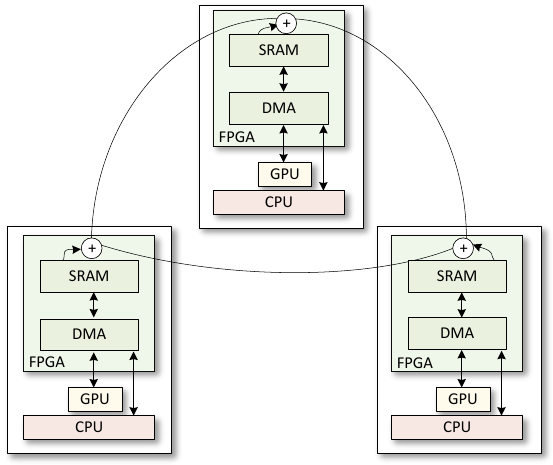}
  \caption{High-level architecture of \cite{itsubo2020accelerating}.}
  \label{fig:tanaka}  
\end{figure} 

Itsubo et al. \cite{itsubo2020accelerating} devised a system that employs in-network gradient aggregation and parameter optimization for neural networks using an FPGA-based SmartNIC. Fig. \ref{fig:itsubo} shows the high-level architecture of \cite{itsubo2020accelerating}. Gradient computation occurs on GPUs and is subsequently transmitted to the FPGA via PCIe, where aggregation takes place. Following aggregation, the FPGA can execute parameter optimization algorithms, including Stochastic Gradient Descent, Adagrad, Adam, and SMORMS3. The proposed framework achieves aggregation at $98.5\%$ line rate and accelerates parameter optimization by $\approx 1.2$ times. 
Tanaka et al. \cite{tanaka2020communication} opt for a different approach to aggregation, employing a ring network topology of FPGA-based SmartNICs using the allreduce algorithm \cite{allreduce-nvidia}. In this setup, each node within the ring network awaits data from the preceding node, aggregates it upon reception using the all-reduce aggregation algorithm, and subsequently forwards it to the succeeding node (as illustrated in Fig. \ref{fig:tanaka}). This strategy not only alleviates CPU load but also establishes a direct memory link between the GPU and the FPGA, thus preserving accuracy, as floating-point operations required by the algorithm are supported at the hardware level. 
Similarly, Ma et al. \cite{ma2022fpga} improve distributed ML training by performing the entirety of allreduce operations on an FPGA-based SmartNIC in a ring network topology. The proposed approach compresses the gradient before they are shared with other nodes, thus reducing bandwidth usage. The aggregation is performed on the SmartNIC of the end-host nodes. 

\subsubsection{ML Inference} 
In the inference phase, various ML models such as decision trees, neural networks, and reinforcement learning algorithms undergo training on a general-purpose CPU. Once trained, these models are translated into rules that can be executed within the data plane of the device. This approach enables accelerated inference, enhancing the efficiency of real-time decision-making processes.

IIsy \cite{ xiong2019switches} explore the feasibility of deploying different classification algorithms on programmable data planes. In particular, IIsy can implement decision trees, K-means, Support Vector Machine (SVM), and Naïve Bayes to perform per-packet classification. The framework converts the code into match-action tables compatible with programmable data planes. IIsy’s prototype is implemented over an FPGA-based SmartNIC using P4 \cite{ibanez2019p4}. Xavier et al. \cite{xavier2021programmable} developed a framework that translates decision trees into a P4-programmable data plane using if-else chain of statements. The proposed framework differs from \cite{xiong2019switches} by introducing per-flow classification.
BaNaNa Split \cite{sanvito2018can} accelerates neural networks inside programmable switches and SmartNICs. This approach leverages the layered structure of neural networks by splitting them between the CPU and the network processor. However, BaNaNa Split necessitates quantization, a process that reduces the precision of neural network weights at the cost of diminishing accuracy.

\subsection{Key-value Stores}
Data centers face a growing demand for collecting and analyzing vast amounts of data. Typically, this data is stored in key-value stores due to their superior performance over traditional relational database systems. Popular key-value stores include Redis \cite{redis} and Memcached \cite{memcached}. As data volumes increase, so does the frequency of reads and writes to these stores, leading to bottlenecks in the traditional network protocol stack and heavy CPU consumption.

The emergence of SmartNICs offers a solution by offloading key-value store operations to accelerate performance and reduce CPU load. One effective method is leveraging RDMA \cite{dragojevic2014farm, mitchell2013using, kalia2014using, kalia2016design, cassell2017nessie}. RDMA allows data to be read from or written to memory without involving the operating system and the traditional network stack. This results in lower latency, reduced CPU overhead, and higher bandwidth compared to traditional networking approaches.

Sun et al. \cite{sun2022skv} implemented SKV, a distributed key-value store accelerated with SmartNIC. The system offloads the data replication and failure detection components. It targets the Redis key-value store and is implemented using the BlueField SmartNIC. The evaluations show that the system reduces the latency by 21\% and increases the throughput by 14\% compared to being implemented fully on the host without SmartNIC acceleration.

\begin{figure}[t]
  \centering
  \includegraphics[width=0.489\textwidth]{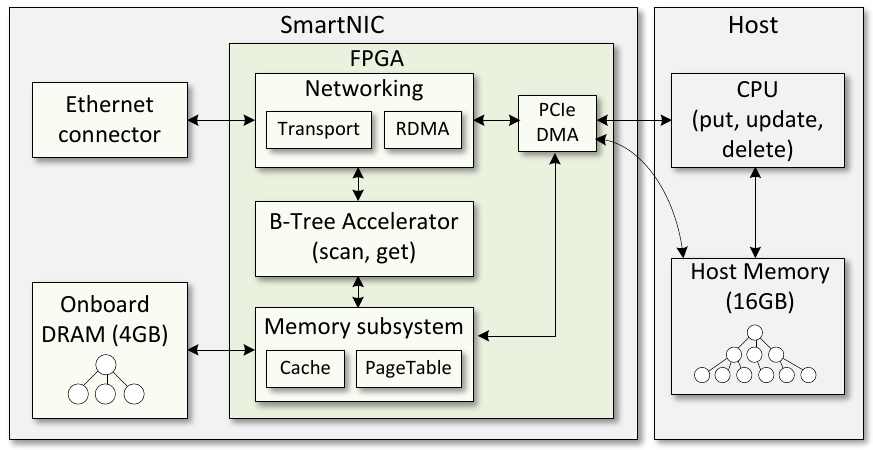}
  \caption{Honeycomb system architecture \cite{liu2023honeycomb}.}
  \label{fig:cache_honeycom}  
\end{figure}

Another aspect of the key-value store that was offloaded to the SmartNIC is the ordering of elements. Ordered key-value stores enable additional applications by allowing an efficient SCAN operation. Liu et al. \cite{liu2023honeycomb} proposed Honeycomb, an FPGA-based system that provides hardware acceleration for an in-memory ordered key-value store. It focuses on the read-dominated workloads. Consider Fig. \ref{fig:cache_honeycom}. The B-Tree accelerator implements the GET and SCAN operations. The CPU executes the PUT, UPDATE, and DELETE operations. The B-Tree is stored on the onboard DRAM in FPGA and on the memory of the host. Storing the B-Tree on the host allows better scalability since its memory is larger than that of the FPGA. The memory subsystem maintains a cache and communicates with the onboard DRAM. It also communicates with the host memory using PCIe. The implementation shows that the system increases the throughput of another ordered key-value store \cite{kalia2019datacenter} by 1.8x. 

Chen et al. \cite{chen2022hkvs} designed a heterogeneous key-value store where a primary instance runs on the host and a secondary instance runs on a SmartNIC. The system identifies the popular items and replicates them to the SmartNIC. The popular items are identified with moving window access counters. The server instance serves the read and write requests of all keys while the SmartNIC instance serves only the read request of popular items. This system targets read-intensive workloads with skewed access. The system was implemented on a BlueField-2 and the results show that the throughput is improved by 1.86x than a standalone RDMA key-value store.

\begin{figure}[b]
  \centering
  \includegraphics[width=0.489\textwidth]{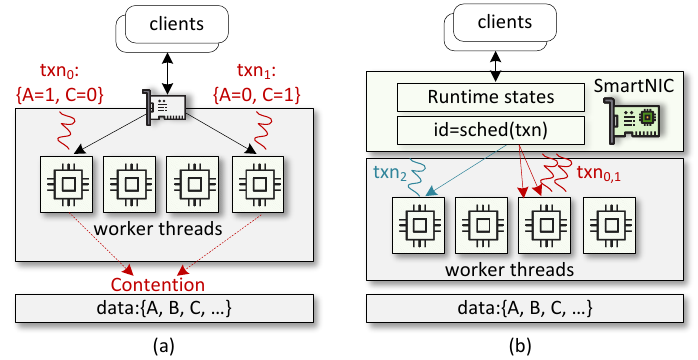}
  \caption{Transaction processing systems. (a) a system without scheduling, (b) scheduling using SmartNIC. Reproduced from \cite{li2022alnico}.}
  \label{fig:scheduling}  
\end{figure} 

\subsection{Transaction Processing}
High-performance transaction processing is important to enable various distributed applications. These systems need to manage a large number of requests from the network efficiently. One crucial aspect is determining how to schedule each transaction request to the most suitable CPU core. 

Consider Fig. \ref{fig:scheduling} (a) which shows the architecture of a transaction processing system without scheduling. A traditional NIC receives requests from the clients and dispatches them to the worker threads. The worker threads then execute the transaction, while considering the contention issues that might happen. Contention in this context means that two workers are accessing the same data and at least one of them is issuing a write. In Fig. \ref{fig:scheduling} (a), two transactions (txn$_0$ and txn$_1$) are writing to the same data blocks A and C. In such a scenario, the transactions are typically aborted, causing the clients to resend the transactions, which degrades the performance. 

Li et al. \cite{li2022alnico} proposed using a SmartNIC to schedule the transactions to the appropriate worker threads. The SmartNIC maintains the runtime states, giving it the flexibility to make accurate scheduling decisions. The SmartNIC queues the transactions belonging to the same worker thread. This avoids having the clients resend the transactions. The system is implemented on an FPGA-based SmartNIC, which further reduces the scheduling overhead. The system was implemented over the Innova-2 SmartNIC and the results show that the throughput is boosted by 2.68x and the latency is reduced by 48.8\% compared to the CPU-based scheduling. 

Schuh et al. \cite{schuh2021xenic} implemented Xenic, a SmartNIC-based system that applies an asynchronous, aggregated execution model to maximize network and core efficiency. It uses a data store on both the SmartNIC and the host. This data store provides fast access to host data via indexing. It also maintains a state to enhance the concurrency and contention issues. Xenic also aggregates work at all inputs and outputs of the SmartNIC to achieve communication efficiency. The system was implemented on a LiquidIO SmartNIC. The results show that Xenic improves the throughput of prior RDMA-based systems by approximately 2x, reduces the latency by up to 59\%, and saves server threads.

\begin{figure}[t]
  \centering
  \includegraphics[width=0.489\textwidth]{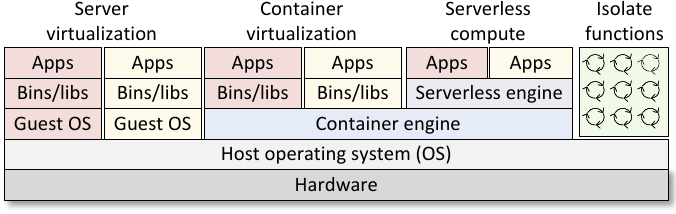}
  \caption{Architectures used in the cloud. Reproduced from \cite{choi2020lambda}.}
  \label{fig:cloud_arch}  
\end{figure} 

\subsection{Serverless Computing}

Figure \ref{fig:cloud_arch} shows the architectures used in the cloud today. The \textit{server virtualization} allows guest operating systems to run on top of a host operating system. The applications and the libraries run on top of the guest OS and are isolated from other operating systems. The trend has shifted towards \textit{container virtualization} where applications and their libraries are isolated but they share the same OS. The containers can be connected through software switches to enable their communications. The complexity and the scale of the cloud make it hard to manage and provision the infrastructure with tasks requiring fine-grained allocation of resources under changing workload demands. This has led to the \textit{serverless compute} architecture, also known as Function as a Service (FaaS). In a serverless architecture, developers write code that represents functions (also known as Lambdas), and these functions are triggered by various events. The users are billed only for the resources consumed during the execution. The serverless workloads, which are targeted by these functions, are typically short-lived with strict computing and memory limits. The cloud provider will manage the infrastructure by creating containers and taking them down when the workload is completed. Examples of serverless computing frameworks include Amazon Lambda \cite{amazon_lambda}, Google Cloud Functions \cite{google_functions}, and Microsoft Azure Functions \cite{ms_functions}.

\begin{figure}[t]
  \centering
  \includegraphics[width=0.489\textwidth]{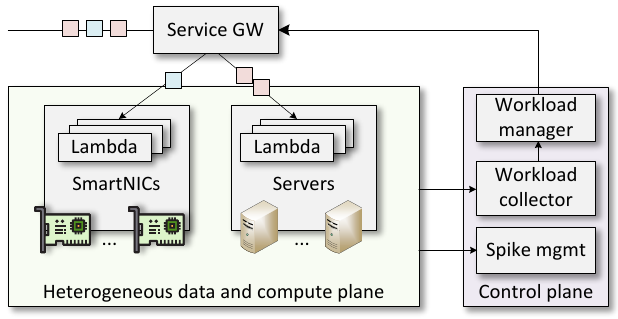}
  \caption{SpikeOffload system architecture. Reproduced from \cite{tootaghaj2022smartnics}.}
  \label{fig:spikeoffload}  
\end{figure} 

Running the serverless computing functions on top of containers that are running on top of an OS incurs processing and networking overhead that increases the latency. Recently, cloud providers have been using the \textit{isolate functions} architecture in which the functions are executed on a bare metal server.

\subsubsection{Executing Lambda Functions on SmartNICs} 
Recent efforts have explored the potential of executing Lambda functions on the SmartNICs. Choi et al. \cite{choi2020lambda} proposed $\lambda$-NIC, a framework where Lambda functions are executed on the SmartNIC. It provides a programming abstraction, which resembles the match-action of the P4 language, to express the lambda functions. The framework analyzes the memory accesses of the functions to map them across the memory hierarchy of the SmartNIC. Because the workloads are short-lived, $\lambda$-NIC assigns a function to a single core on the SmartNIC. The system was implemented on a Netronome Agilio CX, and the results show that $\lambda$-NIC can decrease the average latency by 880x and improve the throughput by 736x.

Tootaghaj et al. \cite{tootaghaj2022smartnics} proposed SpikeOffload, a system that offloads serverless functions to the CPU cores of the SmartNICs, in the presence of transient traffic spikes, see Fig. \ref{fig:spikeoffload}. 
A workload collector module gathers the history of workloads and feeds the summary to the \textit{workload manager} module. The workload manager module predicts the workload spikes based on the service time and the CPU loads of the servers and the SmartNICs. It then configures the service gateway (GW) to distribute the requests to the corresponding device (i.e., servers and SmartNICs) in the compute plane. SpikeOffload predicts the spikes in the workloads using ML. It starts the containers before the actual load arrives to mitigate the containers’ cold start latency. The system was implemented on a BlueField-2, and the results show that the Service Level Agreement (SLA) violations for certain workloads can be reduced by up to 20\%. 

\subsection{Summary and Lessons Learned}
SmartNICs extend their utility beyond infrastructure-related tasks, accelerating various compute functions. The key takeaways include:

\begin{itemize}[leftmargin=*]
    \item Machine learning tasks, encompassing distributed training and inference, experience significant performance enhancements when offloaded to SmartNICs. These devices efficiently aggregate model updates from multiple ML workers and optimize model parameters. Their programmable pipeline also enables the execution of certain ML models directly for line-rate inference. 
    \item Key-value stores operations, which include retrieving and updating data, replicating stores, and detecting failures, can be offloaded to SmartNICs. This would bring notable throughput and latency improvements.
    \item SmartNICs can be used to schedule transactions, aggregate values, and solve contention in distributed systems, improving the latency and throughput.
    \item SmartNICs can execute serverless workloads (lambda functions), which reduces the load on the servers. They can also be used as an additional execution engine in a heterogeneous data and compute cluster. 
    
\end{itemize}

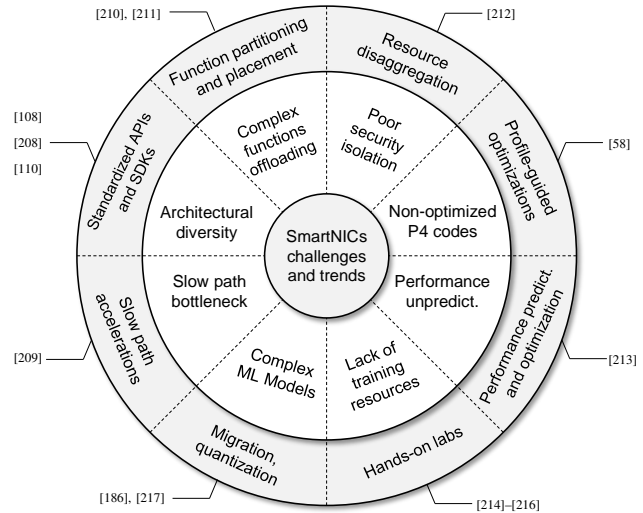
\begin{figure}[t]
\begin{minipage}{5.9in}
\begin{picture}(80,41)
\put(-2.3,0){\includegraphics[scale=0.59]{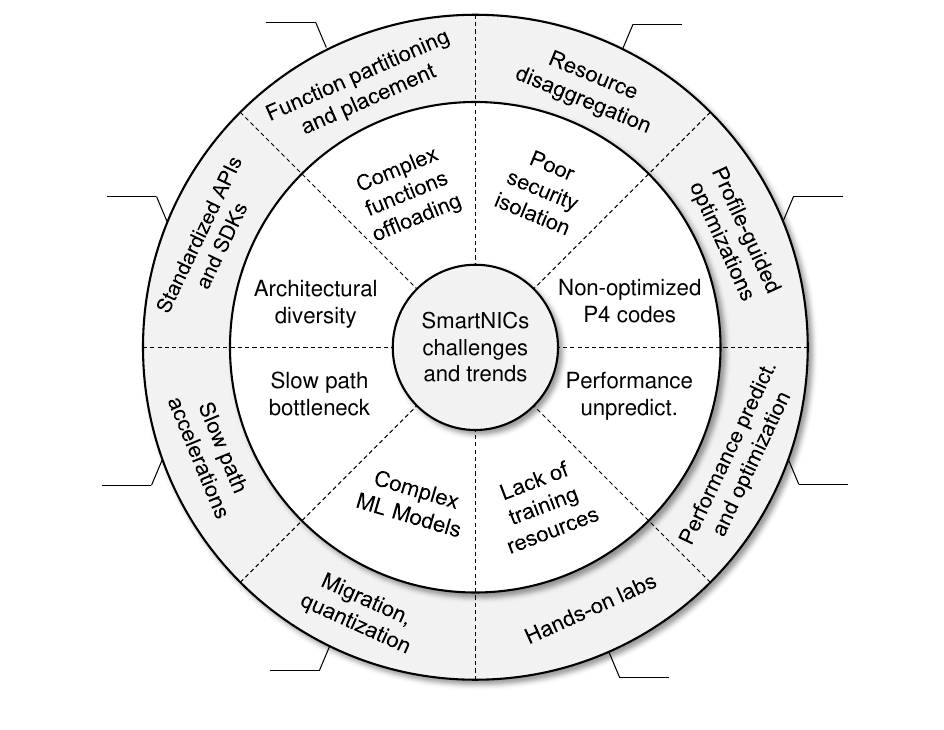}}

\put(0.7,32){\tiny{\cite{opi_project}}}
\put(0.7,30){\tiny{\cite{ipdk}}}
\put(0.7,28){\tiny{\cite{sonic_dash}}}

\put(0.7,13.7){\tiny{\cite{zulfiqar2023slow}}}

\put(6.9,39.8){\tiny{\cite{le2017uno,wang2020smartchain}}}

\put(36.5,39.8){\tiny{\cite{zhou2024smartnic}}}

\put(45.7,29.9){\tiny{\cite{xing2023unleashing}}}

\put(45.7,13.6){\tiny{\cite{qiu2020clara}}}

\put(35.7,2.6){\tiny{\cite{sth, snia_yt, opi_yt}}}

\put(7.2,3.1){\tiny{\cite{simpson2022revisiting, ma2022fpga}}}

\end{picture}
\end{minipage}
\caption{Challenges and future trends. The references represent examples of existing works that tackle the corresponding future trends.}
\label{fig:challenges}
\end{figure}

\section{Challenges and Future Trends}\label{sec:challenges}
In this section, several research and operational challenges that correspond to SmartNICs are outlined. The challenges are extracted after comprehensively reviewing and diving into each work in the described literature. Further, the section discusses and pinpoints several initiatives for future work that could be worthy of being pursued. The challenges and the future trends are illustrated in Fig. \ref{fig:challenges}

\subsection{Architectural Diversity and Vendor Specificity}
SmartNICs can have different architectural models, each requiring unique programming approaches. Even within the same architecture, SmartNICs from different vendors may necessitate proprietary SDKs and distinct programming methods, which present several challenges: 

\begin{itemize}[leftmargin=*]
    \item Vendor Lock-In: Developers may become dependent on a specific vendor's SDK, making it challenging to migrate to alternative SmartNICs or adopt new technologies. For instance, consider the scenario where a developer has written a packet processing logic using DOCA for BlueField SmartNICs. If they were to transfer this logic to Xilinx FPGA-based SmartNICs, they would need to rewrite the entire logic from scratch.

    \item Reduced Collaboration: Proprietary SDKs hinder collaboration and knowledge shared among developers, as expertise gained in one ecosystem may not be easily transferable to another. 

    \item Increased Development Time and Costs: When developers need to tailor their code for each SmartNIC's proprietary SDK, it significantly increases development time and costs. Instead of focusing on advancing the functionality and performance of their applications, developers must spend valuable resources adapting their code to work with different SmartNIC architectures and vendor-specific APIs. This diversion of resources can slow down the pace of innovation within organizations and the industry as a whole.

\end{itemize}

\noindent\textit{Current and Future Initiatives: } 
To address these challenges and foster innovation in the SmartNIC space, there is a growing need for standardized programming interfaces and open-source development frameworks. Standardization efforts could promote interoperability among SmartNICs from different vendors and enable developers to write code that is portable across various architectures. Additionally, open-source initiatives can encourage collaboration, drive community-driven innovation, and provide developers with more flexibility and control over their software stack. Several initiatives (e.g., OPI, IPDK, SONiC-DASH) aim to establish standard APIs for SmartNIC programming and administration, reducing vendor dependency. However, vendor-specific functions remain a challenge for generalization.

\begin{figure}[t]
  \centering
  \includegraphics[width=0.489\textwidth]{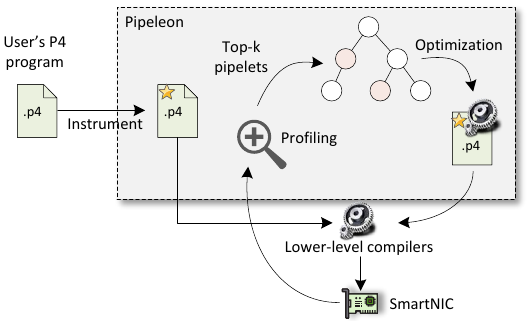}
  \caption{Pipeleon workflow. Reproduced from \cite{xing2023unleashing}.}
  \label{fig:pipeleon}  
\end{figure} 

\subsection{Non-optimized P4 Codes}
Developers have been using low-level optimization to enhance the performance of packet processing in SmartNICs. Recently, vendors are embracing P4 as a uniform programming model for SmartNICs \cite{asic+cpu-dsc2-200, asic+cpu-intel, p42021pna}. While P4 allows ease of programming and offers a high-level standardized model, it does not guarantee the optimal performance on SmartNICs. This is because the P4 compilers were optimized for switch ASICs which have a different execution model than SmartNICs. With switch ASIC, if the program compiles, the packet processing executes at line rate. SmartNICs on the other hand follow the run-to-completion model, where packets are assigned to a particular processing engine during the lifetime. With multicore SmartNICs, the packets may experience variable latencies depending on the complexity of the program and its execution paths. \\
\noindent\textit{Current and Future Initiatives: } A noteworthy work by Xing et al. \cite{xing2023unleashing} presented an automated performance optimization framework (Pipeleon) for P4 programmable SmartNICs, see Fig. \ref{fig:pipeleon}. The framework uses profile-guided optimizations to adapt the P4 program based on the runtime profiles (e.g., traffic patterns, and table entries). The input to this framework is a P4 program which is then partitioned into smaller code snippets called pipelets. The framework leverages the reconfigurability of the SmartNICs (e.g., those that follow the disaggregated dRMT architecture \cite{xing2022runtime, chole2017drmt}) to realize a more efficient implementation. The framework was tested with BlueField2 and Agilio CX SmartNICs and the results show that the optimizations significantly improve the SmartNIC performance in various use cases by up to 5x. Due to such results, it would be beneficial to improve the existing P4 compilers to be tailored to SmartNICs and to consider runtime profiles.

\begin{figure}[b]
  \centering
  \includegraphics[width=0.489\textwidth]{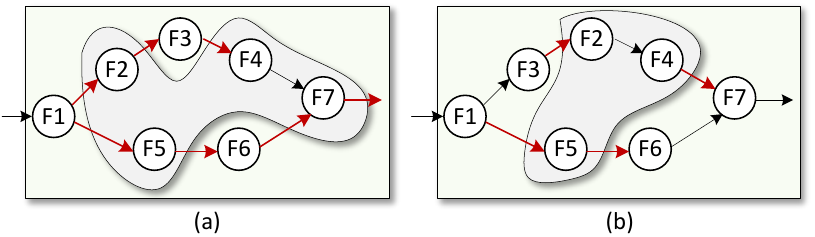}
  \caption{(a) Non-optimized placement; (b) optimized placement. The inter-device transmissions (red arrows) between SmartNIC and CPU lead to additional element graph latency. Reproduced from \cite{wang2020smartchain}.}
  \label{fig:smartchain}  
\end{figure} 

\subsection{Complex Functions Offloading}
Effectively utilizing SmartNICs for running offloaded functions presents several challenges. First, SmartNICs have limited computational and memory resources, which restricts the number of functions that can be accommodated on them. Second, although it is technically possible to host switching exclusively on the SmartNIC, doing so incurs considerable latency costs for packets moving between the functions deployed on the SmartNIC and those on the host. This is due to the overhead caused by the multiple traversals across the host PCI bus. Third, distributing the functions between the host and the SmartNIC introduces management challenges. \\
\noindent\textit{Current and Future Initiatives: } Le et al. \cite{le2017uno} presented UNO, a system that splits switching between the host software and the SmartNIC. It uses linear programming formulation to determine the optimal placement for functions. UNO uses the traffic pattern and the load of the function as input. The experiments show that the savings in processors is up to eight host cores. UNO also reduces power by 2x. Another work by Wang et al. \cite{wang2020smartchain} optimizes the placement of functions according to the processing and the transmission latency. The system analyzes the dependencies and formulates the partition and placement problem using 0-1 linear programming. The system minimizes The inter-device transmissions between the SmartNIC and the CPU, see Fig. \ref{fig:smartchain}.

\begin{figure}[t]
  \centering
  \includegraphics[width=0.489\textwidth]{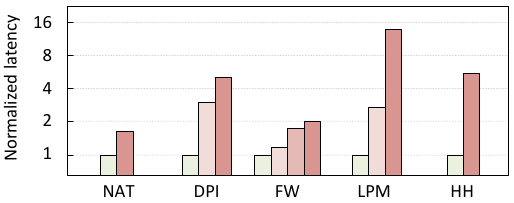}
  \caption{Normalized latency for different implementations of functions: Network Address Translation (NAT), DPI, Firewall (FW), LPM, Heavy Hitter (HH). Reproduced from \cite{qiu2020clara}.}
  \label{fig:latency_diff_implementations}  
\end{figure} 

\subsection{Performance Unpredictability}
When offloading a function to SmartNICs, developers must refactor the core logic to align with the underlying hardware. Determining the optimal offloading strategy may not be straightforward. Moreover, the performance of ported functions can vary among developers, relying heavily on their understanding of NIC capabilities, see Fig. \ref{fig:latency_diff_implementations}. For instance, using the flow cache can offer orders of magnitude improvement in latency compared to DRAM \cite{qiu2020clara}. This is entirely related to how the programmer implements the code. The performance is also influenced by traffic workloads (e.g., flow volumes, packet sizes, arrival rates). Additional functions on the SmartNIC can pose further challenges, particularly with memory-intensive functions potentially impacting cache utilization for others, and compute-intensive functions potentially causing head-of-line blocking at accelerators \cite{qiu2020clara}. All these factors often lead to unexpected performance fluctuations when migrating a function to a SmartNIC. While benchmarking the program will produce performance results, it requires that the program be already developed on the SmartNIC.  
\\
\noindent\textit{Current and Future Initiatives: } Performance prediction can help the developer gain insight prior to porting the code to the hardware. Clara \cite{qiu2020clara} predicts the performance of an unported function on a hypothetical SmartNIC target. Initially, it constructs a model for a given SmartNIC. Then, it creates performance profiles for that SmartNIC by conducting hardware microbenchmarks, which encompass tests on memory latency, accelerator throughput, etc. Clara then creates a code and examines it to identify segments that could be fully offloaded to the SmartNIC. It evaluates the optimal mapping by incorporating constraints derived from the logical NIC model, performance parameters, and code segments. By resolving these constraints, Clara can establish a mapping that optimizes performance after porting. Finally, Clara tests with a PCAP file and assesses how packets would traverse the mapping, thereby providing predictions regarding latency and throughput.

\subsection{Poor Security Isolation}
Commodity SmartNICs suffer from poor isolation between offloaded functions and between functions and data center operators \cite{zhou2024smartnic}. This limitation is a result of the limited access controls on the NIC memory and the absence of virtualization for hardware accelerators. These shortcomings compromise the robustness and security of individual functions, especially in a multi-tenant environment. Additionally, any buggy or compromised code within the NIC poses a risk to all other functions running on it. Concrete attacks on popular SmartNICs including packet corruption, DPI rules stealing, and IO bus denial of service, are presented in \cite{zhou2024smartnic}. \\
\noindent\textit{Current and Future Initiatives: } Zhou et al. \cite{zhou2024smartnic} proposed S-NIC, a hardware design that enforces disaggregation between resources. S-NIC isolates functions at both the ISA level and the microarchitectural level. This ensures integrity and confidentiality, as well as mitigating against side-channel attacks. The design is cost-effective and requires minimal changes to the hardware (e.g., die area). However, it still incurs modest degradation in the performance. Future work could explore alternative architectures that have less impact on performance, or other software-based techniques to isolate the resources.

\begin{figure}[t]
  \centering
  \includegraphics[width=0.489\textwidth]{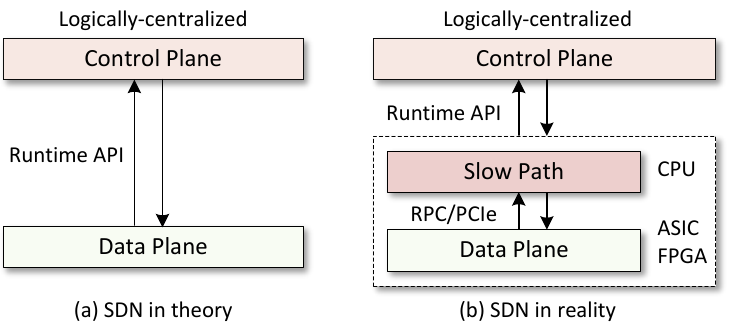}
  \caption{(a) SDN in theory; (b) SDN in reality. Reproduced from \cite{zulfiqar2023slow}.}
  \label{fig:slow_path}  
\end{figure} 

\subsection{Slow Path Bottleneck}
Over recent years, there has been a continuous improvement in the performance of packet-processing data planes, leading to their predominant implementation in hardware such as SmartNICs and programmable switches. Yet, there has been a lack of focus on the slow path, the interface between the data plane and the control plane, which is traditionally considered non-performance critical. The slow path is responsible for handling a subset of traffic that requires special processing (complex control flow, compute, memory resources). These tasks cannot be executed on the data plane, see Fig. \ref{fig:slow_path}. The slow path is executed on the CPU cores, whether on the host or the SmartNIC.

Lately, the slow path is becoming a major bottleneck, driven by the surge in physical network bandwidth and the increasing complexity of network topologies. There is a growth in slow-path traffic in tandem with user traffic. \\
\noindent\textit{Current and Future Initiatives: }  There is a need to re-evaluate the current approach to balancing workload distribution between the data plane and the slow path. Zulfiqar et al. \cite{zulfiqar2023slow} articulated the limitations of the current slow path and argued that the solution is to have a domain-specific accelerator for the slow path. A challenge with creating such an accelerator is to design a generic architecture with common primitives that support most of the slow path use cases. Ideally, the accelerator would have predictable response times, fast table updates, and support large memory pools. Further, the paper advocates extending the match-action model found in most packet processing devices to match-compute for the slow path.

\subsection{ML Offload Complexity}
Offloading the training or the inference in ML from the CPU/GPU to SmartNICs comes with a set of challenges that limit the scalability and innovation of the deployed models.

\begin{itemize}[leftmargin=*]
    \item Accuracy vs Compatibility tradeoff. Some hardware architectures do not support floating-point numbers and complex operations, which are required by advanced ML models, such as neural networks. Workarounds that are proposed to overcome these limitations come at the expense of sacrificing the accuracy of the ML model.

    \item Restriction on the adopted ML algorithm. Despite the continuous exploration of deploying ML models, such as neural networks and decision trees in SmartNICs, a multitude of algorithms, such as Principal Component Analysis (PCA), Genetic Algorithms, are yet to be explored. Additionally, models that are currently deployed are static and any update to the model requires temporarily halting the programmable network device until the new model is compiled and pushed.
    
    \item Flexibility of aggregate functions: In the context of training ML models, the traditional aggregate functions are ‘min’, ‘max’, ‘count’, ‘sum’, and ‘avg’. However, over time, several approaches started adopting and providing user-defined aggregate functions. Implementing such functions over some hardware architectures used in SmartNICs is not straightforward.
\end{itemize}

\noindent\textit{Current and Future Initiatives: } Migration of functionality is one technique that can overcome the restrictions of updating the data plane on the fly. For instance, before the programmable network processor is updated, its functionalities are migrated to another device so that network communication is not interrupted.
To deal with the lack of support of floating-points, approaches such as \cite{simpson2022revisiting} translate floating-point numbers to integers using quantization (i.e., a fixed-point representation of decimal numbers). Such technique can also be used in complex neural network models that need to be simplified to fit in the data plane.
To reduce communication overhead, Ma et al. \cite{ma2022fpga} compresses the parameters (i.e., gradients) before sharing them in the network. Such approaches can enhance network performance, especially when numerous networking devices are cooperating.

\subsection{Lack of Training Resources}
There is an evident lack of detailed documentation and training resources that adequately cover SmartNIC programming and configuration. While some vendors may provide reference applications, basic documentation, and training courses (e.g., \cite{doca_course}), they often fall short of providing the in-depth explanations and hands-on experience that developers need. This makes it difficult for newcomers to understand the intricacies of SmartNIC development and configuration. 

\noindent\textit{Current and Future Initiatives: } 
To address this issue, it is essential for vendors to invest in creating comprehensive training materials, including detailed documentation, tutorials, and hands-on labs. These resources should cover various aspects of SmartNIC programming and configuration, from basic concepts to advanced techniques. Additionally, vendors could offer interactive online courses or workshops led by experienced instructors to provide personalized guidance and support for learners. Some YouTube channels are posting the latest advances and updates on SmartNICs (e.g., STH \cite{sth}, SNIA \cite{snia_yt}, OPI \cite{opi_yt}). However, they are still not comprehensive enough to allow a beginner to start experimenting with SmartNICs.

\section{Conclusion}\label{sec:conclusion}
The evolution of computing has encountered significant challenges with the end of Moore's Law and Dennard Scaling. The emergence of SmartNICs, which combine various domain-specific processors, represents a pivotal shift towards offloading infrastructure tasks and improving network efficiency. This paper has filled a critical void in the literature by providing a comprehensive survey of SmartNICs, encompassing their evolution, architectures, development environments, and applications. The paper has delineated the wide array of functions offloaded to SmartNICs, spanning network, security, storage, and compute tasks. The paper has also discussed the challenges associated with SmartNIC development and deployment, and pinpointed key research initiatives and trends that could be explored in the future. Evidence suggests that SmartNICs are poised to become integral components of every network infrastructure. Smaller networks, which often lack deep technical expertise, can leverage SmartNICs for offloading routine infrastructure tasks. On the other hand, larger and research-oriented networks, with experienced developers, will leverage SmartNICs for offloading complex tasks that are not well-suited for general-purpose CPUs.

\section*{Acknowledgement}
This work is supported by the National Science Foundation (NSF), Office of Advanced Cyberinfrastructure
(OAC), under grant numbers 2118311, 2403360, and 2346726.

\ifCLASSOPTIONcaptionsoff
  \newpage
\fi

\begin{table}[t!]

\caption{Abbreviations used in this article.}
\label{tab:abbrev}
\begin{minipage}{0.01\textwidth}
\begin{tabular}{p{2cm}  p{6cm}}
\hline
\multicolumn{1}{l}{Abbreviation} & \multicolumn{1}{c}{Term} \\
\hline
\hline
ACL & Access Control List\\

AES & Advanced Encryption Standard \\
ALU & Arithmetic Logic Unit\\
ANOVA &  Analysis of Variance\\
API & Application Programming Interface\\
    ASIC & 
    Application Specific Integrated Circuit\\
    BCC & 
    BPF Compiler Collection\\
    BPF & 
    Berkeley Packet Filter\\
    CLI & 
    Command Line Interface\\
    CMS &  
    Count-min Sketch\\
    CPU & 
    Central Processing Unit\\
    DNN & 
    Deep Neural Network \\
    DIP & 
    Dynamic IP\\
    DOCA & 
    Data Center-on-a-Chip Architecture\\
    DPDK & 
    Data Plane Development Kit\\
    DPI & 
    Deep Packet Inspection\\
    DPU & 
    Data Processing Unit\\
    DRAM & 
    Dynamic Random Access Memory\\       
    eBPF & 
    Extended Berkeley Packet Filter\\
    ESnet & 
    Energy Sciences Network\\
    FPGA & 
    Field Programmable Gate Array\\
    GPU & 
    Graphics Processing Units\\
    GRE & 
    Generic Routing Encapsulation\\
    GUI & 
    Graphical User Interface\\
    HDL & 
    Hardware Description Language\\
    HPC & 
    High Performance Computing\\
    IDE & 
    Integrated Development Environment\\
    IDS & 
    Intrusion Detection System\\
    IP & 
    Internet Protocol\\
    IPDK & 
    Infrastructure Programmer Development Kit\\
    IPU & 
    Infrastructure Processing Unit\\
    IPS & 
    Intrusion Prevention System\\
    IPSec & 
    Internet Protocol Security\\
    IT & 
    Information Technology\\
    JBOF & 
    Just a Bunch of Flash\\
    KPI & 
    Key Performance Indicators\\
    kTLS & 
    Kernel TLS\\
    LAN & 
    Local Area Network\\
    LUT & 
    Lookup Table\\
    LPM & 
    Longest Prefix Matching\\
    LSB & 
    Least Significant Bit\\
    MBR & 
    Maximum Bit Rate\\
    ML & 
    Machine Learning\\
    NAS & 
    Network Attached Storage\\
    NAT & 
    Network Address Translation\\
    NFV & 
    Network Function Virtualization\\
    NGFW & 
    Next-Generation Firewall\\
    NIC & 
    Network Interface Card\\
    NLP & 
    Natural Language Processing\\
    NVMe & 
    Non-Volatile Memory Express\\
    NVMe-oF & 
    Non-Volatile Memory Express over Fabric\\
    OFS & 
    Open FPGA Stack\\
    OPAE & 
    Open Programmable Acceleration Engine\\
    OPI & 
    Open Programmable Infrastructure\\

    OS & 
    Operating System\\
    OvS & 
    Open vSwitch\\
    P4 & 
    Programming Protocol-independent Packet Processor\\
    PCIe & 
    Peripheral Component Interconnect Express\\
    PISA & 
    Protocol Independent Switch Architecture\\
    PMD & 
    Poll Mode Driver\\
    PNA & 
    Portable NIC Architecture\\
    PSA & 
    Portable Switch Architecture\\
    QoS & 
    Quality of Service\\
    RAM & 
    Random Access Memory\\
    RAN & 
    Radio Access Network\\
    RDMA & 
    Remote Direct Memory Access\\
    RPC & 
    Remote Procedure Call\\
 
    RSS & 
    Receive Side Scaling\\
    RTL & 
    Register Transfer Level\\
    SAN & 
    Storage Area Network\\

    \hline
\end{tabular}
\end{minipage} \hfill
\end{table}


\begin{table}[t!]
\begin{minipage}{0.5\textwidth}
\begin{tabular}{p{1.5cm}  p{6cm}}
\hline
\multicolumn{1}{l}{Abbreviation} & \multicolumn{1}{c}{Term} \\
\hline
\hline

    SDK & 
    Software Development Kit\\
    SDN & 
    Software Defined Network\\

    SoC & 
    System on a Chip\\
    SPAN & 
    Switched Port Analyzer\\
    SPDK & 
    Storage Performance Development Kit\\
    SSD & 
    Solid State Drives\\
    SVM &
    Support Vector Machine \\
    TCP & 
    Transmission Control Protocol\\
    TLS & 
    Transport Layer Security\\
    TM & 
    Traffic Manager\\
    TRNG & 
    True Random Number Generator\\
    TSO & 
    TCP Segmentation Offload\\
    uBPF & 
    Userspace BPF\\
    UE & 
    User Equipment\\
    UPF & 
    User Plane Function\\
    URL & 
    Uniform Resource Locator\\
    VIP & 
    Virtual IP\\
    VM & 
    Virtual Machine\\
    VPP & 
    Vector Packet Processor\\
    VTEP & 
    VXLAN Tunnel End Point\\
    VXLAN & 
    Virtual Extensible LAN\\
    XDP & 
    eXpress Data Path\\
    xPU & 
    Auxiliary Processing Unit\\
\hline
\end{tabular}
\end{minipage} 
\end{table}



%
\bibliographystyle{ieeetr} 
\bibliography{Ref}

%

\begin{IEEEbiography}
[{\includegraphics[width=1in,clip,keepaspectratio]{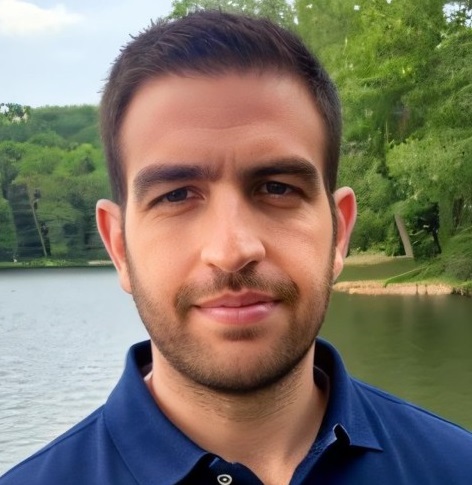}}]{Elie Kfoury}
received the Ph.D. degree in Informatics from the University of South Carolina (USC), in 2023. He is currently an assistant professor in the Integrated Information Technology department at USC. As a member of the Cyberinfrastructure Laboratory, he developed training materials using virtual labs on high-speed networks, TCP congestion control, programmable switches, SDN, and cybersecurity. He is the co-author a book “High-Speed Networks: A Tutorial”, that is being used nationally for deploying, troubleshooting, and tuning Science DMZ networks. His research interests include P4 programmable data planes, computer networks, cybersecurity, and Blockchain. He previously worked as a research and teaching assistant in the computer science department at the American University of Science and Technology in Beirut.
\end{IEEEbiography}

\begin{IEEEbiography}
[{\includegraphics[width=1in,clip,keepaspectratio]{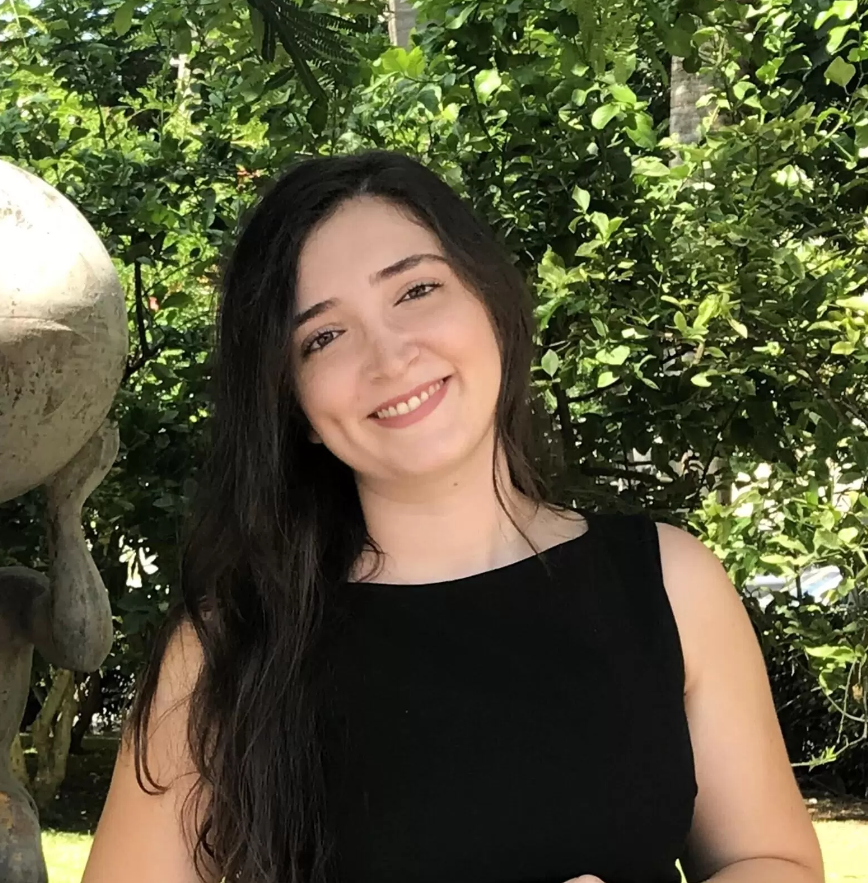}}]{Samia Choueiri}
is a Ph.D. student in the College of Engineering and Computing at the University of South Carolina (USC). Her research interests include SmartNICs, P4 switches, cybersecurity, and robotics.  She received her Masters in Computer and Communications Engineering with emphasis in Mechatronics Engineering from the American University of Science and Technology in Beirut, where she also was a teaching assistant and lab instructor.
 
\end{IEEEbiography}

\begin{IEEEbiography}
[{\includegraphics[width=1in,clip,keepaspectratio]{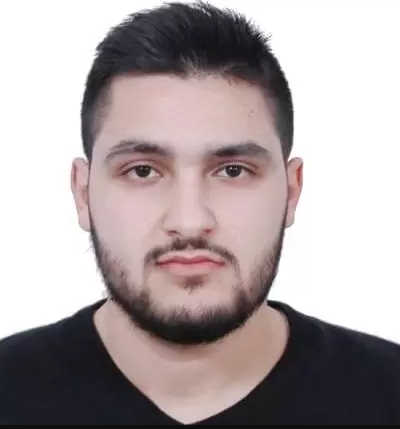}}]{Ali Mazloum}
is a Ph.D. student in the College of Engineering and Computing at the University of South Carolina (USC) in the United States of America. Prior to joining USC,
he received his bachelor’s in computer science from the American University of Beirut (AUB). His research focuses on P4 programmable data planes, SmartNICs, cybersecurity, network measurements, and traffic engineering.
 
\end{IEEEbiography}

\begin{IEEEbiography}
[{\includegraphics[width=1in,clip,keepaspectratio]{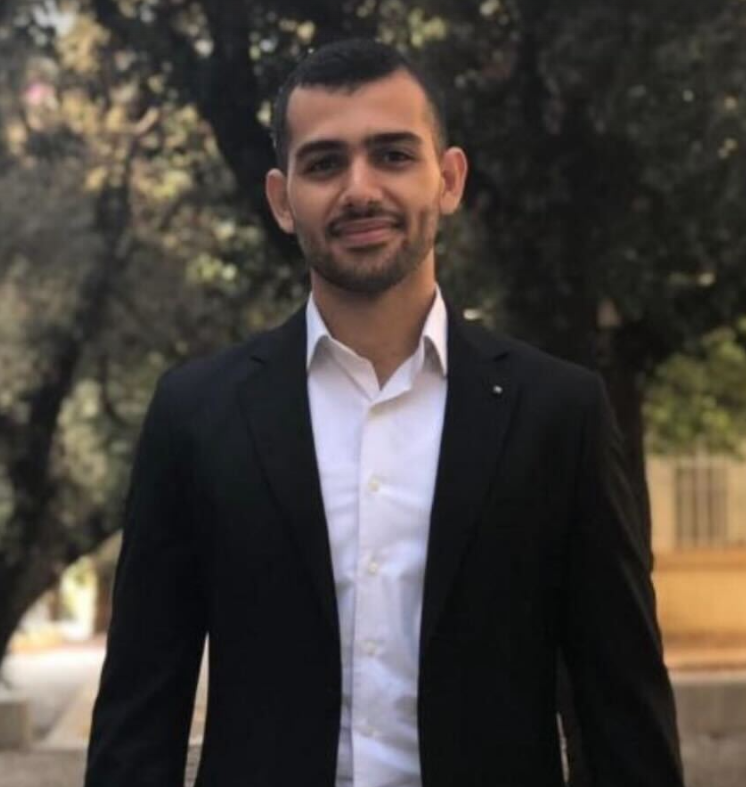}}]{Ali AlSabeh}
is currently a Ph.D. student in the College
of Engineering and Computing at the University of South Carolina, USA. He is a member of the CyberInfrastructure Lab (CI Lab), where he developed training materials for virtual labs on network protocols (BGP, OSPF) and their applications (BGP attributes, BGP hijacking, IP spoofing,
etc.), as well as SDN (OpenFlow, interconnecting SDN with legacy networks, etc.). He previously earned his M.S. degree
in Computer Science from the American University of Beirut, where he also worked as a graduate research assistant and
teacher assistant. His area of research focuses on malware analysis, network security, and P4 programmable switches.
 
\end{IEEEbiography}

\begin{IEEEbiography}
[{\includegraphics[width=1in,clip,keepaspectratio]{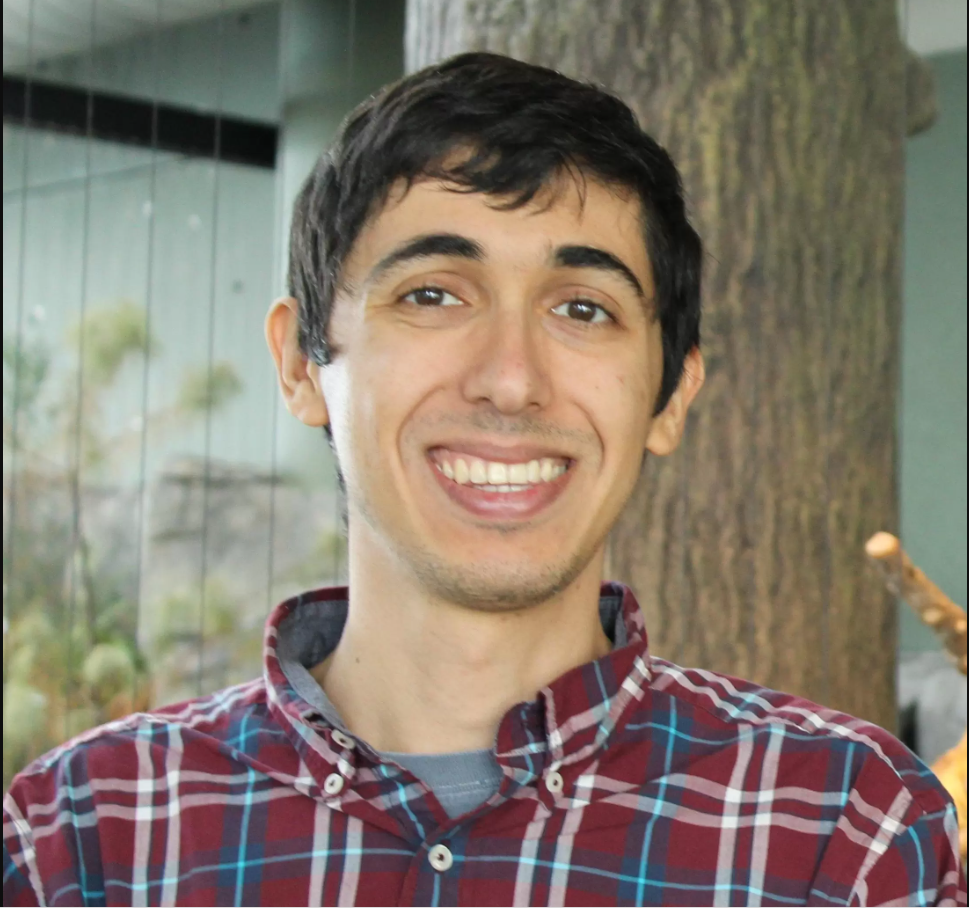}}]{Jose Gomez}
is a Ph.D. student in the College of Engineering and Computing at the University of South Carolina. Jose’s research focuses on P4 programmable data planes, TCP congestion control, passive measurements, and buffer sizing. Currently, Jose is working at the Cyberinfrastructure lab developing a system based on P4 switches to enable programmability in non-programmable networks.
\end{IEEEbiography}

\begin{IEEEbiography}
[{\includegraphics[width=1in,clip,keepaspectratio]{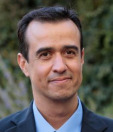}}]{Jorge Crichigno}
is a Professor in the College of Engineering and Computing at the University of South Carolina (USC). He has over 15 years of experience in the academic and industry sectors. Prior to joining USC, Dr. Crichigno was an Associate Professor and Chair of the Department of Engineering at Northern New Mexico College. Dr. Crichigno's research focuses on the practical implementation of high-speed networks and network security. These include the design and implementation of high-speed switched networks, TCP optimization, experimental evaluation of congestion control algorithms tailored for friction-free environments, and scalable flow-based intrusion detection systems. His work has been funded by Google, NSF, and the Department of Energy. He received his PhD in Computer Engineering from the University of New Mexico in 2009, and his Bachelor degree in Electrical Engineering from the Catholic University of Paraguay in 2004.

\end{IEEEbiography}

\end{document}